\documentclass[%
aps,prd,
twocolumn,
amsmath,amssymb,
superscriptaddress,
]{revtex4-2}

\usepackage[T1]{fontenc}
\usepackage[utf8]{inputenc}
\usepackage{amsmath,amsthm,amssymb}
\usepackage{mathtools}
\usepackage{enumerate}
\usepackage{xspace}
\usepackage{longtable}
\usepackage{booktabs}
\usepackage{afterpage}
\usepackage{comment}
\usepackage{color}
\usepackage{lmodern}
\usepackage{array}
\usepackage{graphicx}
\usepackage{float}

\usepackage[pdfusetitle,bookmarksopen]{hyperref}

\graphicspath{{plots/},{figs/}}

\def\ba#1\ea{\begin{align}#1\end{align}}
\newcommand{\nn}{\nonumber}
\newcommand{\ra}{\rangle}
\newcommand{\la}{\langle}

\begin{document}

\title{
Hadronic light-by-light contribution to the muon anomaly from lattice QCD
with infinite volume QED
at physical pion mass
}

\date{\today}

\newcommand\bnl{Physics Department, Brookhaven National Laboratory, Upton, NY 11973, USA}
\newcommand\bnlccs{Computational Science Initiative, Brookhaven National Laboratory, Upton, NY 11973, USA}
\newcommand\cu{Physics Department, Columbia University, New York, NY 10027, USA}
\newcommand\pu{School of Computing \& Mathematics, Plymouth University, Plymouth PL4 8AA, UK}
\newcommand\riken{RIKEN-BNL Research Center, Brookhaven National Laboratory, Upton, NY 11973, USA}
\newcommand\regensburg{Fakult\"at f\"ur Physik, Universit\"at Regensburg, Universit\"atsstra{\ss}e 31, 93040 Regensburg, Germany}
\newcommand\edinb{School of Physics and Astronomy, The University of Edinburgh, Edinburgh EH9 3FD, UK}
\newcommand\uconn{Physics Department, University of Connecticut, Storrs, CT 06269-3046, USA}
\newcommand\soton{School of Physics and Astronomy, University of Southampton,  Southampton SO17 1BJ, UK}
\newcommand\york{Mathematics \& Statistics, York University, Toronto, ON, M3J 1P3, Canada}
\newcommand\cssm{CSSM, University of Adelaide, Adelaide 5005 SA, Australia}
\newcommand\cern{CERN, Theoretical Physics Department, Geneva, Switzerland}
\newcommand\mib{Dipartimento di Fisica, Universit\'a di Milano-Bicocca, Piazza della Scienza 3, I-20126 Milano, Italy}
\newcommand\infn{INFN, Sezione di Milano-Bicocca, Piazza della Scienza 3, I-20126 Milano, Italy}
\newcommand{\ucb}{University of California, Berkeley, CA 94720, USA}
\newcommand{\lbnl}{Lawrence Berkeley National Laboratory, Berkeley, CA 94720, USA}
\newcommand{\eic}{Electron-Ion Collider, Brookhaven National Laboratory, Upton, NY 11973, USA}
\newcommand{\cpthree}{CP$^3$-Origins \& Department of Mathematics and Computer Science, University of Southern Denmark, Campusvej 55, 5230 Odense M, Denmark}

\author{Thomas~Blum}%
\email{thomas.blum@uconn.edu}
\affiliation{\uconn}

\author{Norman~Christ}
\affiliation{\cu}

\author{Masashi~Hayakawa}
\affiliation{Department of Physics, Nagoya University, Nagoya 464-8602, Japan}
\affiliation{Nishina Center, RIKEN, Wako, Saitama 351-0198, Japan}

\author{Taku~Izubuchi}
\affiliation{\bnl}\affiliation{\riken}

\author{Luchang~Jin}
\email{ljin.luchang@gmail.com}
\affiliation{\uconn}
\affiliation{\riken}

\author{Chulwoo~Jung}
\affiliation{\bnl}

\author{Christoph~Lehner}
\affiliation{\regensburg}

\author{Cheng~Tu}
\affiliation{\uconn}

\collaboration{RBC and UKQCD Collaborations}
\noaffiliation
 
\begin{abstract}
    The hadronic light-by-light scattering contribution to the muon anomalous magnetic moment,
    $(g-2$)/2, is computed in the infinite volume QED framework with lattice QCD.
    We report $a_\mu^\text{HLbL}=12.47(1.15)(0.95) \times 10^{-10}$
    where the first error is statistical and the second systematic.
    The result is mainly based on the 2+1 flavor M\"obius domain wall fermion ensemble with
    inverse lattice spacing $a^{-1} = 1.73~\mathrm{GeV}$,
    lattice size $L=5.5~\mathrm{fm}$,
    and $m_\pi = 139~\mathrm{MeV}$, generated by the RBC-UKQCD collaborations.
    The leading systematic error of this result comes from the lattice discretization.
    This result is consistent with previous determinations.
\end{abstract}

\maketitle

\section{Introduction}
\label{sec:intro}

Muons are spin-$1/2$ charged particles with non-zero magnetic moment:
\ba
\boldsymbol{\mu} = - g \frac{e}{2 m} \bf S,
\ea
where $\bf S$ is the particle's spin, $e$ and $m$ are the electric charge and mass, respectively, and $g$ is the Land\'e $g$-factor.
The Dirac equation predicts that $g = 2$, exactly,
so any difference from 2 must arise from interactions.
The magnetic moment of a fermion can be defined in terms of
its electromagnetic form factors in the zero momentum transfer limit.
Lorentz and gauge symmetries tightly constrain the form of the interactions.
In Euclidean space-time:
\ba
\label{eq:ff}
&\langle \mu ({\bf p^\prime},\,s^\prime) | J_\nu(0) |\mu({\bf p},\,s)\rangle
\\
&\hspace{-0.5cm}=
-e \bar u_{s^\prime}({\bf p^\prime})\left(F_1(q^2)\gamma_\nu+i\frac{F_2(q^2)}{4 m}[\gamma_\nu,\gamma_\rho] q_\rho\right)u_s({\bf p}),
\nn
\ea
where $J_\nu$ is the electromagnetic current, and $F_1$ and $F_2$ are form factors, giving the charge and magnetic moment at zero momentum transfer ({$q^2=(p^\prime-p)^2=0$}), or static limit. The ${u}_s(\bf{p})$ and $\bar u_s({\bf p})$ are Dirac spinors. The anomalous part of the magnetic moment is given by $F_2(0)$ alone, and is known as the anomaly,
\ba
a_\mu \equiv F_2(q^2 = 0) =  \frac{g-2}{2}.
\ea

The muon anomalous magnetic moment is one of the most precisely determined quantities in particle physics.
Compared with the electron anomalous magnetic moment, which is determined with higher accuracy,
the muon is expected to be much more sensitive to new physics at very large energy scales due to its heavier mass.
Two experiments Fermilab (E989)~\cite{Muong-2:2021ojo} and J-PARC (E34)~\cite{Abe:2019thb,Sato:2021aor}
are aiming at even higher precision in measuring the muon $g-2$.
The initial result released by Fermilab (E989)~\cite{Muong-2:2021ojo}
confirmed the previously best result obtained by the BNL E821 experiment
\cite{Bennett:2006fi} and reduced the experimental
uncertainty from 0.54~ppm to 0.46~ppm.
The final goal of the Fermilab experiment is to reduce the uncertainty further to approximately
0.14~ppm.
The J-PARC experiment adopts a very different measurement strategy
and will serve as an important cross-check.
Its final accuracy goal is 0.45~ppm for statistical uncertainty and 0.07~ppm for systematic uncertainty.

\begin{figure}
    \centering
    \includegraphics[width=0.2\textwidth]{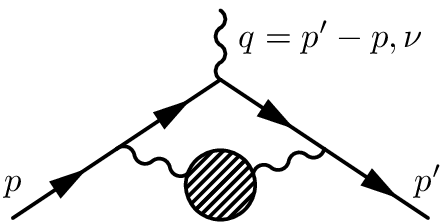}~
    \includegraphics[width=0.2\textwidth]{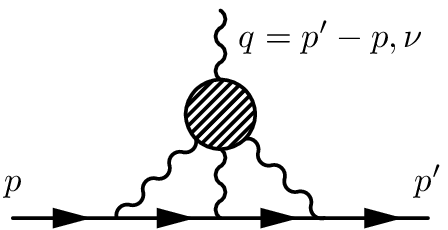}
    \caption{
    \label{fig:hvp-hlbl-blob}The HVP (left) and HLbL (right) contribution to muon $g-2$.
    The shaded circles represent {hadronic interactions}.}
\end{figure}

The Standard Model result provided by the Muon $g-2$ Theory Initiative \cite{Aoyama:2020ynm,Aoyama:2012wk,Aoyama:2019ryr,Czarnecki:2002nt,Gnendiger:2013pva,Davier:2017zfy,Keshavarzi:2018mgv,Colangelo:2018mtw,Hoferichter:2019mqg,Davier:2019can,Keshavarzi:2019abf,Kurz:2014wya,Melnikov:2003xd,Masjuan:2017tvw,Colangelo:2017fiz,Hoferichter:2018kwz,Gerardin:2019vio,Bijnens:2019ghy,Colangelo:2019uex,Blum:2019ugy,Colangelo:2014qya}
currently has an uncertainty of 0.37~ppm and is in $4.2\sigma$ tension with the experimental value.
The two leading sources of uncertainty come from
the leading QCD contributions, hadronic vacuum polarization (HVP) and hadronic light-by-light (HLbL) scattering,
both illustrated in Fig.~\ref{fig:hvp-hlbl-blob}.

For a long time, the HLbL contribution has been estimated by hadronic models~\cite{Prades:2009tw,Nyffeler:2009tw,Jegerlehner:2009ry,Jegerlehner:2017lbd}.
In Ref.~\cite{Green:2015sra,Gerardin:2017ryf},
the hadronic light-by-light scattering amplitudes
estimated by hadronic models
are compared with lattice QCD calculations.
The result from hadronic models is difficult to improve further.
More recently, the HLbL contribution has been calculated with the dispersion relation approach~\cite{%
Melnikov:2003xd,Masjuan:2017tvw,Colangelo:2017fiz,Hoferichter:2018kwz,Gerardin:2019vio,Bijnens:2019ghy,Colangelo:2019uex,Pauk:2014rta,Danilkin:2016hnh,Jegerlehner:2017gek,Knecht:2018sci,Eichmann:2019bqf,Roig:2019reh,Blum:2019ugy,Colangelo:2014qya}.
These results are summarized in the 2020 white paper~\cite{Aoyama:2020ynm},
\ba
a_\mu^\text{HLbL,WP2020,pheno}\times 10^{10}
=
9.2(1.9)
\ea
Within the dispersion relation framework, the pseudo-scalar pole is the leading contribution to HLbL.
This contribution is defined in terms of the pseudo-scalar transition form factors to two (possibly off-shell) photons, $\pi^0,\eta,\eta' \to \gamma^*\gamma^*$.
These form factors can also be calculated with lattice QCD.~\cite{Gerardin:2016cqj,Gerardin:2019vio,Alexandrou:2022qyf,Burri:2022gdg}

The first direct lattice QCD calculation of the HLbL contribution was performed by the RBC-UKQCD collaborations.~\cite{blum:2014oka}
Then, we made many improvements to the calculation method and also included
the contribution from the leading disconnected
diagrams.~\cite{Blum:2015gfa,Blum:2016lnc}
In our previous work~\cite{Blum:2019ugy},
we applied the method developed
to lattice gauge ensembles with different lattice spacings and
spatial volumes, all at physical pion mass.
The calculations incorporate both QCD and QED
in the same finite volume lattice using the QED$_L$ scheme~\cite{hayakawa:2008an}.
We extrapolated the results to infinite volume and the continuum limit.
The final result was
\ba
a_\mu^\text{HLbL,RBC-UKQCD2019}\times 10^{10}
=
7.87(3.06)_\text{stat}(1.77)_\text{syst} \,.
\ea
This was the first lattice QCD result for HLbL with all systematic effects controlled.
The statistical error was still larger than
the phenomenological results obtained with dispersion relations
or the previous hadronic model estimation.

The Mainz group pioneered the QED$_\infty$ approach
and pre-calculated the QED kernel semi-analytically
in infinite volume~\cite{Green:2015sra,Asmussen:2016lse}.
We also explored this strategy and independently calculated
the infinite volume QED weighting function~\cite{Blum:2017cer}.
In that work, we found that while the QED$_\infty$ approach ensures an exponentially suppressed
finite volume error, the size of the finite volume error can still be significant.
However, making use of the current conservation property
of the hadronic four point function,
we designed subtractions to the infinite volume QED weighting functions
to reduce the finite volume errors and also the discretization errors.
The subtracted infinite volume QED weighting function is used in the current work.

The effectiveness of the subtraction is confirmed by the Mainz group
and a different subtraction scheme was adopted in their calculation.~\cite{Chao:2020kwq,Chao:2021tvp,Chao:2022xzg,Asmussen:2016lse}.
Their calculation was performed with pion masses ranging from $200~\mathrm{MeV}$
to $420~\mathrm{MeV}$ and extrapolated to the physical point.
All sub-leading disconnected diagrams were carefully
calculated and were found to be consistent with zero.
The charm quark contribution from the connected and disconnected diagrams
were calculated as well.
Finally, a statistically more precise result was obtained:
\ba
a_\mu^\text{HLbL,Mainz2021}\times 10^{10}
=
10.96(1.59).
\ea

In this work, we present our latest lattice calculation
of the HLbL contribution to muon $g-2$ with subtracted infinite volume
QED weighting function as developed in our previous work~\cite{Blum:2017cer}.
The calculation is directly performed at physical pion mass
and therefore eliminates the systematic uncertainty from
chiral extrapolations.
We find that calculating the HLbL contribution directly at the physical pion mass
is considerably more difficult than a calculation at a heavier pion mass
due the larger contribution and statistical fluctuations
from the long distance region.
Following the suggestion of Ref.~\cite{Chao:2021tvp},
we focus on the calculation of the connected
and leading disconnected diagrams.
The main calculation is performed with a single lattice spacing
where $a^{-1} = 1.73~\mathrm{GeV}$.
While we use M\"obius domain wall fermions (MDWF) to eliminate
$\mathcal O(a)$ discretization errors,
the remaining $\mathcal O(a^2)$ effects
are estimated to be the largest source of systematic uncertainty
of this calculation.

The paper is organized as follows: in section \ref{sec:framework} and \ref{sec:lattice-details},
we review the calculation method developed in our previous works
and describe new techniques used in this calculation.
We then show our numerical results in section \ref{sec:results}
and conclude in section \ref{sec:conclusion}.

\section{Theoretical framework and calculation method}
\label{sec:framework}

We base our current calculation on the framework already set up
in our previous works~\cite{Blum:2015gfa,Blum:2016lnc,Blum:2017cer}.
In this work, we combine the method used in Ref.~\cite{Blum:2015gfa,Blum:2016lnc}
to calculate the hadronic four point function with lattice QCD
with the subtracted infinite volume QED weighting function developed in Ref.~\cite{Blum:2017cer}.
We have tested the QED weighting function by calculating
the leptonic light-by-light contribution to muon $g-2$ and have reproduced
the results obtained by analytical calculation~\cite{Laporta:1992pa,Laporta:1991zw}.
We outline the method below and focus on the improvements
we made in this work.
\begin{figure}[h]
\centering
\includegraphics[width=0.15\textwidth]{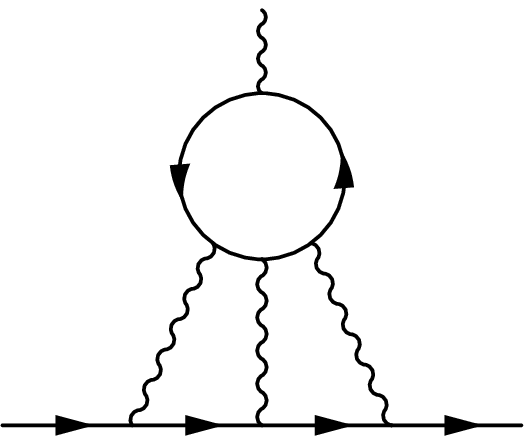}~~
\includegraphics[width=0.15\textwidth]{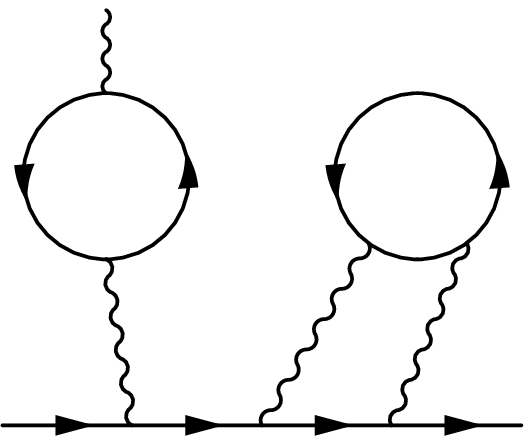}\\
\includegraphics[width=0.13\textwidth]{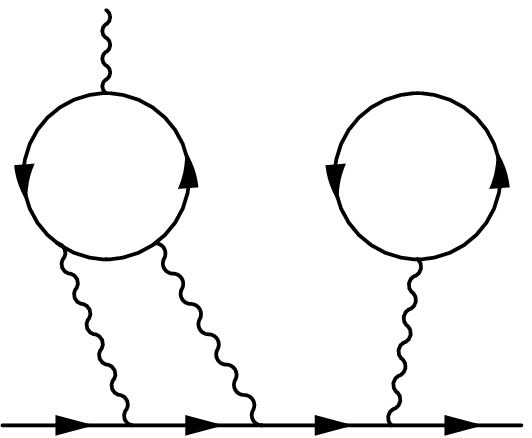}~~
\includegraphics[width=0.13\textwidth]{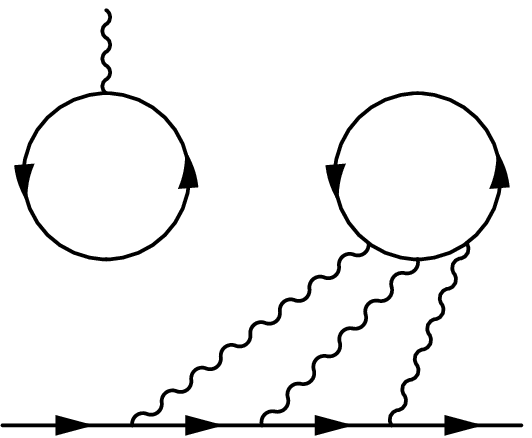}~~
\includegraphics[width=0.13\textwidth]{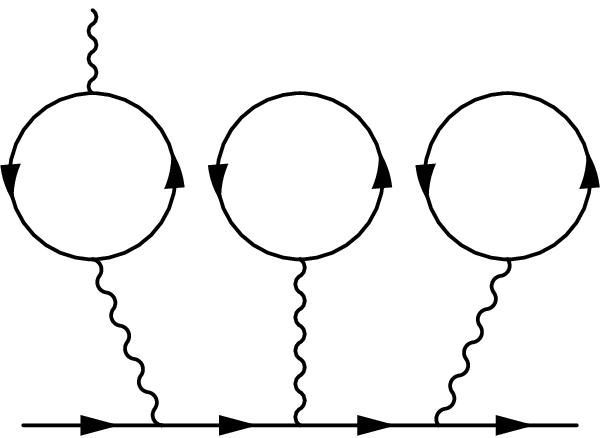}\\
\includegraphics[width=0.13\textwidth]{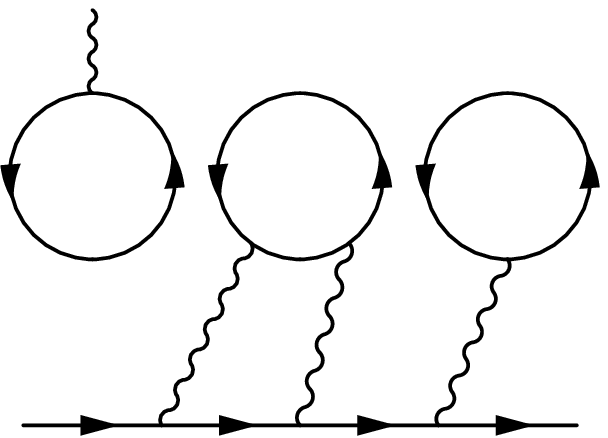}~~
\includegraphics[width=0.15\textwidth]{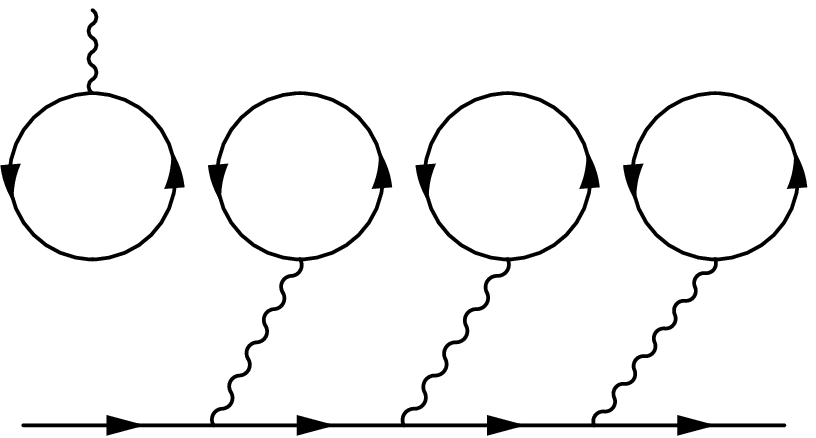}
\caption{\label{fig:hlbl-diagrams}Diagrams contributing to the muon anomaly. The diagrams in the top row are the leading ones, and do not vanish in the $SU(3)$ flavor limit. Strong interactions to all orders, including gluons connecting the quark loops and sea quark loops which are not connected by photons, are not shown.}
\end{figure}

All of the diagrams contributing at $O(\alpha^3)$ to HLbL scattering are shown in Fig.~\ref{fig:hlbl-diagrams}. We use the term (quark-) ``connected diagram'' to refer to the one on the left on the top row and ``disconnected diagram'' to refer to the one on the right. The remaining diagrams, which are all suppressed in the flavor SU(3) limit, are referred to as the ``sub-leading disconnected diagrams''.
We only explicitly draw quark loops that are connected
to photons. Gluons and sea quark loops that are not connected to photons are not shown in the figure
but are included automatically in dynamical lattice QCD calculations.
To make non-zero contributions to the muon $g-2$,
the quark loops in the disconnected diagrams must be connected by gluons.

The contribution to the muon $g-2$ can be calculated with the combination
of the hadronic four-point function $\mathcal H$ and the QED weighting function $\mathcal G$~\cite{Blum:2017cer}:
\ba
  & a_\mu^\text{HLbL} \frac{e}{m}
  \bar u_{s'}(\vec 0) \frac{\Sigma_i}{2} u_s(\vec 0)
  \\
  & \hspace{-0.5cm} =
  \frac{1}{VT}
  \sum_{x_\text{op}}
  \sum_{x,y,z}
  \frac{1}{2} \epsilon_{i, j, k}
  \big(x_\text{op} - x_\text{ref}(x,y,z)\big)_j
  \nn\\&\hspace{0.2cm}  \times
  i^3 e^6
  \mathcal H_{k,\rho,\sigma,\lambda}(x_\text{op},x,y,z)
   \bar u_{s'}(\vec 0) \mathcal{G}_{\rho, \sigma, \lambda} (x, y, z) u_s(\vec 0),
   \nn
\ea
where $\bar u_{s'}(\vec 0)$, $u_s(\vec 0)$ are Dirac spinors for the outgoing and incoming muon
in the diagram, respectively. $V$ stand for the spatial volume of the lattice and $T$ stand for the size of the temporal extent of the lattice. $\Sigma_k = \epsilon_{i,j,k} \gamma_i\gamma_j /(2i)$ is the $4\times 4$ version of the Pauli matrix, $\sigma_k$.
From the spin structure of the muon particle, we can obtain the expression
for $a_\mu^\text{HLbL}$:
\ba
\label{eq:hlbl-master}
a_\mu^\text{HLbL}
&=
\frac{2 m e^2}{3}
\frac{1}{VT}
\sum_{x_\text{op}}
\sum_{x,y,z}
\frac{1}{2}
\epsilon_{i,j,k}
\big(x_\text{op} - x_\text{ref}(x,y,z)\big)_j
\nn\\
&\times (6 e^4) \mathcal H_{k,\rho,\sigma,\lambda}(x_\text{op},x,y,z)
\mathcal M_{i,\rho,\sigma,\lambda}(x,y,z)
\ea
where
\ba
\label{eq:qed-weighting-function}
\mathcal M_{i,\rho,\sigma,\lambda}(x,y,z)
=
\frac{1}{2} %
\mathrm{Tr}
\Big[
\frac{1}{6} %
i^3\mathcal G_{\rho,\sigma,\lambda}(x,y,z)
\Sigma_i
\Big] \,.
\ea
\begin{figure}[h]
\centering
\includegraphics[width=0.35\textwidth]{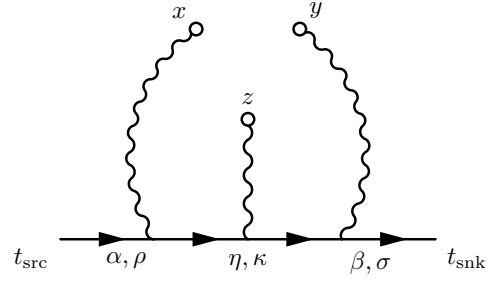}
\caption{\label{fig:qed-weighting-function}Diagramatic representation of the QED weighting function defined in Eq.~\ref{eq:qed-weighting-function}, following Ref.~\cite{Blum:2017cer}.}
\end{figure}
The QED weighting function $\mathcal G$ is shown diagramatically in Fig.~\ref{fig:qed-weighting-function}
and is expressed in terms of the free muon and Feynman gauge photon propagators, $S_{{\mu}}$ and $G$:
\ba
  \mathfrak{G}_{\sigma, \kappa, \rho} (y, z, x)
  & =
  \lim_{t_{\text{src}}
  \rightarrow - \infty, t_{\text{snk}} \rightarrow \infty} e^{m_{{\mu}}
  \left( t_{\text{snk}} - t_{\text{src}} \right)}
  \\&\hspace{-2cm}\times
  \int_{\alpha, \beta, \eta,\vec{x}_{\text{snk}}, \vec{x}_{\text{src}}}
  G(x, \alpha) G (y, \beta) G (z, \eta)
  \nn\\&\hspace{-1.5cm}\times
  S_{{\mu}}
  \left( x_{\text{snk}}, \beta \right) i \gamma_{\sigma} S_{{\mu}} (\beta,
  \eta) i \gamma_{\kappa} S_{{\mu}} (\eta, \alpha) i \gamma_{\rho}
  S_{{\mu}} \left( \alpha, x_{\text{src}} \right) \,.
  \nn
\ea
As is well known, the above expression contains an infrared divergence that
vanishes after projection to its magnetic part.
We can also remove this infrared divergent piece by
the following procedure:
\ba
  \mathfrak{G}^{(1)}_{\sigma, \kappa, \rho} (y, z, x) & =
  \frac{1}{2}
  \mathfrak{G}_{\sigma, \kappa, \rho} (y, z, x)
  \\& + \frac{1}{2}
  [\mathfrak{G}_{\rho, \kappa, \sigma} (x, z, y)]^{\dagger} \,.
  \nn
\ea
In addition we can perform somewhat arbitrary subtractions
to this infinite volume QED weighting function without changing the final result due to vector current conservation
satisfied by the hadronic four point function,
\ba
  \label{eq:qed-weighting-func-subtraction}
  \mathfrak{G}^{(2)}_{\sigma, \kappa, \rho} (y, z, x)
  & =
  \mathfrak{G}^{(1)}_{\sigma, \kappa, \rho} (y, z, x)
  \\&\hspace{-0.5cm}
  -\mathfrak{G}^{(1)}_{\sigma, \kappa, \rho} (z, z, x)
  -\mathfrak{G}^{(1)}_{\sigma, \kappa, \rho} (y, z, z).
  \nn
\ea
Note that $\mathfrak{G}^{(1)}_{\sigma, \kappa, \rho} (z, z, z) = 0$, so
 this subtraction significantly reduces the size of
the QED weighting function when $|x-z|$ or $|y-z|$ is small.
This is the region where the hadronic function from the lattice calculation
has the largest discretization error.
It turns out that this subtraction greatly reduces the discretization error.
This is the major finding of Ref.~\cite{Blum:2017cer}.
We should also note that the subtraction does impact the integrand
and partial sum.
In Ref.~\cite{Chao:2021tvp}, a modified subtraction
scheme is used, so their integrand and partial sum cannot
be directly compared with ours.
Finally, we include all possible permutations of the subtracted QED weighting function which are required for the total contribution to the muon $g-2$:
\ba
  i^3 \mathcal{G}_{\rho, \sigma, \kappa} (x, y, z) & =
  \mathfrak{G}^{(2)}_{\rho,\sigma, \kappa} (x, y, z)
  +\mathfrak{G}^{(2)}_{\sigma, \kappa, \rho} (y, z, x)
  \\ & \hspace{0cm}
  +\mathfrak{G}^{(2)}_{\kappa, \rho, \sigma} (z, x, y)
  +\mathfrak{G}^{(2)}_{\kappa, \sigma, \rho} (z, y, x)
  \nn\\ & \hspace{0cm}
  +\mathfrak{G}^{(2)}_{\rho, \kappa, \sigma} (x, z, y)
  +\mathfrak{G}^{(2)}_{\sigma, \rho, \kappa} (y, x, z).
  \nn
\ea
Another component of the master formula Eq.~\ref{eq:hlbl-master}
is  $x_\text{ref}$, the reference position
for the moment method to calculate the magnetic moment.
Again, due to current conservation,
there are many possible choices for $x_\text{ref}$.
In this work,
we use the following choice for the connected diagram:
\ba
x_\text{ref}(x,y,z)
&= x_\text{ref-far}(x,y,z)
\\
&\hspace{-1cm}
=
\left\{
\begin{array}{l}
    x \qquad \text{if } | y - z | < \min(| x - y |, | x - z |)\\
    y \qquad \text{if } | x - z | < \min(| x - y |, | y - z |)\\
    z \qquad \text{if } | x - y | < \min(| x - z |, | y - z |)\\
    \frac{1}{3}(x + y + z) \qquad \text{otherwise}
\end{array}
\right.
\nn
\ea
We make a slightly different choice of $x_\text{ref}$, Eq.~(\ref{eq:x-ref-discon}), for the disconnected diagrams.
The rationale will be described in the later part of this section.
We use $\mathcal H$ to denote the hadronic four point function:
\ba
\label{eq:hadronic-four-point}
&(6 e^4) \mathcal H_{k,\rho,\sigma,\lambda}(x_\text{op},x,y,z)
\nn\\
&\hspace{2cm}=
\la T J_k(x_\text{op}) J_\rho(x) J_\sigma(y) J_\lambda(z) \ra_\text{QCD}
\ea
\ba
J_\nu(x) = \sum_{q=u,d,s,c} e_q Z_V \bar\psi_q(x) \gamma_\nu \psi_q(x)
\ea
where $Z_V$ is the lattice local vector current renormalization constant.
After Wick contraction, $\mathcal H$ can be expressed
as the sum of different types of diagrams as illustrated in Fig.~\ref{fig:hlbl-diagrams}.
For the disconnected
and sub-leading disconnected diagrams,
we require the quark loops to be connected by gluons.
\ba
\mathcal H
=
\mathcal H^\text{con}
+\mathcal H^\text{discon}
+\mathcal H^\text{sub-leading-discon}
+\mathcal H^\text{charm}
\ea
where $\mathcal H^\text{con}$ includes the light and strange quark connected diagrams,
$\mathcal H^\text{discon}$ the light and strange quark disconnected diagrams,
$\mathcal H^\text{sub-leading-discon}$ the light and strange quark sub-leading disconnected
diagrams (vanish in the flavor SU(3) limit), and
$\mathcal H^\text{charm}$ all diagrams involving charm quark loops.

\begin{figure}[h]
\centering
\includegraphics[width=0.35\textwidth]{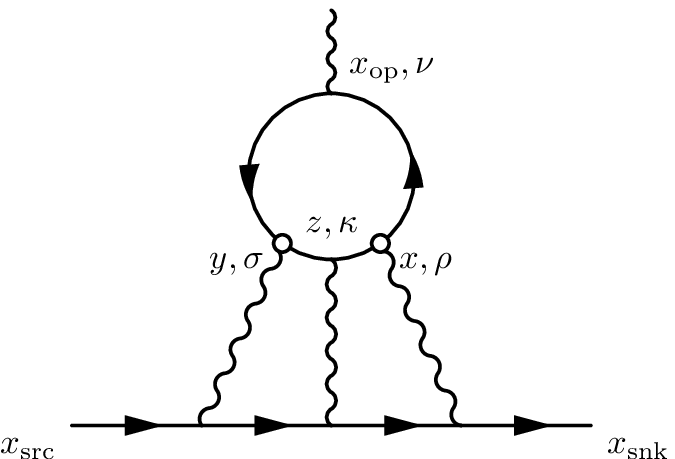}
\includegraphics[width=0.35\textwidth]{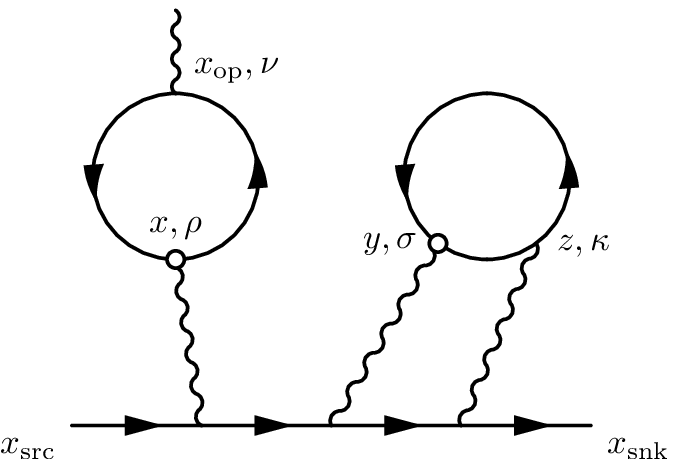}
\caption{\label{fig:hlbl-psrc}Two point-source propagators are used to calculate the hadronic four-point function.
The locations of the point-sources are indicated in the diagrams as small circles.}
\end{figure}

Naturally, after the Wick contraction,
$\mathcal H$ defined in Eq.~(\ref{eq:hadronic-four-point})
can be expressed in terms of quark propagators and includes all permutations of $x$, $y$, and $z$.
However, note that all other terms
in the master formula Eq.~(\ref{eq:hlbl-master})
are symmetric with respect to permutations of $x$, $y$, and $z$ (along with its Lorentz indices). Therefore, we are allowed to calculate only a subset of the diagrams generated by the Wick contractions in Eq.~(\ref{eq:hadronic-four-point})
and multiply them with appropriate factors.
\ba
\label{eq:hadronic-four-point-con}
&\hspace{0.5cm}
6 e^4 \mathcal H^\text{con}_{\nu,\rho,\sigma,\kappa}(x_\text{op},x,y,z)
\nn\\&
\Rightarrow
6 e^4 \mathcal H^\text{con-no-perm}_{\nu,\rho,\sigma,\kappa}(x_\text{op},x,y,z)
\\&
=-6
\Big\la
\mathrm{Re}
\sum_{q=u,d,s} e_q^4
\mathrm{Tr}\Big(
\gamma_\nu S_q(x_\text{op},x)
\gamma_\rho S_q(x,z)
\nn\\&\hspace{2.0cm}\times
\gamma_\kappa S_q(z,y)
\gamma_\sigma S_q(y,x_\text{op})
\Big)
\Big\ra_\text{QCD},
\nn
\ea
\ba
\label{eq:hadronic-four-point-discon}
&\hspace{0.5cm}
6 e^4 \mathcal H^\text{discon}_{\nu,\rho,\sigma,\kappa}(x_\text{op},x,y,z)
\nn\\&
\Rightarrow
6 e^4 \mathcal H^\text{discon-no-perm}_{\nu,\rho,\sigma,\kappa}(x_\text{op},x,y,z)
\\&
=
3
\Big\la
\sum_{q'=u,d,s} e_{q'}^2
\mathrm{Tr}\Big(\gamma_\nu S_{q'}(x_\text{op},x)
\gamma_\rho S_{q'}(x,x_\text{op})
\Big)
\nn\\&\hspace{1.0cm}\times
\sum_{q=u,d,s} e_{q}^2
\mathrm{Tr}\Big(\gamma_\kappa S_{q}(z,y)
\gamma_\sigma S_{q}(y,z)
\nn\\&\hspace{2.0cm}
-\big\la
\gamma_\kappa S_{q}(z,y)
\gamma_\sigma S_{q}(y,z)
\big\ra_\text{QCD}
\Big)
\Big\ra_\text{QCD},
\nn
\ea
where $S_q$ denotes the quark propagator.
As shown in Fig.~\ref{fig:hlbl-psrc},
we calculate the hadronic four-point function with two point-source propagators.
The above trick can make the evaluation more efficient~\cite{Blum:2015gfa,Chao:2020kwq}.
We note that
$\mathcal H^\text{discon-no-perm}$,
the disconnected diagram
of the hadronic four-point function
without permutation,
still satisfies the current conservation
condition in the continuum limit
(this is not true for
$\mathcal H^\text{con-no-perm}$)
which leads to the following relation
in the infinite volume limit:
\ba
\sum_{x_\text{op}}
\mathcal H^\text{discon-no-perm}_{k,\rho,\sigma,\lambda}(x_\text{op},x,y,z)
= 0.
\ea
We are therefore allowed
to alter the choice of $x_\text{ref}$
for the disconnected diagram:
\ba
\label{eq:x-ref-discon}
x_\text{ref-discon}
= x.
\ea
This choice allows the summation
over $x_\text{op}$ to be performed
independently of coordinates $y$ and $z$.
This choice also has the benefit
of suppressing the contribution
from the region where $|x-x_\text{op}|$
is small and the quark loop
is large before subtracting the
vacuum expectation value.
Also, we note that the new choice
$x_\text{ref-discon}$ is the same as
the initial choice
in the long distance region where
the contribution is dominated by
$\pi^0$-exchange and
$|y-z| \ll \min (|x-y|, |x-z|)$.
This property guarantees that
the connected and disconnected diagrams
have exactly the same QED weighting function
in the long distance region,
which leads to Eq.~\ref{eq:discon-con-ratio}
before taking the continuum limit.
This is the main reason for the choice
of $x_\text{ref}$ for the connected diagrams.

\section{Lattice details}
\label{sec:lattice-details}

The calculation presented here is performed on ensembles of gauge fields generated by the RBC and UKQCD collaborations~\cite{RBC:2014ntl}.
The main calculation is carried out using the 48I ensemble, a physical-mass ensemble generated with 2+1 flavors of M\"obius domain wall fermions (MDWF).
We also use a few other ensembles to calculate various corrections and estimate systematic effects. The relevant information about the 48I ensemble and other ensembles is listed in Table~\ref{tab:ensembles}.
We always use the MDWF action.
For most of the ensembles, the quarks have their physical masses.

\begin{table}[h]
\begin{center}
\begin{tabular}{c|ccccc}
\hline
& 48I & 64I & 24D & 32D & 24DH  \\
\hline
$m_\pi$ (MeV) & 139 & 135 & 142 & 142 &  341 \\
$a^{-1}$ (GeV) & 1.730 & 2.359 & 1.015 & 1.015 & 1.015 \\
$a$ (fm) & 0.114 &0.084 & 0.194 & 0.194  & 0.194 \\
$L$ (fm)  & 5.47 & 5.38 & 4.67 & 6.22 & 4.67 \\
$L_s$ & 24 & 12 & 24 & 24  & 24 \\
\# meas & 113 & 64 & 31 & 70 & 37 \\
\hline
\end{tabular}
\caption{2+1 flavors of MDWF gauge field ensembles generated by the RBC-UKQCD collaborations~\cite{RBC:2014ntl}. 
The labels indicate the lattice size in lattice units and the QCD gauge action,
where ``I'' stands for Iwasaki, and ``D'' stands for Iwasaki+DSDR.
DSDR stands for dislocation-suppressing-determinant-ratio, which is an additional term
in the action to soften explicit chiral symmetry breaking effects, needed in particular for very coarse lattices~\cite{Renfrew:2008zfx}.
The pion mass $m_\pi$, lattice spacing $a$\footnote{After the analysis for this work was completed, we updated the lattice spacing for the 24D and 32D ensembles. The new lattice spacing is $a^{-1} = 1.023~\mathrm{GeV}$ and makes no material difference to the results presented here.}, spatial extent $L$, extra fifth dimension size $L_s$, and the number of QCD configurations used are listed.
}
\label{tab:ensembles}
\end{center}
\end{table}

For our main calculation
on the 48I ensemble,
we use 113 configurations.
On each configuration, 
we randomly sample
2048 uniformly distributed
distinct points
and calculate light quark point-source
propagators for each of the selected points.

Among the above 2048 points
we randomly select 1024 points and also
calculate strange quark propagators.
These 2048 + 1024 point-source propagators are computed using the (deflated-) conjugate gradient algorithm with a sloppy stopping condition.
To further speed up the sloppy propagator calculation, we use the zM\"obius domain wall fermion formulation~\cite{PhysRevD.96.014501} with reduced
fifth dimension size $L_s = 14$ 
to approximate the unitary MDWF action used in the gauge ensemble generation ($L_s = 24$).
For deflation of the Dirac operator we reuse the eigenvectors generated for the calculation of the hadronic vacuum polarization contribution
to the muon $g-2$~\cite{RBC:2018dos,Blum:2023qou} using the locally-coherent Lanczos approach \cite{clark:2017wom}.
Then, we use a two-level all-mode-averaging (AMA) method~\cite{Blum:2012uh}
to correct the bias caused by the inexact propagators.
The AMA method requires a portion of the sloppy propagators be computed again with higher precision.
In the first step a more precise version of the light and strange quark propagators are calculated on a randomly selected set of 64 points among the 1024 points. In the second step
among the 64 points, we randomly select 16 points and calculate these propagators to full precision.
These propagators are used to calculate both the connected and the disconnected diagrams in the bias correction.
By using two steps, we are able to reduce the statistical error coming from the bias correction to a level well below the statistical error for the sloppy part.

For the connected diagrams, we sample two-point pairs among all the possible $2048 \times (2048 - 1) / 2 + 2048 = 2098176$ combinations (including the case where $x = y$).
For each of the two-point pairs, we can perform the contraction as described in Eq.~(\ref{eq:hadronic-four-point-con}) in the previous section.
As the number of all possible pairs is enormous, the cost to perform all the contractions is not affordable.
Therefore, we only calculate the contraction for a subset of the available two-point pairs.
We sample the subset with the empirical probability $p(r)$, which is a function of the distance between the two points, $r = |x - y|$,
\ba
p(r) =
\left\{
\begin{array}{ll}
\frac{N_\text{psrc} - 1}{2 (L^3 T - 1)} & \text{if }r = 0
\\
1 & \text{if } 8 \ge r > 0
\\
\frac{1}{(r/8)^3} & \text{if } L \ge r > 8
\\
0 & \text{if } r > L
\end{array}
\right.,
\ea
where $L = 48$, $T=96$, and the numbers are in lattice units.
On average, about $57,000$ light quark point-source propagator pairs per configuration
are sampled to calculate the connected diagrams.
The reason we sample long-distance point-pairs less often is because the connected hadronic four point function and its statistical fluctuation
decrease with distance for long distances.
For each pair, due to the sampling procedure, the probability of that pair taking a particular relative coordinate
is a function of $r$.
The total connected diagram contribution is equal to the
expectation value of the contraction for the point pairs multiplied by the inverse probabilities.
We refer to the inverse probability as the weight $w(r)$: 
\ba
w(r) =
\left\{
\begin{array}{ll}
w_0 & \text{if }r = 0
\\
w_0 / p(r) & \text{if } r > 0
\end{array}
\right.,
\ea
\ba
w_0 = \sum_x p(|x|).
\ea
To further reduce the contraction cost and the storage cost of saving these propagators,
we use field sparsening techniques~\cite{Detmold:2019fbk,Li:2020hbj}.
We randomly sample $1/16$ points among all the points of the lattice and only perform contractions
on these randomly selected points.
Since we only need propagator values on these points, we only save these values to disk.
To accommodate the sparsening, we multiply a factor of $16^2$ for the summation over $z$ and $x_\text{op}$ in Eq.~(\ref{eq:hlbl-master}),
except when $z = x_\text{op}$, where we multiply by $16$.
In this way the sampling procedure
does not introduce any systematic effects on our central value.

For the disconnected diagrams,
the two point-source locations are also indicated in Fig.~\ref{fig:hlbl-psrc}.
Different from the connected diagrams, the statistical
fluctuations of the disconnected hadronic functions
are not suppressed when $r=|x -y|$ increases.
Therefore, we use all possible combinations of point pairs of $x$ and $y$ as long as $|x - y| \le L$
to estimate the summation over $x$ and $y$ in Eq.~(\ref{eq:hlbl-master}).
Thanks to the choice of $x_\text{ref} = x_\text{ref-discon} = x$ in Eq.~(\ref{eq:x-ref-discon}),
the summation over $x_\text{op}$ can be performed for each point-source propagator with source location at $x$,
the result can be used for all possible values of $y$ locations.
However, the summation of $z$ depends on the location of $x$ and $y$ due to the QED weighting function.
Therefore, we need to perform the summation over $z$ for each pair of points.

To make the contractions of all point pairs affordable,
we aggressively sparsen when summing over $z$. Fortunately, the vertex $z$ and vertex $y$ are connected by two quark propagators, and
the four point function is exponentially suppressed when the distance between
$z$ and $y$ increases.

Naturally, we can sample the $z$ locations based on the distance between $z$ and $y$,
similar to the sampling of $x$ and $y$ combinations in the calculation of the connected diagrams.
However, for this disconnected diagram, we discovered a much more efficient
adaptive sampling scheme as follows.
For each point-source location $y$, we calculate the following square norm $n(z,y)$ of the quark loop
for all $z$:
\ba
n(z, y)
&=
\sum_{\kappa,\sigma}
\Big|\mathrm{Tr}\Big(\gamma_\kappa S_{q}(z,y)
\gamma_\sigma S_{q}(y,z)
\nn\\&\hspace{1.0cm}
-\big\la
\gamma_\kappa S_{q}(z,y)
\gamma_\sigma S_{q}(y,z)
\big\ra_\text{QCD}
\Big)\Big|^2
\nn
\ea
Unlike the full contraction, the above norm is independent of the location of $x$, and the location of $z$ is sampled with the following (empirical) probability
\ba
p_y(z)
=
\left\{
\begin{array}{ll}
1 & \text{if } n(z, y) \ge t_0^2 \text{ and } |z-y| \le L \\
\sqrt{n(z,y)} / t_0  & \text{if } n(z, y) < t_0^2 \text{ and } |z-y| \le L \\
0  & \text{if } |z-y| > L
\end{array}
\right.,
\ea
where $t_0$ is the threshold parameter to control the overall sample frequency.
With this probability distribution, we sample more often points where
the (norm of) the quark loop is large which is much more efficient than basing the probability on the distance $|z - y|$. We choose $t_0 = 5 \times 10^{-5}$ in this work for the 48I ensemble.

For the light quark loop, there are $9,156,431$ points within the $|z-y|\le L$ range. Among these, we sample about $17,000$ points for $z$ on average.
For each of the sampled $z$, we then loop over all possible $x$ locations
with $|x -y| \le L$
and perform the full contraction.
Due to the sampling procedure, we multiply the result
of the contraction by the inverse of the sample probability $1 / p_y(z)$
to obtain an unbiased final result.
While the sampled $z$ points amount to less than $0.2\%$ of the possible $z$ points in the allowed range, we expect the final statistical precision
to be almost unaffected by the adaptive sampling procedure.
That is, we expect the final statistical error would be almost
the same even if we had calculated the contraction using all the $z$ locations.
This expectation is based on a quick test run with much lower threshold
$t_0 = 10^{-3}$, which corresponds to about a factor of $20$ smaller number of sampled points; we find the final statistical error is roughly the same.

In contrast to the connected diagram, we do not sparsen the propagators.
To save disk space, we temporarily save the following intermediate contraction
for each point-source propagator:
\ba
\mathrm{Tr}\Big(\gamma_\kappa S_{q}(z,y)
\gamma_\sigma S_{q}(y,z)
\Big).
\ea
For each propagator on each lattice site,
the above contraction corresponds to $4\times 4$ complex numbers
while the full propagator takes $12\times 12$ complex numbers.

\section{Results}
\label{sec:results}

In this section we display several figures of the summand (integrand) or its partial-sum as a function of $R_\text{max}$:
\ba
\label{eq:rmax-def}
R_\text{max} = \text{max}(|x - y|, |y - z|, |x - z|),
\ea
where $x$, $y$, and $z$ represent the positions of the three internal  quark-photon vertices. %
Partial-sums are obtained by summing the corresponding summand from $0$ up to the specified value of $R_\text{max}$ (inclusive). The rightmost value in a partial-sum plot corresponds to the total contribution.

Since the summand decreases exponentially at large $R_\text{max}$, we expect the partial-sum to approach a plateau for large enough $R_\text{max}$. We usually use the result of the partial-sum at $4~\mathrm{fm}$, which includes all contributions from the region $R_\text{max} < 4~\mathrm{fm}$.
In the lattice calculation, we calculate the partial sum with respect to $R_\text{max}$ and
save the partial sum for all integer values (in lattice units) of the upper limit of $R_\text{max}$.
To compare with results calculated using ensembles with different lattice spacings,
we can linearly interpolate the partial sum for any upper limit of $R_\text{max}$.
In this work, we choose to interpolate at integer multiples $\times\, 0.1~\mathrm{fm}$.
The summand is obtained at half-integer multiples $\times\, 0.1~\mathrm{fm}$ by taking the difference of the linearly interpolated partial sum.
Therefore, the data points of the plots shown in this work always have spacing $0.1~\mathrm{fm}$,
and the data points of the integrand plots always have its x-axis value equal to half-integer times $0.1~\mathrm{fm}$.
In this work, we present our results for the anomalous magnetic moment in the unit of $10^{-10}$.

\subsection{Light quark contribution}
\label{sec:light-contrib}

\begin{figure}[h]
\centering
\includegraphics[width=0.4\textwidth]{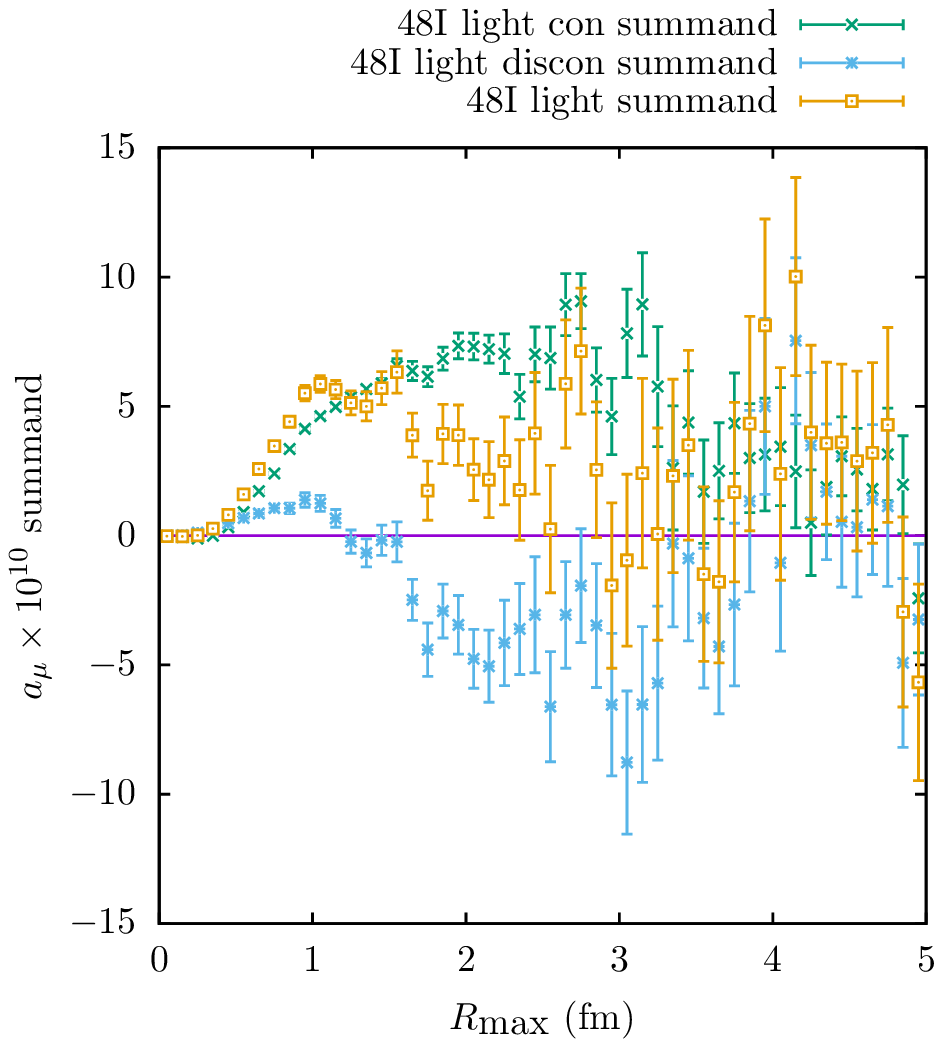}
\includegraphics[width=0.4\textwidth]{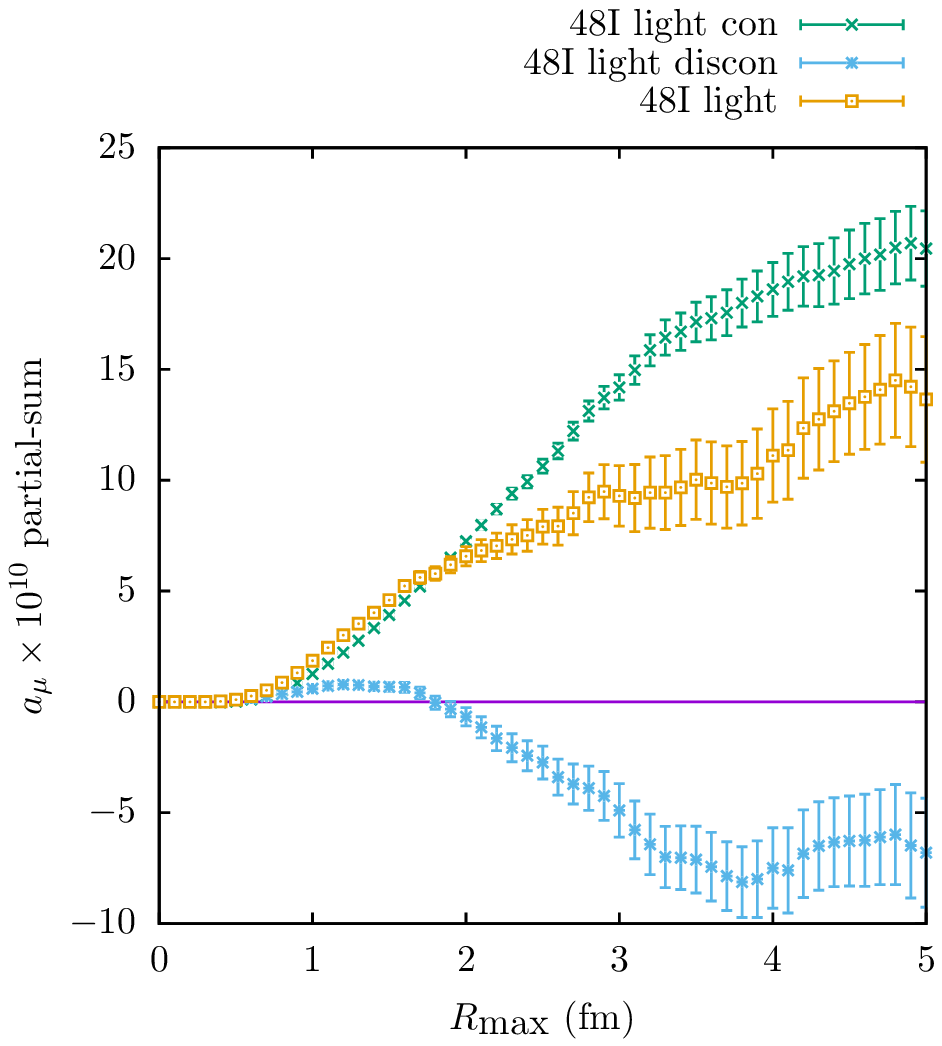}
\caption{\label{fig:48I-light}Light quark contributions computed on the 48I ensemble from the connected diagrams,
the disconnected diagrams, and the total. The upper plot shows the corresponding summands and the lower plot shows the partial sum.
}
\end{figure}

\begin{figure}[h]
\centering
\includegraphics[width=0.4\textwidth]{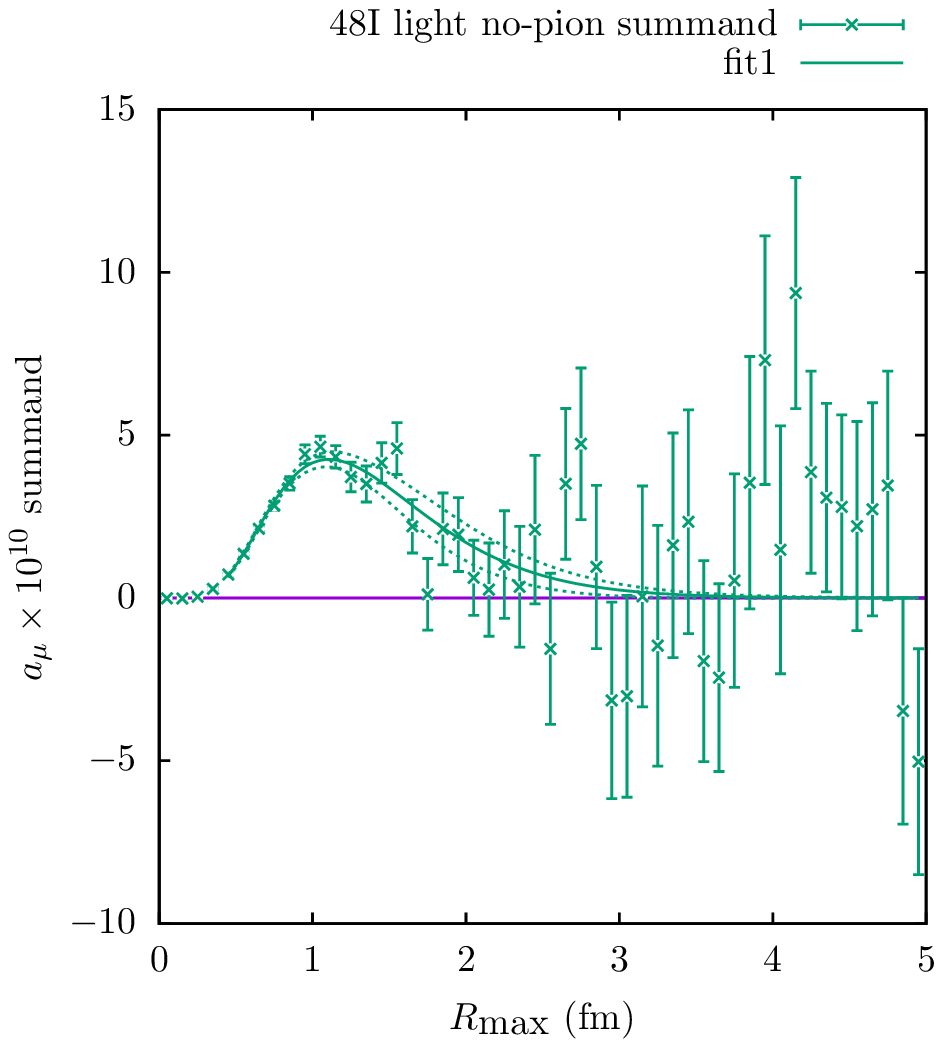}
\includegraphics[width=0.4\textwidth]{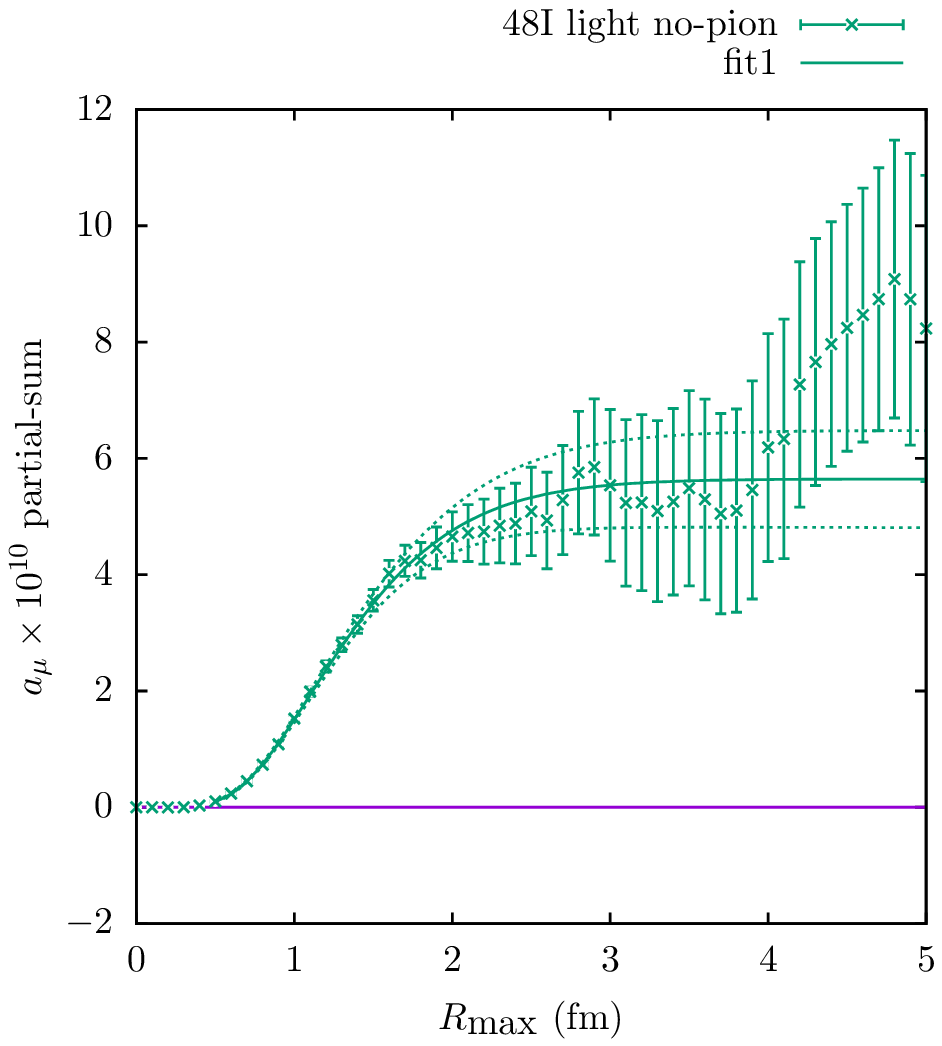}
\caption{\label{fig:48I-light-no-pion}Light quark contributions from a special combination
of the connected and disconnected diagrams to cancel the long-distance $\pi^0$ exchange
contribution (see Eq.~\ref{eq:a-mu-no-pion}).
The upper plot shows the summand and the lower plot shows the partial sum.
The combined summand vanishes much faster than the light quark diagrams by themselves.
Solid curves represent a fit of the data to Eq.~(\ref{eq:fit-form}).
The fit starts at $0.5~\mathrm{fm}$.
The dashed lines denote the statistical uncertainty of the fit.
}
\end{figure}

The contributions from the connected diagrams, the disconnected diagrams, and the sum of the two contributions are shown in Fig.~\ref{fig:48I-light}. As can be seen from the figure, the statistical error grows as $R_\text{max}$ increases. However, for both the connected and the disconnected diagrams, there are non-negligible contributions that come from regions where $R_\text{max}$ is as large as 3 or 4 fm. The statistical error is quite significant. Making the situation worse, the contributions from the long-distance region of the connected and the disconnected diagrams are opposite in sign, while the statistical error of the connected and the disconnected diagrams are largely independent. As a result, the relative uncertainty on the sum is much larger than that for the individual contributions.
Results are given in Tab.~\ref{tab:48I-light} as ``48I light con $R_\text{max}<4\mathrm{fm}$'', ``48I light discon $R_\text{max}<4\mathrm{fm}$'', and ``48I light $R_\text{max}<4\mathrm{fm}$''.

To improve the situation, note that the large contribution and the cancellation between the connected and disconnected diagrams at long distance are due to $\pi^0$ exchange and are well-understood theoretically~\cite{Bijnens:2016hgx,Jin:2016rmu,Chao:2021tvp}. It has been shown that, at large $R_\text{max}$, the ratio of the disconnected and the connected hadronic four-point function is $-25 / 34$.
In our present computational setup, we use the same infinite volume QCD weighting function for both the connected and disconnected diagrams, and use the same variable $R_\text{max}$ to study the partial-sum of the connected and disconnected diagrams.\footnote{One can use different setup for the connected and disconnected diagrams, in which case this relation will not hold exactly. For example, Ref.~\cite{Chao:2021tvp} uses different setups.}
Therefore the same ratio applies to the contribution to $a_\mu$. Formally, we have:
\ba
\label{eq:discon-con-ratio}
\lim_{R\to\infty}
\frac{a_\mu^\text{discon}(R_\text{max} > R)}{a_\mu^\text{con}(R_\text{max} > R)}
=
-\frac{1}{2} \cdot \frac{(e_u^2 + e_d^2)^2}{e_u^4 + e_d^4}
=
-\frac{25}{34}.
\ea
Here, $R_\text{max}$ is defined in Eq.~\eqref{eq:rmax-def}, being a function of the three vertex locations $x$, $y$, $z$ as shown in Fig.~\ref{fig:hlbl-psrc}.
The long-distance contributions for the connected and disconnected diagrams are denoted as $a_\mu^\text{con}(R_\text{max} > R)$ and $a_\mu^\text{discon}(R_\text{max} > R)$.
They can be calculated with the same summation as in Eq.~\eqref{eq:hlbl-master}
but with the additional constraint that $R_\text{max} > R$, where $R$ is the long-distance cutoff.
Note we take the $R\to\infty$ limit in the above equation.
This ratio is exact in this limit and is not affected by lattice artifacts or finite volume effects.
Therefore, we can construct the following combination:
\ba
\label{eq:a-mu-no-pion}
a_\mu^\text{no-pion}
=
a_\mu^\text{discon} + \frac{25}{34} a_\mu^\text{con},
\ea
where the $\pi^0$ exchange contribution to $a_\mu^\text{no-pion}$ vanishes in the long distance. We plot the summand and partial-sum of $a_\mu^\text{no-pion}$ in Fig.~\ref{fig:48I-light-no-pion}. Indeed, the partial sum of $a_\mu^\text{no-pion}$ reaches the plateau much earlier than the connected or disconnected diagrams.
This trick of combining the connected and disconnected with appropriate factors
to obtain a faster plateau was employed in Ref.~\cite{Zhao:2022njd}.
For $a_\mu^\text{no-pion}$, we will use $2.0~\mathrm{fm}$ and $2.5~\mathrm{fm}$ as the upper limit of $R_\text{max}$.
The results are shown in Tab.~\ref{tab:48I-light} as ``48I light no-pion $R_\text{max}<2.5\mathrm{fm}$'' or ``48I light no-pion $R_\text{max}<2\mathrm{fm}$''.

In the upper panel of Fig.~\ref{fig:48I-light-no-pion}, we fit the summand to the following empirical form:
\ba
\label{eq:fit-form}
f(R_\text{max}) = A / \text{fm}^4 \frac{R_\text{max}^6}{R_\text{max}^3 + (C~\text{fm})^3} e^{-B R_\text{max} / (\text{fm}\cdot\text{GeV})}
\ea
The fit range is from 0.5 fm to 4 fm. We use the result of the fit to estimate the long-distance contribution to $a_\mu^\text{no-pion}$.
Since this is a completely empirical fit, we will assign a $100\%$ systematic uncertainty to it.
The ``no-pion" results are also collected in Tab.~\ref{tab:48I-light}.

\begin{table*}[t]
\centering
\begin{tabular}{l|rl}
\hline
Contribution name & $a_\mu$ & $\times 10^{10}$ \\
\hline
48I light con $R_\text{max}<2\mathrm{fm}$ & $7.24$ & $(0.15)_\text{stat}$ 
\\
48I light con $R_\text{max}<2.5\mathrm{fm}$ & $10.64$ & $(0.31)_\text{stat}$ 
\\
48I light con $R_\text{max}<4\mathrm{fm}$ & $18.61$ & $(1.22)_\text{stat}$ 
\\
48I light con $R_\text{max}>4\mathrm{fm}$ & $7.56$ & $(0.42)_\text{stat}(1.06)_\text{syst}~[1.14]$ 
\\
48I light con FV-corr & $-1.79$ & $\phantom{(0.00)_\text{stat}}(0.42)_\text{syst}$ 
\\
48I light con $m_\pi$-corr & $1.32$ & $(0.27)_\text{stat}(0.66)_\text{syst}~[0.72]$ 
\\
48I light con $a^2$-corr & $0.00$ & $\phantom{(0.00)_\text{stat}}(1.49)_\text{syst}$ 
\\
light con & $25.70$ & $(1.33)_\text{stat}(1.99)_\text{syst}~[2.39]$  %
\\
\hline
48I light no-pion $R_\text{max}<2.5\mathrm{fm}$ & $5.09$ & $(0.76)_\text{stat}$ 
\\
48I light no-pion $R_\text{max}<2\mathrm{fm}$ & $4.65$ & $(0.42)_\text{stat}$ 
\\
48I light no-pion $R_\text{max}>2.5\mathrm{fm}$ & $0.31$ & $(0.22)_\text{stat}(0.31)_\text{syst}~[0.38]$ 
\\
48I light no-pion $R_\text{max}>2\mathrm{fm}$ & $0.88$ & $(0.46)_\text{stat}(0.88)_\text{syst}~[0.99]$ 
\\
\hline
48I light discon $R_\text{max}<2\mathrm{fm}$ & $-0.67$ & $(0.41)_\text{stat}$ 
\\
48I light discon $R_\text{max}<2.5\mathrm{fm}$ & $-2.73$ & $(0.74)_\text{stat}$ 
\\
48I light discon $R_\text{max}<4\mathrm{fm}$ & $-7.49$ & $(1.82)_\text{stat}$ 
\\
48I light discon $R_\text{max}<4\mathrm{fm}$ hybrid-2.5fm & $-8.28$ & $(1.31)_\text{stat}(0.31)_\text{syst}~[1.35]$ 
\\
48I light discon $R_\text{max}<4\mathrm{fm}$ hybrid-2fm & $-8.15$ & $(1.24)_\text{stat}(0.88)_\text{syst}~[1.51]$ 
\\
48I light discon $R_\text{max}>4\mathrm{fm}$ & $-5.56$ & $(0.31)_\text{stat}(0.78)_\text{syst}~[0.84]$ 
\\
48I light discon FV-corr & $1.31$ & $\phantom{(0.00)_\text{stat}}(0.31)_\text{syst}$ 
\\
48I light discon $m_\pi$-corr & $-0.98$ & $(0.20)_\text{stat}(0.49)_\text{syst}~[0.53]$ 
\\
48I light discon $a^2$-corr & $0.00$ & $\phantom{(0.00)_\text{stat}}(0.66)_\text{syst}$ 
\\
light discon & $-12.71$ & $(1.87)_\text{stat}(1.17)_\text{syst}~[2.20]$  %
\\
light discon hybrid-2.5fm & $-13.50$ & $(1.36)_\text{stat}(1.21)_\text{syst}~[1.82]$  %
\\
light discon hybrid-2fm & $-13.37$ & $(1.29)_\text{stat}(1.46)_\text{syst}~[1.95]$  %
\\
\hline
48I light $R_\text{max}<2\mathrm{fm}$ & $6.57$ & $(0.43)_\text{stat}$ 
\\
48I light $R_\text{max}<2.5\mathrm{fm}$ & $7.90$ & $(0.78)_\text{stat}$ 
\\
48I light $R_\text{max}<4\mathrm{fm}$ & $11.11$ & $(2.11)_\text{stat}$ 
\\
48I light $R_\text{max}<4\mathrm{fm}$ hybrid-2.5fm & $10.32$ & $(0.99)_\text{stat}(0.31)_\text{syst}~[1.04]$ 
\\
48I light $R_\text{max}<4\mathrm{fm}$ hybrid-2fm & $10.46$ & $(0.89)_\text{stat}(0.88)_\text{syst}~[1.25]$ 
\\
48I light $R_\text{max}>4\mathrm{fm}$ & $2.00$ & $(0.11)_\text{stat}(0.28)_\text{syst}~[0.30]$ 
\\
48I light FV-corr & $-0.47$ & $\phantom{(0.00)_\text{stat}}(0.11)_\text{syst}$ 
\\
48I light $m_\pi$-corr & $0.35$ & $(0.07)_\text{stat}(0.17)_\text{syst}~[0.19]$ 
\\
48I light $a^2$-corr & $0.00$ & $\phantom{(0.00)_\text{stat}}(0.83)_\text{syst}$ 
\\
light total & $12.99$ & $(2.11)_\text{stat}(0.90)_\text{syst}~[2.29]$  %
\\
light total hybrid-2.5fm & $12.20$ & $(1.01)_\text{stat}(0.95)_\text{syst}~[1.38]$  %
\\
light total hybrid-2fm & $12.33$ & $(0.90)_\text{stat}(1.25)_\text{syst}~[1.55]$  %
\\
\hline
\end{tabular}
\caption{\label{tab:48I-light}Light quark contributions computed on the 48I ensemble. Values are different contributions to $a_\mu \times 10^{10}$ where $ a_\mu = (g_\mu - 2) / 2$. The numbers in the square bracket are the statistical and systematic uncertainty combined in quadrature. The tag ``hybrid-'' indicates the same quantity obtained using the hybrid approach.}
\end{table*}

With the definition of $a_\mu^\text{no-pion}$, we obtain:
\ba
\label{eq:discon-no-pion}
a_\mu^\text{discon} &= a_\mu^\text{no-pion} - \frac{25}{34} a_\mu^\text{con},\\
\label{eq:total-no-pion}
a_\mu^\text{total} &= a_\mu^\text{no-pion} + \frac{9}{34} a_\mu^\text{con}.
\ea
The advantage comes in using the early plateau value of $a_\mu^\text{no-pion}$
and combining it with $a_\mu^\text{con}$, which plateaus at much larger $R_\text{max}$.
This is a ``hybrid'' approach to calculate $a_\mu^\text{discon}$ and $a_\mu^\text{total}$, which has a much smaller statistical error than the direct combination.
For $a_\mu^\text{con}$, we will use $4~\mathrm{fm}$ as the upper limit of $R_\text{max}$.
The results are shown in Tab.~\ref{tab:48I-light} as entries that start with ``48I light discon $R_\text{max}<4\mathrm{fm}$ hybrid-'' and ``48I light $R_\text{max}<4\mathrm{fm}$ hybrid-''.

The fitting function form in Eq.~(\ref{eq:fit-form}) is inspired by the function in Eq.~(20) of Ref.~\cite{Chao:2021tvp}.
The functional form used here differs due to the meanings of $R_\text{max}$ and the variable ``$|y|$'' in Ref.~\cite{Chao:2021tvp}. Also, note that the subtraction scheme of the QED weighting function is also different~\cite{Blum:2017cer,Chao:2020kwq,Chao:2021tvp,Asmussen:2022oql}.
We tried using this fit function to fit other contributions. The results are shown in Appendix~\ref{sec:fit-forms}.

The statistical errors are estimated by assuming the results
from each 48I configuration that we performed the measurements are
statistically independent. We also tried to bin the data with different lengths in MD time units.
The results are shown in Table.~\ref{tab:auto-corr}.
We can see that the statistical error does not significantly depend on the binning sizes.
More interestingly, the statistical error calculated by assuming all the individual samples in each configuration are independent is only a little bit smaller than the statistical error
calculated from the fluctuation of different configurations.
This implies that the current statistical errors still mainly come from the sampling of the points
from each configuration, instead of from the gauge field fluctuations sampled by different gauge configurations.
Also, this shows that the strategy using different combinations of the point source propagators as point pairs to calculate the connected diagrams is very effective.

\begin{table*}[t]
    \centering
    \begin{tabular}{l|rl|c|c|c|c}
    \hline
    Contribution name & $a_\mu$ & $\times 10^{10}$ & Bin(20) & Bin(30) & Bin(40) & Bin(sample) \\
    \hline
    48I light con $R_\text{max}<4\mathrm{fm}$
        & 18.61 & (1.22) & 1.30 & 1.15 & 1.24 & 1.15 \\
    48I light discon $R_\text{max}<2.5\mathrm{fm}$
        & -2.73 & (0.74) & 0.79 & 0.84 & 0.75 & 0.65 \\
    \hline
    \end{tabular}
    \caption{Two light quark contributions computed on the 48I ensemble. The table lists the statistical errors calculated with different binning sizes in MD time units in the HMC evolutions.
    The separation between the configurations we performed the measurements is 10 MD time units.
    The last column named ``Bin(sample)'' list the statistical error calculated by assuming
    all the individual samples in each configuration are independent.
    The samples for the connected diagrams are the point pairs, where we have about 57,000 pairs
    for each configuration.
    The samples for the disconnected diagrams are the points for $y$ in Fig.~\ref{fig:hlbl-psrc} (summation over $x$ is already performed), where we have about 2048 points for each configuration.
    }
    \label{tab:auto-corr}
\end{table*}

The remaining finite volume, pion mass, and non-zero lattice spacing corrections will be studied in
Sections~\ref{sec:fv-corr-contrib}, \ref{sec:mpi-corr-contrib}, and~\ref{sec:a2-corr-contrib}.

\subsection{Strange quark contribution}
\label{sec:strange-contrib}

\begin{figure}[h]
\centering
\includegraphics[width=0.4\textwidth]{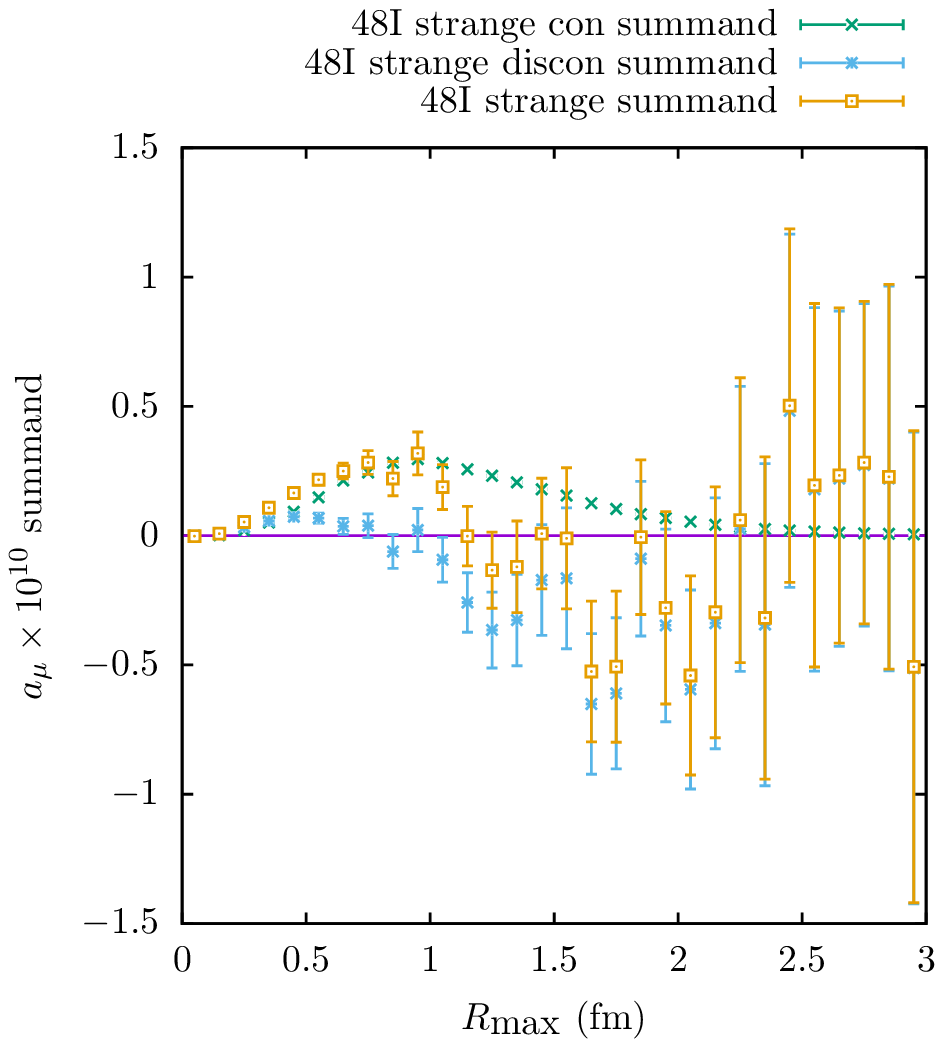}
\includegraphics[width=0.4\textwidth]{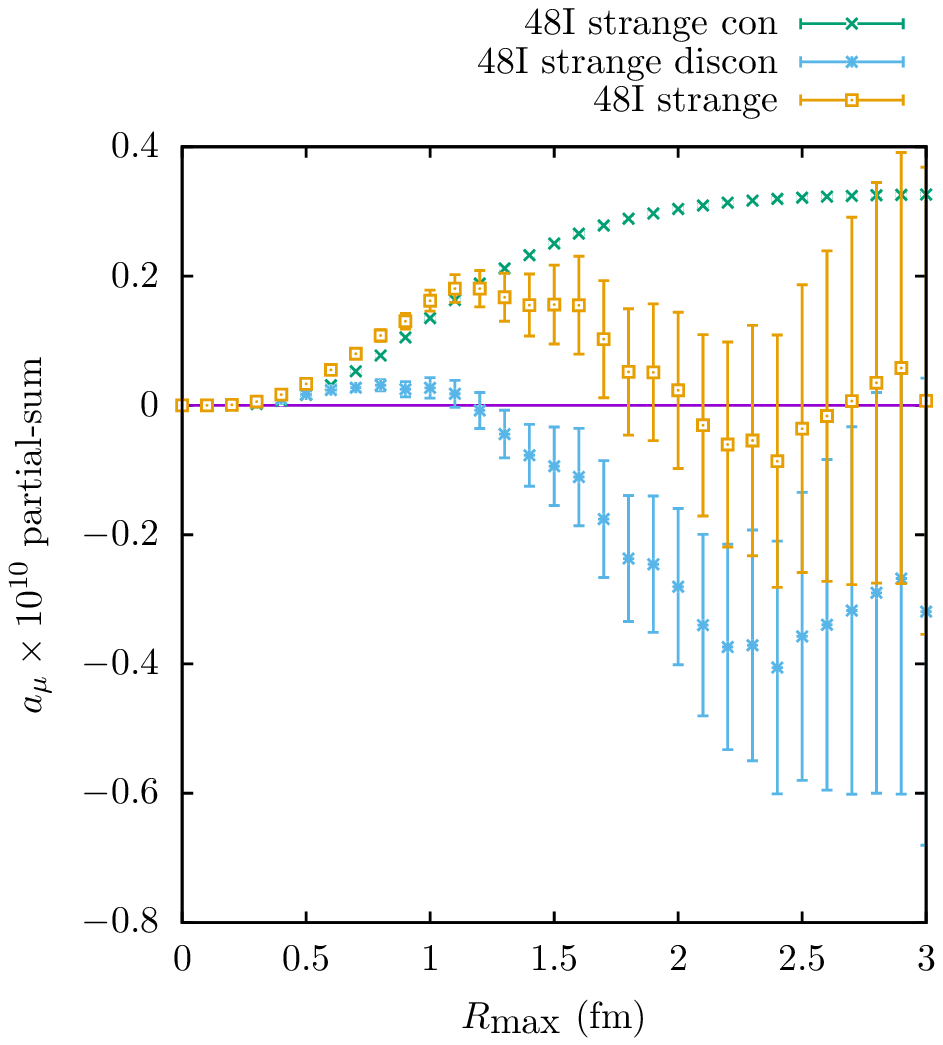}
\caption{\label{fig:48I-strange}Strange quark contributions computed on the 48I ensemble from the connected diagrams,
the disconnected diagrams, and the total. The upper plot shows the corresponding summands and the lower plot shows the partial sum.
}
\end{figure}

The contributions from the strange quark connected diagrams, the disconnected diagrams, and the sum of the two contributions are plotted in Fig.~\ref{fig:48I-strange}. Note the strange quark disconnected diagrams include diagrams where one or both loops are strange quark loops, while the light quark disconnected diagrams discussed in the previous section  contain only light quarks. 
As seen in the figure, the contribution from the strange quark-connected diagrams is very precise and very small. It can also be clearly seen that the strange quark contribution vanishes much faster at long-distance compared to the light quark contribution.

The disconnected diagrams, on the other hand, are much noisier. However, we still expect the signal to vanish faster than the light quark contribution due to the absence of the $\pi^0$ exchange contribution.
Therefore, we can treat the strange quark disconnected diagrams as similar to $a_\mu^\text{no-pion}$.
We listed the results with $4.0~\mathrm{fm}$, $2.5~\mathrm{fm}$, and $2~\mathrm{fm}$ as the upper limit of $R_\text{max}$.
These results are shown in Tab.~\ref{tab:48I-strange} as entries that start with ``48I strange discon $R_\text{max}<$''.
We estimate the systematic error caused by truncating the strange quark disconnected diagram integration using the size of the long-distance contribution from the strange quark connected diagrams, which is also shown in Tab.~\ref{tab:48I-strange} as entries that start with ``48I strange con $R_\text{max}>$''.
The size of the long-distance connected contribution is used as the estimation for the systematic uncertainty for the corresponding disconnected contributions.

The remaining finite volume, pion mass, and non-zero lattice spacing corrections will be studied in
Section.~\ref{sec:fv-corr-contrib}, \ref{sec:mpi-corr-contrib}, and~\ref{sec:a2-corr-contrib}.

\begin{table*}[t]
\centering
\begin{tabular}{l|rl}
\hline
Contribution name & $a_\mu$ & $\times 10^{10}$ \\
\hline
48I strange con $R_\text{max}<4\mathrm{fm}$ & $0.327$ & $(0.002)_\text{stat}$ 
\\
48I strange con $a^2$-corr & $0.026$ & $(0.008)_\text{stat}$ 
\\
strange con & $0.353$ & $(0.007)_\text{stat}$  %
\\
\hline
48I strange con $R_\text{max}>2.5\mathrm{fm}$ & $0.006$ & $(0.000)_\text{stat}$ 
\\
48I strange con $R_\text{max}>2\mathrm{fm}$ & $0.024$ & $(0.000)_\text{stat}$ 
\\
\hline
48I strange discon $R_\text{max}<4\mathrm{fm}$ & $-0.380$ & $(0.607)_\text{stat}$ 
\\
48I strange discon $R_\text{max}<2.5\mathrm{fm}$ & $-0.357$ & $(0.223)_\text{stat}$ 
\\
48I strange discon $R_\text{max}<2\mathrm{fm}$ & $-0.280$ & $(0.121)_\text{stat}$ 
\\
48I strange discon $a^2$-corr & $0.000$ & $\phantom{(0.000)_\text{stat}}(0.029)_\text{syst}$ 
\\
strange discon & $-0.380$ & $(0.607)_\text{stat}(0.029)_\text{syst}~[0.608]$  %
\\
strange discon hybrid-2.5fm & $-0.357$ & $(0.223)_\text{stat}(0.029)_\text{syst}~[0.225]$  %
\\
strange discon hybrid-2fm & $-0.280$ & $(0.121)_\text{stat}(0.037)_\text{syst}~[0.126]$  %
\\
\hline
48I strange $R_\text{max}<4\mathrm{fm}$ & $-0.053$ & $(0.607)_\text{stat}$ 
\\
48I strange $R_\text{max}<4\mathrm{fm}$ hybrid-2.5fm & $-0.030$ & $(0.222)_\text{stat}(0.006)_\text{syst}~[0.223]$ 
\\
48I strange $R_\text{max}<4\mathrm{fm}$ hybrid-2fm & $0.047$ & $(0.121)_\text{stat}(0.024)_\text{syst}~[0.123]$ 
\\
48I strange $a^2$-corr & $0.026$ & $(0.008)_\text{stat}(0.029)_\text{syst}~[0.030]$ 
\\
strange total & $-0.027$ & $(0.607)_\text{stat}(0.029)_\text{syst}~[0.608]$  %
\\
strange total hybrid-2.5fm & $-0.004$ & $(0.223)_\text{stat}(0.029)_\text{syst}~[0.225]$  %
\\
strange total hybrid-2fm & $0.073$ & $(0.121)_\text{stat}(0.037)_\text{syst}~[0.127]$  %
\\
\hline
\end{tabular}
\caption{\label{tab:48I-strange}Strange quark contributions computed on the 48I ensemble. Values are different contributions to $a_\mu \times 10^{10}$ where $ a_\mu = (g_\mu - 2) / 2$. The numbers in the square bracket are the statistical and systematic uncertainty combined in quadrature. The ``strange discon'' contribution includes disconnected diagrams where one or two loops are strange quark loops. The tag ``hybrid-'' indicates the same quantity obtained using the hybrid approach.}
\end{table*}

\subsection{Long-distance $\pi^0$-exchange contribution}
\label{sec:long-distance-contrib}

\begin{figure}[h]
\begin{center}
\includegraphics[width=0.4\textwidth]{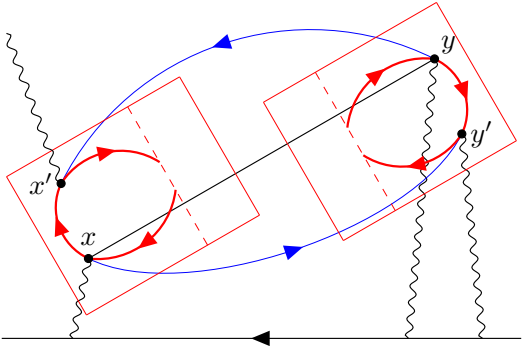}
\caption{The long-distance HLbL contribution to the muon $g-2$ associated with $\pi^0$ exchange. The amplitudes inside the boxes are calculated in lattice QCD while the pion propagator linking them (solid lines) is given by the analytic, infinite-volume, continuum expression.}
\label{fig:long-distance-pi0}
\end{center}
\end{figure}

The dominant source of the HLbL contribution is expected to be the so-called ``$\pi^0$-pole'' contribution~\cite {Aoyama:2020ynm}.
Due to the small mass of the $\pi^0$, this contribution can be non-negligible for large $R_\text{max}$ in a Euclidean space-time lattice calculation, even where long-distance contributions and finite volume effects are exponentially suppressed
since it is only suppressed as $e^{-m_\pi R_\text{max}}$.

We separately calculate the long-distance $\pi^0$-exchange contribution as illustrated in Fig.~\ref{fig:long-distance-pi0}.
We use the lattice calculation of the $\pi^0$ transition form factor:
\ba
\label{eq:pi0-tff}
\mathcal{F}_{\mu,\nu} (x, p) & = \langle 0 | T J_{\mu}(x) J_{\nu}(0) | \pi^0(\vec{p}) \rangle
\\
& = \epsilon_{\mu,\nu,\rho,\sigma} x_\rho p_\sigma \mathcal{F}(x^2,p\cdot x),
\ea
where
$p$ is the Euclidean four-momentum of the $\pi^0$ state.
We use $p=(i m_\pi, \vec 0) = i m_\pi \hat t$ in the numerical lattice calculation. So we can obtain the numerical values of
\ba
\label{eq:pi0-tff-zero}
\mathcal{F}_{\mu,\nu} \big(x, i m_\pi \hat t \big)
\ea
for all possible Euclidean space-time locations $x$ directly from lattice QCD calculations,
without assuming any particular function form for the transition form factor.
This form factor can then be used to construct approximations for infinite volume hadronic correlation functions with large separation.
A detailed derivation is given in Appendix~\ref{sec:pi0-long-distance}.
Here, we briefly describe the idea. First, we introduce a properly normalized $\pi^0$ interpolating field, e.g.
\ba
\pi^0(x) = Z_\pi^{1/2} \frac{i}{\sqrt{2}} \big(\bar u(x) \gamma_5 u(x) - \bar d(x) \gamma_5 d(x) \big) \,.
\ea
The normalization constant $Z_\pi$ is determined by the following requirement:
\ba
\la 0 | \pi^0 (x) | \pi^0(\vec p) \ra = e^{i p \cdot x}.
\ea
We can then rewrite Eq.~\eqref{eq:pi0-tff} in a slightly different form:
\ba
\label{eq:pi0-tff-rewrite}
&\langle 0 | T J_{\mu}(x) J_{\nu}(y) | \pi^0(\vec{p}) \rangle
\nn\\
&\qquad =
\mathcal{F}_{\mu,\nu} (x - y, p)  \la 0 | \pi^0(y) | \pi^0(\vec{p}) \ra \,.
\ea
The above relation suggest that the time-ordered product $T J_{\mu}(x) J_{\nu}(y)$,
when used in between the vacuum state and a single $\pi^0$ state, can be viewed
as a properly weighted $\pi^0$ interpolating field: $\mathcal{F}_{\mu,\nu} (x - y, p)  \pi^0(y)$.
This property can be used to calculate the following three-point function:
\ba
\label{eq:pi0-tff-3pt-1}
\langle T J_{\mu}(x) J_{\nu}(y) \pi^0(z) \rangle
\ea
For very large $m_\pi |y-z|$ and $y-z$ along the positive time direction, i.e. $y-z = (|y-z|, \vec 0)$,
the intermediate states between $T J_{\mu}(x) J_{\nu}(y)$ and $\pi^0(z)$ will be mostly $\pi^0$ states with $\vec p \approx \vec 0$.
We can employ Eq.~\eqref{eq:pi0-tff-rewrite} with $p \approx (i m_\pi, \vec 0)$ and obtain:
\ba
\label{eq:pi0-tff-3pt-example}
&\hspace{-0.5cm}
\langle T J_{\mu}(x) J_{\nu}(y) \pi^0(z) \rangle
\nn
\\
&\approx
\mathcal{F}_{\mu,\nu} \Big(x - y, i m_\pi \frac{y-z}{|y - z|}\Big) \langle T \pi^0(y) \pi^0(z) \rangle.
\ea
Note that we have expressed the above relation in an $O(4)$ rotationally covariant form,
so the equation is also valid for $y-z$ in directions other than the time direction.
Then, we can introduce the infinite-volume free scalar propagator $D_{\pi^0}(z)$, with physical pion mass. For sufficiently large $|x - y|$, we have:
\ba
\la T \pi^0(x) \pi^0(y)\ra = D_{\pi^0}(x - y).
\ea
Now, the relation between Eq.~(\ref{eq:pi0-tff}) and Eq.~(\ref{eq:pi0-tff-3pt-example})
is established.
We can approximate the infinite volume hadronic four-point function
illustrated in Fig.~\ref{fig:long-distance-pi0} in the large $|x-y|$ region
in a similar approach:
\ba
\label{eq:pi0-exch-approx}
&\hspace{-1cm}\langle T  J_{\mu'}(x') J_{\mu}(x)  J_{\nu'}(y') J_{\nu}(y) \rangle
\approx
D_{\pi^0}(x - y)
\nn\\
&\times\mathcal{F}_{\mu',\mu}\Big(x'-x, i m_\pi \frac{x-y}{|x-y|}\Big)
\nn\\
&\times\mathcal{F}_{\nu',\nu}\Big(y'-y, i m_\pi \frac{y-x}{|y-x|}\Big).
\ea
{In the above approximation, $|x-x'|$ and $|y - y'|$ will not  be too large in order to create localized pion states, so the form factor $\mathcal F$ can be computed using a reasonably sized lattice.}
As described in Eq.~\eqref{eq:pi0-tff} and Eq.~\eqref{eq:pi0-tff-zero}, we only directly
calculate the transition form factor $\mathcal F$ for a zero-momentum pion state.
To obtain the form factor needed in  Eq.~\eqref{eq:pi0-exch-approx} above,
we need to perform $O(4)$ rotations:
\ba
\label{eq:pi0-form-factor-rotation}
&\mathcal{F}_{\mu,\nu}\Big(\widetilde x, i m_\pi \hat n \Big)
=
\Lambda_{\mu,\mu'}
\Lambda_{\nu,\nu'}
\mathcal{F}_{\mu',\nu'}\Big(\widetilde x', i m_\pi \hat t \Big),
\ea
where $\tilde x$ can be either $x'-x$ or $y'-y$ in Eq.~\eqref{eq:pi0-exch-approx}
and $\hat n$ can be $\pm (x-y)/|x-y|$.
The $O(4)$ rotation matrix $\Lambda = \Lambda(\hat n)$ and Euclidean space-time coordinate $\widetilde x'$ satisfy:
\ba
\delta_{\mu,\nu} &= \Lambda_{\mu,\mu'} \Lambda_{\nu,\nu'}\delta_{\mu',\nu'},
\\
\hat n_\mu &= \Lambda_{\mu,\mu'} \hat t_{\mu'},
\\
\widetilde x_\mu &= \Lambda_{\mu,\mu'} \widetilde x'_{\mu'}.
\ea

We would like to emphasize again that we do not assume any particular form of the
$\pi^0$ transition form factor and perform fits. The inputs to Eq.~\eqref{eq:pi0-exch-approx}
can be directly obtained from the $O(4)$ rotation in Eq.~\eqref{eq:pi0-form-factor-rotation}
and the lattice QCD calculation of the position-space matrix elements
$\mathcal F_{\mu,\nu}(x,i m_\pi \hat t)$ for a zero-momentum $\pi^0$ state
defined in Eq.~\eqref{eq:pi0-tff} and Eq.~\eqref{eq:pi0-tff-zero}.
The only approximation is made in Eq.~\eqref{eq:pi0-exch-approx}, which is based on
large $m_\pi |x-y|$.

Using the 48I ensemble to calculate $\mathcal F$
and making this approximation,
we obtained the long-distance part corresponding to $R_\text{max} > 4~\mathrm{fm}$:
\ba
&a_\mu(R_\text{max} > 4~\mathrm{fm}) \times 10^{10}
\\
&\approx
a_\mu^{\pi^0\text{-exch}}(R_\text{max} > 4~\mathrm{fm}) \times 10^{10}
= 2.00(11)_\text{stat}(28)_\text{syst},
\nn
\ea
where we use the same subtracted infinite-volume QED weighting function as in the direct calculation.
We estimate the relative systematic uncertainty from the approximation employed in Eq.~(\ref{eq:pi0-exch-approx})
due to not including the charged loop contribution as:
\ba
\frac{e^{-2 m_\pi (4\,\mathrm{fm})}}{e^{-m_\pi (4\,\mathrm{fm})}} \approx 6\%
\ea
and the error from approximating the $\pi^0$ propagation direction to be along $x-y$ to be:
\ba
\frac{0.5\,\mathrm{fm}}{4\,\mathrm{fm}} \approx 13\%
\ea
where
we assume the typical separation for $|x-x'|$ and $|y-y'|$ to be $0.5~\mathrm{fm}$
and $|x-y|$ to be $4~\mathrm{fm}$.
Combining these two estimates in quadrature, we assigned 14\% total systematic uncertainty for the long distance part ($R_\text{max} > 4~\mathrm{fm}$) contribution.
Using the ratio between the connected and disconnected contributions as described in Eq.~\ref{eq:discon-con-ratio},
we can also properly assign this contribution
to the connected and disconnected diagrams.
The results are shown in Tab.~\ref{tab:48I-light} with the labels ``$\dots R_\text{max}>4\mathrm{fm}$''.

\subsection{Finite volume corrections}
\label{sec:fv-corr-contrib}

The long-distance part of the HLbL contribution usually suffers the most significant finite volume effects.
However, the long-distance $\pi^0$ exchange contribution calculated in Section~\ref{sec:long-distance-contrib}
is performed in infinite volume and is free of finite volume effects.
Therefore, we only need to correct for the finite volume effects for the relatively
short-distance region, where $R_\text{max} < 4~\mathrm{fm}$.
These finite volume effects are expected to be quite small due to the constraint $R_\text{max} < 4~\mathrm{fm}$.
The finite volume effects can be estimated with the ``$\pi^0$-pole'' contribution as defined in Ref.~\cite{Hoferichter:2018kwz}.
Similar to Ref.~\cite{Feng:2012ck,Gerardin:2019vio}, we define the $\pi^0$ transition form factor in Euclidean space-time as:
\ba
\label{eq:pi0-tff-1}
& \mathcal{F}_{\mu,\nu} (x, p)
 = \langle 0 | T J_{\mu}(x) J_{\nu}(0) | \pi^0(\vec{p}) \rangle
\nn\\
= & \int\frac{d^4 q_1}{(2\pi)^4} e^{i q_1\cdot x} \frac{-i}{4\pi^2 F_\pi} \epsilon_{\mu,\nu,\rho,\sigma}
{q_1}_\rho {q_2}_\sigma F_{\pi^0\gamma\gamma}(q_1^2, q_2^2)
\ea
where $p = q_1 + q_2$ and $p^2 = - m_\pi^2$. We can convert the form factor into coordinate space via the following:
\ba
\int d^4 u\, e^{i p\cdot u} \widetilde{\mathcal F}_{\mu,\nu}(u, x, y) = \mathcal F_{\mu,\nu}(x-y,p).
\ea
The solution to the above condition is not unique. Based on the momentum space form factor $F_{\pi^0\gamma\gamma}(q_1^2, q_2^2)$ in Eq.~(\ref{eq:pi0-tff-1}),
we obtain
\ba
\widetilde{\mathcal F}_{\mu,\nu}(u, x, y)
&=
\frac{i}{4\pi^2 F_\pi} \epsilon_{\mu,\nu,\rho,\sigma}
\partial^{x}_\rho \partial^{y}_\sigma
\\&\hspace{-1.5cm}\times
\int\frac{d^4 q_1}{(2\pi)^4} e^{i q_1\cdot (x - u)}
\int\frac{d^4 q_2}{(2\pi)^4} e^{i q_2\cdot (y - u)}
F_{\pi^0\gamma\gamma}(q_1^2, q_2^2).
\nn
\ea
Note this definition of the position space form factor
is different from the inverse Fourier transformation
of the {Euclidean space} hadronic matrix element $\mathcal F_{\mu,\nu}(x,p)$,
since the on-shell condition $p^2 = - m_\pi^2$
for the physical $\pi^0$ state cannot be satisfied always in the inverse Fourier transformation.
With the form factor defined above, we can construct the $\pi^0$-pole
contribution to the hadronic four-point function:
\ba
\label{eq:pi0-pole}
&\hspace{-1cm}\langle T J_{\mu}(x) J_{\mu'}(x') J_{\nu}(y) J_{\nu'}(y') \rangle
\\
&\hspace{-0.5cm}\approx
\int d^4 u \int d^4 v \,
D_{\pi^0}(u - v)
\nn\\
&\times\Big(
\widetilde{\mathcal F}_{\mu,\mu'}(u, x, x')
\widetilde{\mathcal F}_{\nu,\nu'}(v, y, y')
\nn\\
&\hspace{1cm}+
\widetilde{\mathcal F}_{\mu,\nu}(u, x, y)
\widetilde{\mathcal F}_{\mu',\nu'}(v, x', y')
\nn\\
&\hspace{1cm}+
\widetilde{\mathcal F}_{\mu,\nu'}(u, x, y')
\widetilde{\mathcal F}_{\mu',\nu}(v, x', y)
\Big)
\nn
\ea
where $D_{\pi^0}(x - y)$ is the infinite volume free scalar propagator with physical pion mass.
In this calculation, we use the Lowest Meson Dominance (LMD) model \cite{Gerardin:2016cqj,Moussallam:1994xp,Knecht:1999gb} for the $\pi^0$ transition form factors.
\ba
&\hspace{-1.0cm}F_{\pi^0\gamma\gamma}(q_1^2, q_2^2)
\approx
F^\text{LMD}_{\pi^0\gamma\gamma}(q_1^2, q_2^2)
\nn\\
=&
F^\text{VMD}_{\pi^0\gamma\gamma}(q_1^2, q_2^2)
+
\frac{8\pi^2 F_\pi^2}{3m_V^2}
\\
&\hspace{0.2cm}
\times\Big(
F^\text{TE}_{\pi^0\gamma\gamma}(q_1^2, q_2^2)
-
F^\text{VMD}_{\pi^0\gamma\gamma}(q_1^2, q_2^2)
\Big),
\nn
\ea
where
\ba
F^\text{VMD}_{\pi^0\gamma\gamma}(q_1^2, q_2^2)
=
\frac{m_V^2}{q_1^2 + m_V^2}
\frac{m_V^2}{q_2^2 + m_V^2}
\ea
\ba
F^\text{TE}_{\pi^0\gamma\gamma}(q_1^2, q_2^2)
=
\frac{m_V^2/2}{q_1^2+m_V^2}
+ \frac{m_V^2/2}{q_2^2+m_V^2}.
\ea
The parameters used in the calculation are $m_\pi = 134.9766~\mathrm{MeV}$, $F_\pi = 92~\mathrm{MeV}$, $m_V = 770~\mathrm{MeV}$.
We discretize the model and calculate it with lattices of different sizes using the same
subtracted infinite volume QED weighting as the direct calculation.
The results are given in Tab.~\ref{tab:lmd-model}.
Note the physical size of the ``Match'' and ``Coarse'' lattices are the same as
the 48I ensemble, which we used to perform our main lattice QCD calculation.
The spatial size is $L = 5.5~\mathrm{fm}$.
The ``Large'' lattice has the same lattice spacing as the ``Coarse'' lattice
but has a much larger physical size, $L=16.4~\mathrm{fm}$.
We use the difference between the ``Large'' and the ``Coarse'' results of
$a_\mu(R_\text{max} < 4~\mathrm{fm}) \times 10^{10}$ as the finite volume correction.
Comparing the results of ``Match'' and ``Coarse'',
we estimate the uncertainty due to the non-zero lattice spacing effects of this model lattice calculation
can be about 12\%.
Combined with the additional 20\% uncertainty due to the inaccuracy of the model itself
(and potential finite volume corrections of other heavier intermediate states),
we obtain our final estimate of the finite volume correction:
\ba
a_\mu^\text{FV-corr}(R_\text{max} < 4~\mathrm{fm}) \times 10^{10} = -0.47(11)_\text{syst}.
\ea
Similar to the long-distance $\pi^0$-exchange contribution,
we use the ratio between the connected and disconnected contributions as described in Eq.~\ref{eq:discon-con-ratio}
to assign this contribution to the connected and disconnected diagrams.
The results are shown in Tab.~\ref{tab:48I-light} as ``48I light con FV-corr'',
``48I light discon FV-corr'',
and ``48I light FV-corr''.

Also, note that we can calculate the long-distance $\pi^0$-pole contribution using this LMD model.
In Tab.~\ref{tab:lmd-model}, we also list the contribution to the region $R_\text{max}>4~\mathrm{fm}$.
Using the LMD model results calculated with the ``Large'' lattice, we obtain
\ba
a_\mu^{\pi^0\text{-pole};\text{LMD}}(R_\text{max} > 4~\mathrm{fm}) \times 10^{10} = 2.34.
\ea
This result agrees with the long-distance $\pi^0$-exchange contribution calculated
with $\pi^0$ transition form factors from the previously described lattice QCD calculation and long-distance approximation.

\begin{table*}[t]
\centering
\begin{tabular}{ccc|cc}
\hline
Label & Size & $a^{-1} / \mathrm{GeV}$ & $a_\mu(R_\text{max} < 4~\mathrm{fm}) \times 10^{10}$ & $a_\mu(R_\text{max} > 4~\mathrm{fm}) \times 10^{10}$ \\
\hline
Match & $48^3\times 96$ & 1.73 & 5.19 & 0.22 \\
Coarse & $24^3\times 48$ & 0.865 & 4.65 & 0.20 \\
Large & $64^3\times 128$ & 0.865 & 4.18 & 2.34 \\
\hline
\end{tabular}
\caption{\label{tab:lmd-model}
Lattice calculation of the $\pi^0$-pole contribution using the LMD model. Note that this is not a lattice QCD calculation.}
\end{table*}

\subsection{Pion mass extrapolation}
\label{sec:mpi-corr-contrib}

While the calculation is performed very near the physical pion mass, there is a slight mis-tuning of the lattice
parameters which leads to a slightly heavier $m_\pi = 139~\mathrm{MeV}$ compared to the physical
pion mass $m_\pi = 134.9766~\mathrm{MeV}$.
We correct this small difference using the 32D and 24DH ensembles.
These two ensembles have the same lattice spacing but different pion masses.
The results of these two ensembles are listed in Tab.~\ref{tab:pion-mass-dependence}.
Note that for the long-distance contribution ($R_\text{max} > 4~\mathrm{fm}$)
for the 32D physical pion mass ensemble, we use the same long-distance $\pi^0$-exchange contribution
calculated with ensemble 48I as described in Section~\ref{sec:long-distance-contrib}.
Similar to the 48I calculation, we calculate the ``no-pion'' contribution
and obtain the ``hybrid-2fm'' results for the ``discon'' and ``total'' results.
To reduce the statistical error, we use the ``hybrid-2fm'' results to calculate
the pion mass correction.
\begin{table*}[t]
\centering
\begin{tabular}{cc|rlrl}
\hline
name & $R_\text{max}$ limit & $a_\mu(\text{32D})$ & $\times 10^{10}$ & $a_\mu(\text{24DH})$ & $\times 10^{10}$ \\
\hline
con & 2 fm & $8.17$ & $(0.32)_\text{stat}$ & $8.29$ & $(0.10)_\text{stat}$ \\
con & 4 fm & $21.29$ & $(3.37)_\text{stat}$ & $12.35$ & $(0.24)_\text{stat}$ \\
con & $\infty$ & $28.84$ & $(3.39)_\text{stat}$ & $12.35$ & $(0.24)_\text{stat}$ \\
no-pion & 2 fm & $3.91$ & $(0.63)_\text{stat}$ & $4.18$ & $(0.57)_\text{stat}$ \\
no-pion & 4 fm & $4.96$ & $(4.46)_\text{stat}$ & $6.74$ & $(4.18)_\text{stat}$ \\
discon & 2 fm & $-2.10$ & $(0.59)_\text{stat}$ & $-1.91$ & $(0.55)_\text{stat}$ \\
discon & 4 fm & $-10.70$ & $(3.84)_\text{stat}$ & $-2.34$ & $(4.18)_\text{stat}$ \\
discon & $\infty$ & $-16.25$ & $(3.84)_\text{stat}$ & $-2.34$ & $(4.18)_\text{stat}$ \\
discon hybrid-2fm & 4 fm & $-11.74$ & $(2.52)_\text{stat}$ & $-4.90$ & $(0.54)_\text{stat}$ \\
discon hybrid-2fm & $\infty$ & $-17.30$ & $(2.53)_\text{stat}$ & $-4.90$ & $(0.54)_\text{stat}$ \\
total & 2 fm & $6.07$ & $(0.66)_\text{stat}$ & $6.38$ & $(0.57)_\text{stat}$ \\
total & 4 fm & $10.59$ & $(4.97)_\text{stat}$ & $10.01$ & $(4.18)_\text{stat}$ \\
total & $\infty$ & $12.59$ & $(4.98)_\text{stat}$ & $10.01$ & $(4.18)_\text{stat}$ \\
total hybrid-2fm & 4 fm & $9.55$ & $(1.12)_\text{stat}$ & $7.45$ & $(0.59)_\text{stat}$ \\
total hybrid-2fm & $\infty$ & $11.55$ & $(1.12)_\text{stat}$ & $7.45$ & $(0.59)_\text{stat}$ \\
\hline
\end{tabular}
\caption{\label{tab:pion-mass-dependence}Study of the pion mass dependence using the 32D and 24DH ensembles.
For the 32D (physical pion mass ensemble) results, we apply the 48I long-distance $\pi^0$-exchange
contribution in the $R_\text{max} > 4~\mathrm{fm}$ region.
For the 24DH ensemble ($m_\pi=341~\mathrm{MeV}$), we assume the contribution
is negligible in the $R_\text{max} > 4~\mathrm{fm}$ region.
}
\end{table*}
To obtain the pion mass correction to our main 48I results, we assume the following
pion mass dependence for the ``no-pion'' contribution:
\ba
a_\mu^\text{no-pion}(m_\pi) = C_1 + C_2 m_\pi^2.
\ea
Due to the pion pole piece in the quark-connected diagram, we assume a more singular
pion mass dependence for the ``con'' contribution:
\ba
a_\mu^\text{con}(m_\pi) = C_1 + \frac{C_2}{m_\pi^2}.
\label{eq:con fit}
\ea
The disconnected diagram contribution can be obtained as a combination of the
``con'' and ``no-pion'' contribution as given in Eq.~(\ref{eq:discon-no-pion})((\ref{eq:total-no-pion})).

The fitting form in Eq.~(\ref{eq:con fit})(plus an additional $m_\pi^2$ term) was used for the chiral extrapolation
of the connected diagram in Ref.~\cite{Chao:2021tvp} and seems to provide the most
accurate description of the connected {diagram's mass dependence}.
The correction from the chiral extrapolation is $0.35(7)\times 10^{-10}$ with the above fitting form.
In Ref.~\cite{Prades:2009tw}, the $\pi^0$-pole contribution is calculated to behave as
$C_1 + C_2 \log^2(m_\pi^2)$,
If we estimate the pion mass correction with this form instead, the size of the correction
is $0.20(4)\times 10^{-10}$.
In Ref.~\cite{Ramsey-Musolf:2002gmi}, the HLbL is calculated with Chiral effective theory,
we have calculated the pion mass dependence with Eq.~(13) of this reference.
The shift of $a_\mu$ due to the pion mass change from the $139~\mathrm{MeV}$ used
in the lattice calculation to the physical point $135~\mathrm{MeV}$
is $0.25\times 10^{-10}$.
Here, we do not include the correction to the charged pion loop contribution of $a_\mu$
due to the pion mass shift, as the charged pion loop contribution
are significantly reduced compared with the scalar QED estimate due to the pion dressing~\cite{Kinoshita:1984it,Hayakawa:1995ps,Bijnens:2015jqa}.
We, therefore, estimate a 50\% systematic uncertainty to these corrections to account
for possible inaccuracies of
the empirical chiral extrapolations, plus any systematic effects
in obtaining the 32D and 24DH results, including the effects caused by the hybrid method used to calculate the long-distance part of the disconnected diagrams, the discretization effects, the finite volume effects, and so on.
The final results for the correction due to the slight mismatch of the pion mass is:
\ba
&a_\mu^\text{$m_\pi$-corr} \times 10^{10}
\\
&= (a_\mu(m_\pi = 135~\mathrm{MeV}) - a_\mu(m_\pi = 139~\mathrm{MeV})) \times 10^{10}
\nn\\
&= 0.35(0.07)_\text{stat}(0.17)_\text{syst}.
\nn
\ea
We also show the results in Tab.~\ref{tab:48I-light} as ``48I light con $m_\pi$-corr'', ``48I light discon $m_\pi$-corr'', and ``48I light $m_\pi$-corr''.

\subsection{Continuum limit extrapolation}
\label{sec:a2-corr-contrib}

In our previous work~\cite{Blum:2019ugy}, we used finite-volume lattices for the entire calculation, including QED.
The QED$_L$ scheme was used for the photon propagators, and
sizable discretization effects were observed.
We took the continuum limit with the 48I and 64I ensembles.
These two ensembles have different lattice spacings but all other properties are almost identical.
They have the same gauge and fermion actions,
and almost the same pion mass and volume in physical units.
We used the results from these two ensembles to perform the continuum extrapolation.
The continuum limit value is $30(12)\%$ higher compared to the value on the 48I ensemble for the connected piece
and $58(27)\%$ higher for the disconnected piece.
The numbers in parentheses are statistical uncertainties.

In this work, we have mainly used the 48I ensemble to perform the calculation with the
infinite volume, continuum QED weighting function described in Ref.~\cite{Blum:2017cer}.
Compared with QED$_L$, the infinite volume QED weighting function reduces the finite volume effects from power-law to exponential.

One may expect similar discretization effects for these two calculations
since they should be the same in the large volume limit and only differ by finite volume effects.
However, there are two reasons that the discretization errors in
the current work should be much smaller.
First, the infinite volume QED weighting function is calculated in the continuum
with a semi-analytic method, which eliminates the discretization errors
in the QED$_L$ weighting function which is calculated on a discrete lattice.
Second, and more importantly, we apply the subtraction scheme
as described in Ref.~\cite{Blum:2017cer} and repeated in Eq.~\ref{eq:qed-weighting-func-subtraction}.
In a test calculation of the leptonic light-by-light contribution to the muon $g-2$,
we found about a factor of $5$ reduction in the discretization error after
this subtraction is applied.
We expect a similar reduction in discretization error in the hadronic case, and
this is indeed the case in our calculation of the strange quark connected diagrams.

\begin{figure}[h]
\centering
\includegraphics[width=0.4\textwidth]{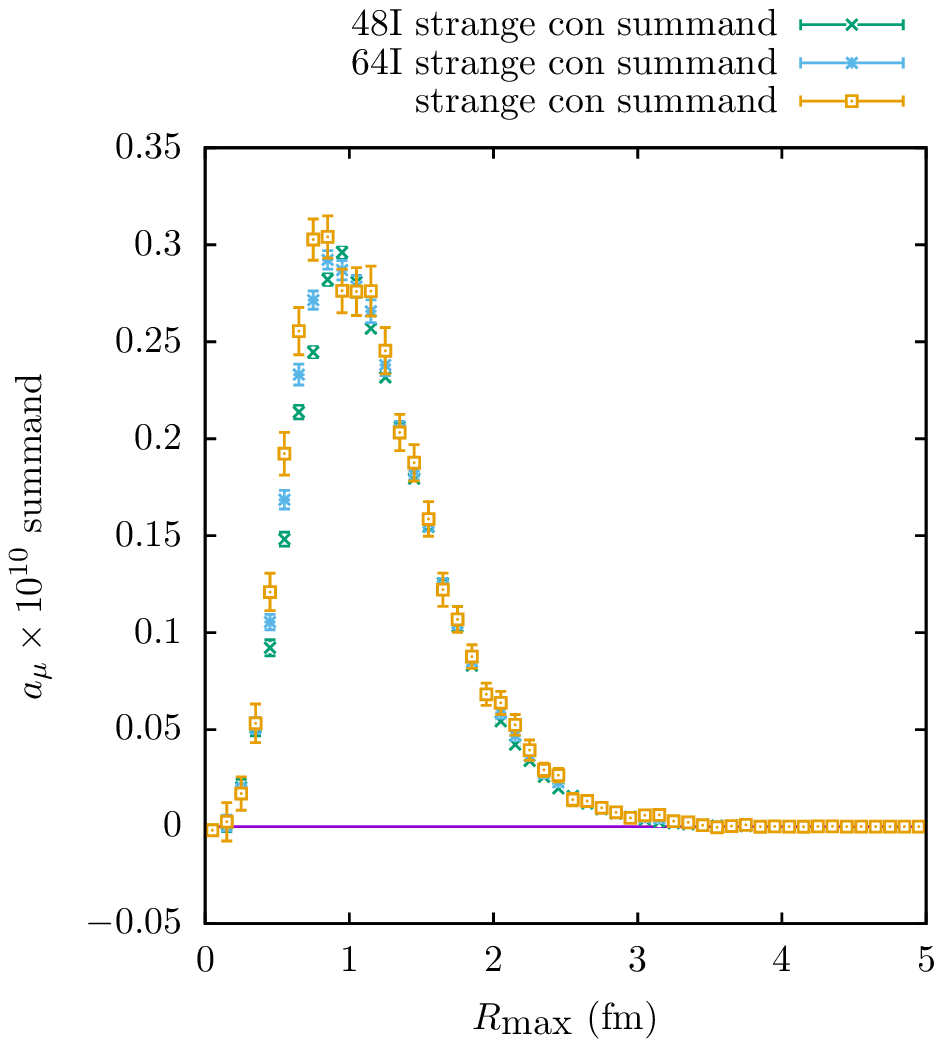}
\includegraphics[width=0.4\textwidth]{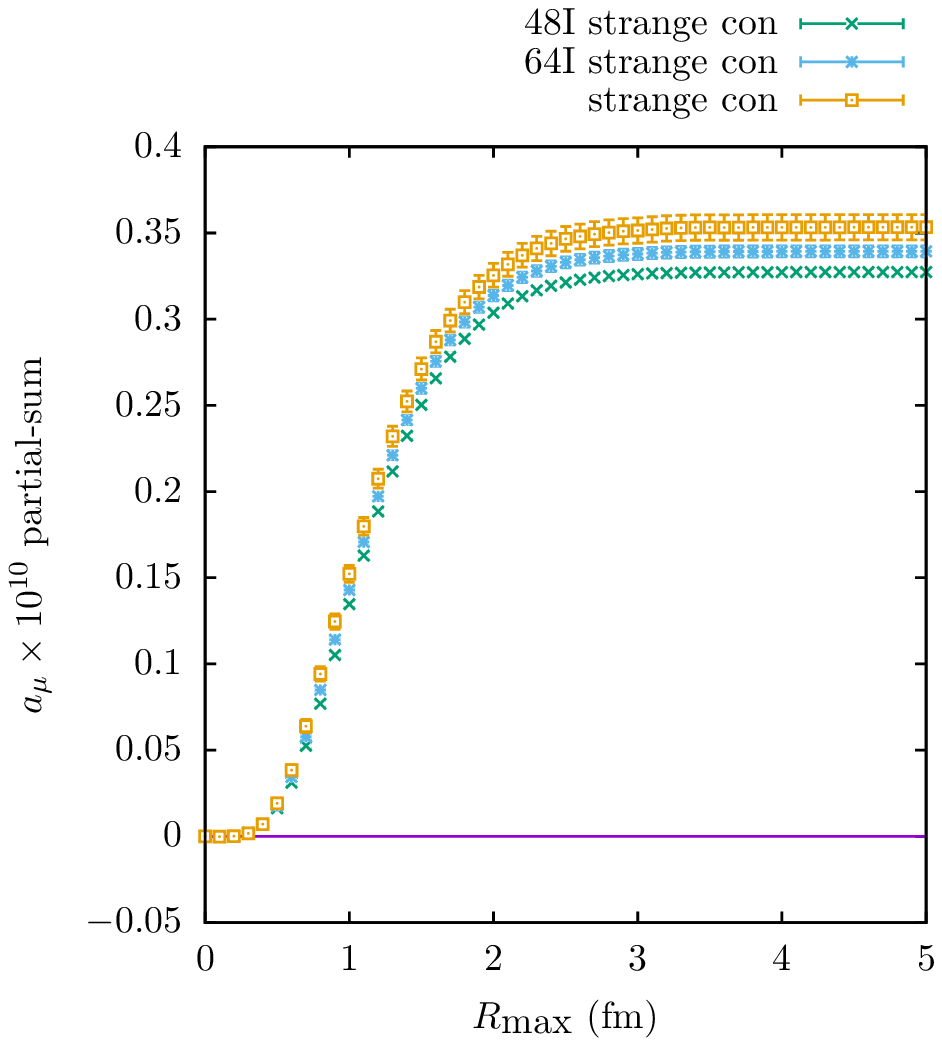}
\caption{\label{fig:48I-64I-strange-con}Strange quark connected part of $a_\mu^{\rm HLbL}$ (lower panel). Continuum limit (upper curve) and finite lattice spacing results on 48I and 64I ensembles. Corresponding summands (upper panel).
}
\end{figure}

The strange quark connected contribution is shown in Fig.~\ref{fig:48I-64I-strange-con}.
Compared with the light quark, due to the heavier strange quark mass,
the statistical noise in the long-distance region is very small.
The overall signal-to-noise ratio is much better than the light quark case which allows a precise continuum limit extrapolation.
As is shown in the figure, the limit is 7.9(2.6)\% higher than the 48I results.
The continuum limit is obtained by performing an $\mathcal O(a^2)$ extrapolation to $a\to0$,
\ba
a_\mu(a^2) = C_1 + C_2 a^2.
\ea
Compared with the observed $\sim 30$\% discretization error in the QED$_L$ calculation
of the light quark connected diagrams,
there is indeed a significant reduction in the relative discretization error
with the subtracted infinite volume weighting function,
despite the heavier strange quark mass, which usually leads to larger discretization errors.

The study of the continuum limit of the strange quark connected diagrams strongly suggests
the light quark connected diagrams, in particular in the relatively short distance region
where we performed a reliable continuum extrapolation for the strange quark connected diagrams,
is under 8\%.
We also have the same 64I dataset for the light quark connected diagram.
We have listed the comparison in
Tab.~\ref{tab:light-con-a2-correction}.
The results from the 64I ensemble are statistically consistent with the 48I ensemble.
Unforunately, the statistical error for 64I is relatively larger due to the absence of
the improvement we made for the new 48I calculation.
We cannot constrain the continuum extrapolation for the light quark connected diagram
better than the 8\% estimate above based on the strange quark connected results.

\begin{table*}[t]
\centering
\begin{tabular}{c|r|r|c}
\hline
Observable & 48I & 64I & Percentage correction \\
\hline
        $a_\mu^\text{con}(R_\text{max} < 1.0~\mathrm{fm})\times 10^{10}$
        & $1.27(0.02)$
        & $1.32(0.05)$
        & $9.2(8.6)$
        \\
        $a_\mu^\text{con}(R_\text{max} < 1.5~\mathrm{fm})\times 10^{10}$
        & $3.92(0.05)$
        & $4.04(0.23)$
        & $7.0(12.6)$
        \\
        $a_\mu^\text{con}(R_\text{max} < 2.0~\mathrm{fm})\times 10^{10}$
        & $7.24(0.15)$
        & $7.19(0.66)$
        & $-1.4(20.0)$
        \\
        $a_\mu^\text{con}(R_\text{max} < 2.5~\mathrm{fm})\times 10^{10}$
        & $10.64(0.31)$
        & $9.61(1.40)$
        & $-20.8(29.2)$
        \\
\hline
\end{tabular}
\caption{\label{tab:light-con-a2-correction}
The last column is the percent correction to obtain the continuum limit relative to the value of the 48I ensemble.
The values in parentheses are the statistical uncertainty of the corresponding quantity.
}
\end{table*}

In the long distance (large $R_\text{max}$) region,
the contributions are mostly coming from the $\pi^0$-exchange contribution.
As can be seen from Eq.~(\ref{eq:pi0-exch-approx}),
the hadronic four-point function in the HLbL diagrams in the long-distance part
can be approximated by the product of two such transition form factors.
Therefore, the discretization effects of the direct calculation of the four-point function
in the long distance region
will be similar to the discretization effects of the calculation of the square of the transition form factor.
In the $\pi^0\to e^+ e^-$ calculation of Ref.~\cite{Christ:2022rho},
this relevant $\pi^0\to \gamma\gamma$ transition form factor is studied with the same 48I ensemble
as this calculation. In addition, the continuum limit is calculated with a parallel 64I ensemble
calculation. We quote the corrections obtained in the calculation in Tab.~\ref{tab:piee-correction}.
Note that the correction is zero consistent given the statistical uncertainty.
In Fig.~4 of Ref.~\cite{Christ:2022rho}, the partial sum of these quantities
for both 48I and 64I are plotted.
More statistically precise agreement is observed at smaller time separation of the two electromagnetic vector currents.

\begin{table}[H]
\centering
\begin{tabular}{c|c}
\hline
Observable & Percentage correction \\
\hline
$(\mathrm{Im} \mathcal A)^2$ & -11.4(11.6)\% \\
$(\mathrm{Re} \mathcal A)^2$ & -16.8(15.8)\% \\
\hline
\end{tabular}
\caption{\label{tab:piee-correction}
Percent correction to obtain the continuum limit relative to the value computed on the 48I ensemble in the
$\pi^0 \to e^+ e^-$ calculation in Ref.~\cite{Christ:2022rho}.
The values in parentheses are the statistical uncertainty of the corresponding quantity.
The square of the amplitude is used to estimate the lattice spacing error of the long-distance $\pi^0$-exchange contribution
to the muon $g-2$ in this work.}
\end{table}

For the light quark disconnected diagrams, we again separately discuss the short and long distance regions.
In Fig.~\ref{fig:48I-light}, we observe that in the short distance region ($R_\text{max} \lesssim 2~\mathrm{fm}$),
the disconnected diagram contribution is much smaller than the connected contribution.
This is expected as the disconnected diagrams involve two quark loops and the exchange of at least two gluons.
Therefore, the disconnected diagrams are suppressed by
$\mathcal O(\alpha_s^2)$ and $1/ N_c$ relative to the connected diagrams.
Due to the smallness of the contributions from the short-distance disconnected diagrams,
we expect that the same reasoning applies to the discretization effects.

For the long distance region, we again have Eq.~(\ref{eq:discon-con-ratio}),
which mandates the relationship with the connected diagrams, even for non-zero lattice spacing and finite volume
(so long that the lattice is large enough to contain the correlation function).
Therefore, we expect the long distance part of the disconnected diagrams to
have the same relative discretization effects as the connected diagrams
and also the combined total.
Note that the situation is different from the Mainz group's work in
Ref.~\cite{Chao:2020kwq,Chao:2021tvp}. While the argument presented here
should also apply to the hadronic function in their calculation, the
different QED kernels
(due to the subtraction scheme of the QED kernel and the choice of $x_\text{ref}$)
used for the connected and disconnected diagrams
lead to different shapes of the summands and also
different discretization errors.

Without a 64I ensemble calculation of similar precision,
the continuum limit cannot be reliably taken.
However, based on the evidence presented above,
in particular the size of the correction in the strange quark connected diagrams,
we estimate an 8\% uncertainty from non-zero lattice spacing for the contributions from the light quark connected, disconnected, and total contributions, and the strange quark disconnected diagrams.
The corrections for light quark are shown in Tab.~\ref{tab:48I-light} as ``48I light con $a^2$-corr'', ``48I light discon $a^2$-corr'', and ``48I light $a^2$-corr''.
The corrections for the strange quark are shown in Tab.~\ref{tab:48I-strange} as ``48I strange con $a^2$-corr'', ``48I strange discon $a^2$-corr'', and ``48I strange $a^2$-corr''.

\subsection{Sub-leading disconnected diagrams}

\begin{figure}[h]
\centering
\includegraphics[width=0.4\textwidth]{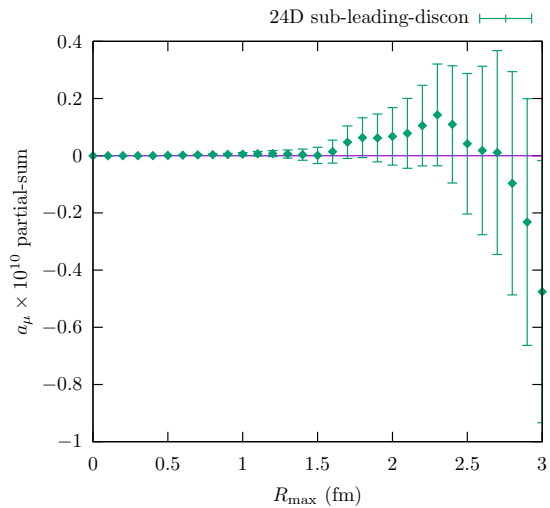}
\caption{\label{fig:sub-leading-discon}Sub-leading disconnected diagrams computed on the 24D ensemble.~\cite{Blum:2019ugy}}
\end{figure}

So far, our discussions have centered on the first two diagrams in Fig.~\ref{fig:hlbl-diagrams}.
The remaining diagrams, which are expected to be small due to the suppression by
flavor SU(3) and quark electric charge factors,
are not yet included.
In our earlier work~\cite{Blum:2019ugy}, we have already calculated
the third diagram in Fig.~\ref{fig:hlbl-diagrams}, which is expected
to be the largest within the remaining the sub-leading disconnected diagrams
based on the flavor SU(3) and quark electric charge factor counting.
To estimate the contribution from this diagram, we employ the same subtracted infinite volume QED weighting function
and the 24D ensemble for the quark loops.
The result, shown in Fig.~\ref{fig:sub-leading-discon}, is zero consistent, and we estimate this contribution to be $0.0(0.5)\times 10^{-10}$.

In the more recent work by the Mainz group~\cite{Chao:2021tvp},
all the sub-leading diagrams have been calculated.
The result is still zero consistent but a more stringent bound is obtained.
Therefore, in this work, we use their result to account
for the contribution from all the sub-leading disconnected diagrams.
The value is shown in Tab.~\ref{tab:final-results} as ``sub-leading discon''.

\subsection{Charm quark contributions}

The charm quark contribution is also expected to be very small
due its heavy mass relative to the light quark and the strange quark.
In the 2020 white paper~\cite{Aoyama:2020ynm}, its contribution
is included as $0.3(0.1)\times 10^{-10}$, calculated based on perturbation theory
and consideration of charm meson resonances~\cite{Colangelo:2019uex}.
The value is similar to the strange quark connected diagram contribution
due to the heavier mass of the charm quark and larger electric charge.
In a recent work by the Mainz group~\cite{Chao:2022xzg},
the charm quark contribution is calculated with lattice QCD,
including both the charm quark connected and disconnected diagrams.
The result for the total charm quark contribution is $0.28(0.05)\times 10^{-10}$,
where the uncertainty is mostly
due to the systematic effects from modeling the lattice spacing
and $M_{\eta_c}$ dependence.
In this work, we use this more recent lattice calculated result to account
for the contribution from the charm quark.
The values are shown in Tab.~\ref{tab:final-results} as ``charm con'', ``charm discon'' ``charm''.

\begin{table*}[t]
\centering
\begin{tabular}{l|rl}
\hline
Contribution & $a_\mu$ & $\times 10^{10}$ \\
\hline
light con & $25.70$ & $(1.33)_\text{stat}(1.99)_\text{syst}~[2.39]$  %
\\
light discon & $-12.71$ & $(1.87)_\text{stat}(1.17)_\text{syst}~[2.20]$  %
\\
light discon hybrid-2.5fm & $-13.50$ & $(1.36)_\text{stat}(1.21)_\text{syst}~[1.82]$  %
\\
light discon hybrid-2fm & $-13.37$ & $(1.29)_\text{stat}(1.46)_\text{syst}~[1.95]$  %
\\
light total & $12.99$ & $(2.11)_\text{stat}(0.90)_\text{syst}~[2.29]$  %
\\
light total hybrid-2.5fm & $12.20$ & $(1.01)_\text{stat}(0.95)_\text{syst}~[1.38]$  %
\\
light total hybrid-2fm & $12.33$ & $(0.90)_\text{stat}(1.25)_\text{syst}~[1.55]$  %
\\
\hline
strange con & $0.35$ & $(0.01)_\text{stat}$  %
\\
strange discon & $-0.38$ & $(0.61)_\text{stat}(0.03)_\text{syst}~[0.61]$  %
\\
strange discon hybrid-2.5fm & $-0.36$ & $(0.22)_\text{stat}(0.03)_\text{syst}~[0.22]$  %
\\
strange discon hybrid-2fm & $-0.28$ & $(0.12)_\text{stat}(0.04)_\text{syst}~[0.13]$  %
\\
strange total & $-0.03$ & $(0.61)_\text{stat}(0.03)_\text{syst}~[0.61]$  %
\\
strange total hybrid-2.5fm & $-0.00$ & $(0.22)_\text{stat}(0.03)_\text{syst}~[0.23]$  %
\\
strange total hybrid-2fm & $0.07$ & $(0.12)_\text{stat}(0.04)_\text{syst}~[0.13]$  %
\\
\hline
sub-leading discon & $0.00$ & $\phantom{(0.00)_\text{stat}}(0.07)_\text{syst}$ \cite{Chao:2021tvp}
\\
\hline
charm con & $0.31$ & $\phantom{(0.00)_\text{stat}}(0.04)_\text{syst}$ \cite{Chao:2022xzg}
\\
charm discon & $-0.03$ & $\phantom{(0.00)_\text{stat}}(0.02)_\text{syst}$ \cite{Chao:2022xzg}
\\
charm total & $0.28$ & $\phantom{(0.00)_\text{stat}}(0.05)_\text{syst}$ \cite{Chao:2022xzg}
\\
\hline
con & $26.36$ & $(1.33)_\text{stat}(1.99)_\text{syst}~[2.39]$  %
\\
discon & $-13.12$ & $(2.30)_\text{stat}(1.18)_\text{syst}~[2.59]$  %
\\
discon hybrid-2.5fm & $-13.89$ & $(1.47)_\text{stat}(1.22)_\text{syst}~[1.91]$  %
\\
discon hybrid-2fm & $-13.68$ & $(1.35)_\text{stat}(1.47)_\text{syst}~[1.99]$  %
\\
total & $13.24$ & $(2.53)_\text{stat}(0.90)_\text{syst}~[2.68]$  %
\\
{\bf total hybrid-2.5fm} & $12.47$ & $(1.15)_\text{stat}(0.95)_\text{syst}~[1.49]$  %
\\
total hybrid-2fm & $12.68$ & $(0.98)_\text{stat}(1.26)_\text{syst}~[1.59]$  %
\\
\hline
\end{tabular}
\caption{\label{tab:final-results}Summary of final results. Values are different contributions to $a_\mu \times 10^{10}$ where $ a_\mu = (g_\mu - 2) / 2$. The numbers in the square bracket (except references) are the statistical and systematic uncertainty combined in quadrature. The ``strange discon'' contribution includes disconnected diagrams where one or two loops are strange quark loops. The tag ``hybrid-'' indicates the same quantity obtained using the hybrid approach as described in Section \ref{sec:light-contrib} and \ref{sec:strange-contrib}. The result for ``total hybrid-2.5fm'' is used as our final result.}
\end{table*}

\section{Conclusion}
\label{sec:conclusion}

We summarized all the results discussed above in Tabs.~\ref{tab:48I-light}, \ref{tab:48I-strange}, and \ref{tab:final-results}.
Adding all the individual contributions and corrections, we obtain our final result
for the HLbL contribution to the muon $g-2$:
\ba
\label{eq:final-results}
a_\mu^\text{HLbL} \times 10^{10} = 12.47(1.15)_\text{stat}(0.95)_\text{syst}~[1.49],
\ea
where the systematic errors were added in quadrature. We also give the HLbL contributions to the muon $g-2$ from the connected and disconnected pieces separately:
\ba
\label{eq:final-con-discon-results}
a_\mu^\text{HLbL,con} \times 10^{10} &= 26.36(1.33)_\text{stat}(1.99)_\text{syst}~[2.39],
\\
a_\mu^\text{HLbL,discon} \times 10^{10} &= -13.89(1.47)_\text{stat}(1.22)_\text{syst}~[1.91].
\ea
Numbers in square square brackets denote total erorrs, combining the statistical
uncertainty (``stat'') and systematic ones (``sys'') in quadrature.

Here, we summarize techniques used in this calculation that we believe are important to obtain the precision
for HLbL scattering at the physical pion mass:
\begin{itemize}
    \item The subtracted infinite volume QED weighting function developed in our previous work~\cite{Blum:2017cer}.
    \item Use all combinations of the point source propagators to calculate the HLbL diagrams as described
    in section~\ref{sec:lattice-details}.
    \item The adaptive sampling scheme used for the disconnected diagram calculation as described in section~\ref{sec:lattice-details}.
    \item The rearrangement of the connected and disconnected diagrams
    in Eq.~(\ref{eq:a-mu-no-pion},\ref{eq:total-no-pion})
    based on Eq.~(\ref{eq:discon-con-ratio}).
    \item The AMA method~\cite{Blum:2012uh} and efficient
    GPU solver for MDWF propagators from Grid and GPT.
\end{itemize}

\begin{table}[htbp]
\centering
\begin{tabular}{l|rl}
\hline
 & $a_\mu$ & $\times 10^{10}$ \\
\hline
all con QED$_L$ & $24.46$ & $(2.35)_\text{stat}(5.11)_\text{syst}~[5.62]$ \\
all con diff & $1.90$ & $(2.76)_\text{stat}(5.48)_\text{syst}~[6.14]$ \\
\hline
all discon QED$_L$ & $-16.45$ & $(2.09)_\text{stat}(3.99)_\text{syst}~[4.50]$ \\
all discon diff & $2.56$ & $(2.57)_\text{stat}(4.17)_\text{syst}~[4.90]$ \\
\hline
total QED$_L$ & $8.17$ & $(3.03)_\text{stat}(1.77)_\text{syst}~[3.51]$ \\
total diff & $4.30$ & $(3.25)_\text{stat}(2.01)_\text{syst}~[3.82]$ \\
\hline
\end{tabular}
\caption{\label{tab:comparison}Comparison of this work with
our previous QED$_L$ results~\cite{Blum:2019ugy}
plus the charm quark contribution from~\cite{Aoyama:2020ynm,Colangelo:2019uex}.
The field ``all con'' includes all connected diagram contributions, including the charm quark.
The field ``all discon'' includes all the disconnected diagram contributions, including the sub-leading disconnected diagrams and disconnected diagrams including charm quarks.
The fields ``all con diff'', ``all discon diff'', ``total diff'' shows
the difference obtained by subtracting the QED$_L$ results from the new results presented in this work.
The statistical and systematic uncertainties of the two works are almost independent.
We therefore add them in quadrature for the difference.
The numbers in the square bracket are the statistical and systematic uncertainty combined in quadrature.
}
\end{table}

\begin{figure}[htbp]
\centering
\includegraphics[width=0.48\textwidth]{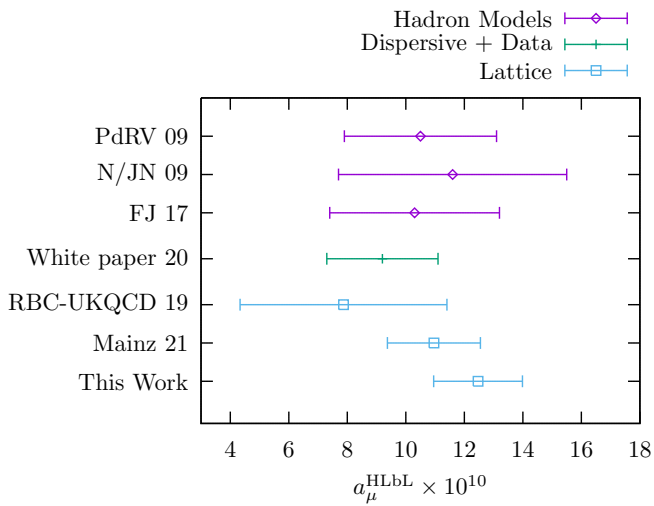}
\caption{\label{fig:compariosn}Comparison of our result with values in the literature.
The hadronic model values are from Refs.~\cite{Prades:2009tw,Nyffeler:2009tw,Jegerlehner:2009ry,Jegerlehner:2017lbd}.
The dispersive data driven result is compiled in Ref.~\cite{Aoyama:2020ynm}.
The lattice results include Refs.~\cite{Blum:2019ugy,Chao:2021tvp,Chao:2022xzg} and this work.
}
\end{figure}

We can compare these results with our previous work~\cite{Blum:2019ugy} based on the finite volume QED$_L$ formulation.
The comparison is summarized in Table~\ref{tab:comparison}.
We can see that the results for both the connected and disconnected diagrams
are in good agreement.
For the total, the current result is $1.12$ standard deviations higher
than the previous results, possibly due to a slightly larger statistical fluctuation.
We also compare the final result in this work with the existing literature in Fig.~\ref{fig:compariosn}.
The new result is consistent with previous determinations.

\begin{acknowledgments}
We thank our RBC and UKQCD collaborators for helpful discussions
and critical software and hardware support.
T.B. and L.J. have been supported under US DOE grant DE-SC0010339.
L.J. is also supported by US DOE Office of Science Early Career Award DE-SC0021147.
N.C. is supported by US DOE grant DE-SC0011941.
M.H. is supported by Japan Grants-in-Aid for Scientific Research, No20K03926.
T.I., C.J., and C.L. were supported in part by US DOE Contract DESC0012704(BNL).
T.I. and C.J. were supported in part by the Scientific Discovery through Advanced Computing (SciDAC) program LAB 22-2580.
T.I. is also supported by the Department of Energy, Laboratory Directed Research and Development (LDRD No. 23-051) of BNL and RIKEN BNL Research Center, and by JSPS KAKENHI under grant numbers JP26400261, JP17H02906.
C.L. has been supported by a DOE Office of Science Early Career Award.
We developed the computational code used for this work based on
the \hyperlink{https://github.com/RBC-UKQCD/CPS_public}{Columbia Physics System} (CPS),
\hyperlink{https://github.com/paboyle/Grid}{Grid}, \hyperlink{https://github.com/lehner/gpt}{GPT}, and \hyperlink{https://github.com/jinluchang/Qlattice}{Qlattice}.
Computations were performed mainly under the ALCC Program of the US DOE
on the SUMMIT computer at the Oak Ridge Leadership Computing Facility,
a DOE Office of Science Facility supported under Contract No. DE-AC05-00OR22725.
These calculations used gauge
configurations and propagators created using resources
of the Argonne Leadership Computing Facility, which is
a DOE Office of Science User Facility supported under
Contract DE-AC02-06CH11357.
We also acknowledge computer resources at the Oakforest-PACS
supercomputer system at Tokyo University,
partly through the HPCI System Research Project (hp180151, hp190137),
the BNL SDCC computer clusters at Brookhaven National Lab as well as computing resources provided through USQCD at Brookhaven and Jefferson National Labs.

\end{acknowledgments}

\appendix

\section{Notation}
\label{sec:notations}

We use
$S_{{\mu}}$ and $G$ to denote free muon and photon propagators:
\ba
  S_{{\mu}} (x, y)
  & = \int \frac{d^4 p}{(2 \pi)^4}  \frac{1}{i\! \not{\! p}+ m} e^{ip \cdot (x - y)}
  \\&
  = \left( -\!\not{\! \partial}_x + m \right) \int
  \frac{d^4 p}{(2 \pi)^4}  \frac{1}{p^2 + m^2} e^{ip \cdot (x - y)},
\ea
\ba
  G (x, y) & = \int \frac{d^4 p}{(2 \pi)^4}  \frac{1}{p^2} e^{ip \cdot (x -  y)}
  \\
  &= \frac{1}{4 \pi^2}  \frac{1}{(x - y)^2} .
\ea
The $\gamma_\mu$ matrices satisfy the Euclidean space-time metric
\ba
\gamma_\mu \gamma_\nu + \gamma_\nu \gamma_\mu = 2 \delta_{\mu,\nu} \,,
\ea
and
\ba
\gamma_5 = \gamma_x \gamma_y \gamma_z \gamma_t \,.
\ea

\section{$\pi^0$ long-distance contribution}
\label{sec:pi0-long-distance}
In Eq.~(\ref{eq:pi0-exch-approx}),
we replaced the QCD, Euclidean space-time, four-current connected Green's function
with the product of two amplitudes,
each coupling a pair of currents to an on-shell $\pi^0$.
These two amplitudes are joined by a pion propagator and all amplitudes are expressed in position space, so they can be directly inserted in our standard position-space evaluation of the HLbL amplitude.  Since the final expression involves two independent factors evaluating the $\pi\gamma\gamma$ coupling which are connected by an analytic, position-space pion propagator, this QCD part of the HLbL amplitude can be evaluated in a ``QCD volume'' of arbitrary size.  In particular, this volume could be much larger than that of the gauge configurations used to compute each $\pi\gamma\gamma$ vertex function.
Here we work out a concrete derivation of this formula that can be used to evaluate the long-distance part of the $\pi^0$ exchange contribution to the leading order in $1/L$.
We leave open the possibility that this approach could be developed further to systematically capture terms falling with higher powers of $1/L$ if the large volume $\pi^0$ contribution is expressed as a power series in $1/L^n$ where $L$ is the size of the QCD volume. We assume the QED volume to be infinite.

We begin with the Euclidean-space Green's function defined in Eq.~(\ref{eq:hadronic-four-point}):
\ba
\label{eq:A_def}
&\hspace{-0.5cm}
6 e^4 \mathcal{H}_{\mu',\mu,\nu',\nu}(x',x,y',y)
\\&
=
\bigl\langle T\bigl(J_{\mu'}(x') J_{\mu}(x) J_{\nu'}(y') J_{\nu}(y) \bigr) \bigr\rangle.
\nn
\ea
We will choose $x'$ and $x$ close to each other as are $y'$ and $y$.
However, we will assume that the $(x', x)$ pair is far from the $(y', y)$ pair.
Define:
\ba
\widetilde{x} &= x'-x ,\\
\widetilde{y} &= y'-y.
\ea
Since Euclidean space-time is rotational invariant,
without loss of generality,
we can choose $x - y$ to be along the Euclidean time direction,
with $ x - y = (t, \vec 0)$ and $t > 0$.

We now follow the usual steps to obtain a variant of the K\"allen-Lehman representation of the amplitude but keep only the single $\pi^0$ intermediate state since, as the lightest particle, its exchange will dominate this Green's function when $x$ and $y$ are far separated:
\ba
\label{eq:pi0_1} 
&\hspace{-0.5cm}
6 e^4 \mathcal{H}^{\pi^0}_{\mu',\mu,\nu',\nu}(x',x,y',y)
\\=&
\int\! \frac{d^3 p}{(2\pi)^3} \frac{1}{2E_{\pi,\vec p}}
\bigl\langle 0\bigl|T\bigl(J_{\mu'}(x') J_{\mu}(x)\bigr)\bigr|\pi^0(\vec p)\bigr\rangle
\nn
\\&\hspace{2cm}\times
\bigl\langle\pi^0(\vec p) \bigl|T\bigl(J_{\nu'}(y') J_{\nu}(y) \bigr)\bigr|0\bigr\rangle,
\nn
\ea
where the superscript $\pi^0$ indicates that only the $\pi^0$ contribution to $\mathcal{H}$ is represented.

Next we use four-dimensional translational invariance to remove the variables $x$ and $y$ from the two current-current-$\pi^0$ amplitudes.  We will also replace these two $O(4)$-covariant current-current-$\pi^0$ amplitudes by functions of the four momentum
$p = (i E_{\pi,\vec p}, \vec p)$. Given the factors of $\exp{(i p\cdot x)}$ and $\exp{(-i p\cdot y)}$ which appear, we can then replace the four vector $p_\mu$ by $-i\partial/\partial {x}_\mu$ or $i\partial/\partial {y}_\mu$ as needed.  For example,
\begin{eqnarray}
\label{eq:pi0-tff-appendix}
\bigl\langle 0\bigl|T\bigl(J_{\mu'}(x') J_{\mu}(x)\bigr)\bigr|\pi^0(\vec p)\bigr\rangle \nonumber&&\\
&&\hskip -1.5 in =\bigl\langle 0\bigl|T\bigl(J_{\mu'}(\widetilde{x}) J_{\mu}(0)\bigr)\bigr|\pi^0(\vec p)\bigr\rangle e^{ip\cdot x} \nonumber\\
&&\hskip -1.5 in = \mathcal{F}_{\mu',\mu}(\widetilde{x},p) e^{ip\cdot x}\nonumber \\
&&\hskip -1.5 in = \mathcal{F}_{\mu',\mu}\left(\widetilde{x},-i\frac{\partial}{\partial {x}}\right) e^{ip\cdot x}.
\end{eqnarray}
where the definition of the $\pi^0$ transition form factors follows Eq.~\eqref{eq:pi0-tff}.

Using this approach to remove the explicit dependence of the two amplitudes on $p_\mu$, we can then perform the final step of the usual K\"allen-Lehman derivation and introduce the free pion propagator:
\ba
&\hspace{-0.5cm}
\int\! \frac{d^3 p}{(2\pi)^3} \frac{1}{2E_{\pi,\vec p}}
e^{ip \cdot (x-y)}
\nn\\&=
\int\! \frac{d^3 p}{(2\pi)^3} \frac{1}{2E_{\pi,\vec p}}
e^{i \vec p \cdot (\vec x- \vec y)} e^{-E_{\pi,\vec p} ({x}_t - {y}_t)}
\\&=
\int\! \frac{d^4 p}{(2\pi)^4}  \frac{e^{ip\cdot (x-y)}}{p^2 + m_\pi^2}
\\& = D_{\pi^0}(x-y)
\ea
where $D_{\pi^0}(x-y)$ is the free Euclidean-space propagator for a scalar particle of mass $m_\pi$.

We can then rewrite Eq.~\eqref{eq:pi0_1} in terms of $\mathcal{F}$ and $D_{\pi^0}$ to obtain the result:
\ba
\label{eq:pion-pole}
&6 e^4 \mathcal{H}^{\pi^0}_{\mu',\mu,\nu',\nu}(x',x,y',y)
\\
&=
\mathcal{F}_{\mu',\mu}\left(\widetilde{x},-i\frac{\partial}{\partial {x}_\mu}\right) 
\mathcal{F}_{\nu',\nu}\left(\widetilde{y},-i\frac{\partial}{\partial {y}_\mu}\right)
D_{\pi^0}(x-y).
\nn
\ea
At this step, we have made explicit the $O(4)$-covariance of the right-hand side of the equation.
So this equation should hold for an arbitrary orientation of $x - y$.

So far all of the steps taken have been exact.  We expect the quantity computed, the contribution of a single pion exchange, to dominate the long distance limit in which $|x-y|$
is large, i.e. $|x - y| \gtrsim L$.
Now we will evaluate the derivatives with respect to $x$ and $y$ which appear in these equations but keep only the leading term in an expansion in powers of $1/L$.  For example, to leading order in $1/|x-y|$:
\ba
&\quad\prod_{i=1}^N \left(\frac{\partial}{\partial {x}_{\rho_i}}\right) D_{\pi^0}(x-y)
\nn\\
&=
\prod_{i=1}^N \left(\frac{\partial}{\partial {x}_{\rho_i}}\right) \frac{c\, e^{-m_\pi|x-y|}}{|x-y|^{3/2}}
\\
&\approx\prod_{i=1}^N \left(-m_\pi \frac{(x-y)_{\rho_i}}{|x-y|} \right) \frac{c\, e^{-m_\pi|x-y|}}{|x-y|^{3/2}},
\ea
where the relation holds for arbitrary $N$ and $\rho_i$ labels the component of $x$ appearing in the $i^{th}$ derivative in the product of $N$ derivatives.
Here we use the large distance behavior of $D_{\pi^0}(x-y)$:
\ba
D_{\pi^0}(x - y) \approx
\frac{c\, e^{-m_\pi |x - y|}}{|x - y|^{3/2}}.
\ea
Introducing this approximation into Eqs.~\eqref{eq:pion-pole},
we obtain
\ba
\label{eq:pion-exch-approx-1}
&6 e^4 \mathcal{H}^{\pi^0}_{\mu',\mu,\nu',\nu}(x',x,y',y)
\\
&\approx
\mathcal{F}_{\mu',\mu}\left(\widetilde{x},im_\pi \hat n\right) 
\mathcal{F}_{\nu',\nu}\left(\widetilde{y},-im_\pi \hat n\right)
D_{\pi^0}(x-y).
\nn
\ea
where $\hat n_\mu$ is the Euclidean unit vector $(x-y)_{\mu}/|x-y|$.  
This is finally the equation of interest, Eq.~\eqref{eq:pi0-exch-approx}, which appeared earlier.
Since the two quantities expressed in terms of the form factor $\mathcal F$ can be evaluated from independent lattice QCD ensemble averages, the expression on the right hand side of Eq.~\eqref{eq:pion-exch-approx-1} can be evaluated from a standard lattice QCD ensemble average followed by an $O(4)$ rotation as defined in Eq.~\eqref{eq:pi0-form-factor-rotation}, which we repeat here:
\ba
\label{eq:pi0-form-factor-rotation-1}
&\mathcal{F}_{\mu',\mu}\Big(\widetilde x, i m_\pi \hat n \Big)
=
\Lambda_{\mu',\rho'}
\Lambda_{\mu,\rho}
\mathcal{F}_{\rho',\rho}\Big(\widetilde{x}', i m_\pi \hat t \Big),
\ea
The $O(4)$ rotation matrix $\Lambda = \Lambda(\hat n)$ and Euclidean space-time coordinate $\tilde x'$ satisfy:
\ba
\delta_{\mu,\nu} &= \Lambda_{\mu,\mu'} \Lambda_{\nu,\nu'}\delta_{\mu',\nu'},
\\
\hat n_\mu &= \Lambda_{\mu,\mu'} \hat t_{\mu'},
\\
\widetilde x_\mu &= \Lambda_{\mu,\mu'} \widetilde x'_{\mu'}.
\ea
The rotated form factor $\mathcal{F}_{\rho',\rho}\Big(\widetilde{x}', i m_\pi \hat t \Big)$ follows the definition
Eq.~\eqref{eq:pi0-tff-appendix} or Eq.~\eqref{eq:pi0-tff} with zero momentum $\pi^0$ state:
\ba
&\vspace{-1.0cm}\mathcal{F}_{\rho',\rho}\Big(\widetilde{x}', i m_\pi \hat t \Big)
\nn\\
&=
\la 0 | T J_{\rho'}(\widetilde{x}') J_{\rho}(0) | \pi^0(\vec{p} = \vec 0) \ra
\nn\\
&=
\sqrt{2 m_\pi L^3}
\lim_{t_\text{tsep} \to +\infty}
\frac
{\la T J_{\rho'}(\widetilde{x}') J_{\rho}(0) \pi^0(-t_\text{sep}) \ra}
{\sqrt{\la T \pi^0(t_\text{sep}) \pi^0(-t_\text{sep}) \ra}},
\ea
where $\pi^0(t)$ is a zero momentum $\pi^0$ interpolating operator.
We use Coulomb fixed gauge wall source pion operator in our numerical lattice calculation.
The other factor $\mathcal{F}_{\nu',\nu}\left(\widetilde{y},-im_\pi \hat n\right)$ can be calculated similarly.

The above procedure allows a direct calculation of the $\pi^0$ contribution to the long distance part the HLbL amplitude
without relying on any parameterization or modelling of the $\pi^0$ transition form factor.

In this calculation, we employ this long-distance $\pi^0$-exchange approximation in the following region:
\ba
R_\text{max} = \text{max}(|x - y|, |y' - y|, |x - y'|) \geq 4~\mathrm{fm},
\ea
as defined in Eq.~\eqref{eq:rmax-def}.  This long-distance hadronic function is then combined with infinite volume QED weighting function for HLbL
as illustrated in Fig.~\ref{fig:long-distance-pi0}. Note that we need to sum over all permutations of $x, y, y'$.

\section{Summand fitting form}
\label{sec:fit-forms}
In section \ref{sec:light-contrib}, we have combined the disconnected diagrams and connected diagrams with a certain fraction multiplied
to form $a_\mu^\text{no-pion}$ according to Eq.~(\ref{eq:a-mu-no-pion}).
We have fitted the long distance behavior of this combination with Eq.~(\ref{eq:fit-form}) and the results
are shown in Fig.~\ref{fig:48I-light-no-pion}.
In this appendix, we will explore other possible fitting forms and will also
apply these fitting functions to other contributions.

The fitting forms are
\begin{itemize}
    \item fit1: This is the form used in the main calculation, same as Eq.~(\ref{eq:a-mu-no-pion}).
    \ba
    \label{eq:fit-form-1}
    f(R_\text{max}) = A / \text{fm}^4 \frac{R_\text{max}^6}{R_\text{max}^3 + (C~\text{fm})^3} e^{-B R_\text{max} / (\text{fm}\cdot\text{GeV})}
    \ea
    \item fit2:
    \ba
    \label{eq:fit-form-2}
    f(R_\text{max}) =  A / \text{fm}^4 R_\text{max}^3 e^{-B R_\text{max} / (\text{fm}\cdot\text{GeV})}
    \ea
    \item fit3:
    \ba
    \label{eq:fit-form-3}
    f(R_\text{max}) =  A / \text{fm} \frac{R_\text{max}^6}{(R_\text{max}^3 + (C~\text{fm})^3)^2} e^{-B R_\text{max} / (\text{fm}\cdot\text{GeV})}
    \ea
    \item fit4:
    \ba
    \label{eq:fit-form-4}
    f(R_\text{max}) = A / \text{fm}^4 \frac{R_\text{max}^6}{R_\text{max}^3 + (C~\text{fm})^3} e^{-m_\eta R_\text{max} / (\text{fm}\cdot\text{GeV})}
    \ea
    \item fit5:
    \ba
    \label{eq:fit-form-5}
    f(R_\text{max}) =  A / \text{fm}^4 R_\text{max}^3 e^{-m_\eta R_\text{max} / (\text{fm}\cdot\text{GeV})}
    \ea
    \item fit6:
    \ba
    \label{eq:fit-form-6}
    f(R_\text{max}) =  A / \text{fm} \frac{R_\text{max}^6}{(R_\text{max}^3 + (C~\text{fm})^3)^2} e^{-m_\eta R_\text{max} / (\text{fm}\cdot\text{GeV})}
    \ea
    \item fit7:
    \ba
    \label{eq:fit-form-7}
    f(R_\text{max}) =  A / \text{fm} \frac{R_\text{max}^6}{(R_\text{max}^3 + (C~\text{fm})^3)^2} e^{-2 m_\pi R_\text{max} / (\text{fm}\cdot\text{GeV})}
    \ea
    \item fit8:
    \ba
    \label{eq:fit-form-8}
    f(R_\text{max}) =  A / \text{fm} \frac{R_\text{max}^6}{(R_\text{max}^3 + (C~\text{fm})^3)^2} e^{-m_\pi R_\text{max} / (\text{fm}\cdot\text{GeV})}
    \ea
\end{itemize}
where we set $m_\pi = 139~\mathrm{MeV}$ and $m_\eta = 550~\mathrm{MeV}$. For ``fit1'', ``fit2'', ``fit3'', we always constrain the $B$ parameter to be larger or equal to $m_\pi$.

\begin{figure*}[t]
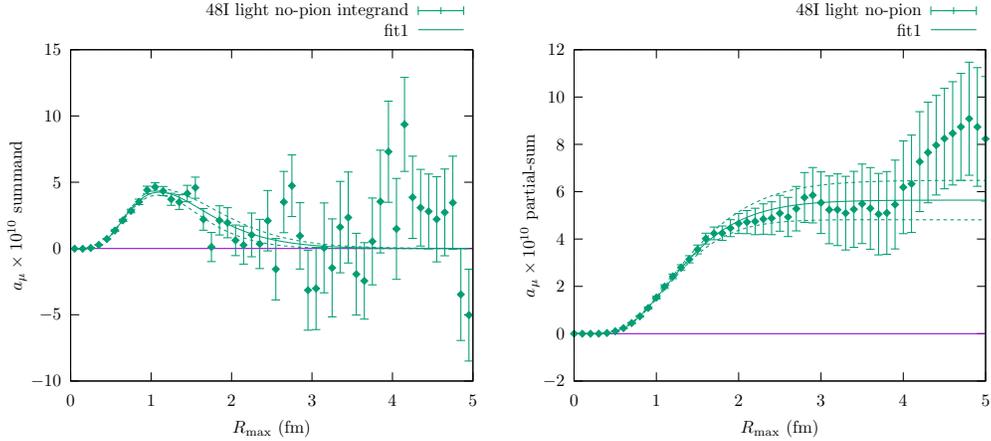

\centering
\includegraphics[width=0.35\textwidth]{fig_48I_light_no-pion_1.eps}
~~~
\includegraphics[width=0.35\textwidth]{fig_48I_light_no-pion_2.eps}
\caption{\label{fig:48I-light-no-pion-1}Same as Fig.~\ref{fig:48I-light-no-pion}, reproduced for easy comparison with the other figures in this appendix. The fit function is Eq.~(\ref{eq:fit-form-1}). The fit starts at $0.5~\mathrm{fm}$.}
\end{figure*}

\begin{figure*}[t]
\centering
\includegraphics[width=0.35\textwidth]{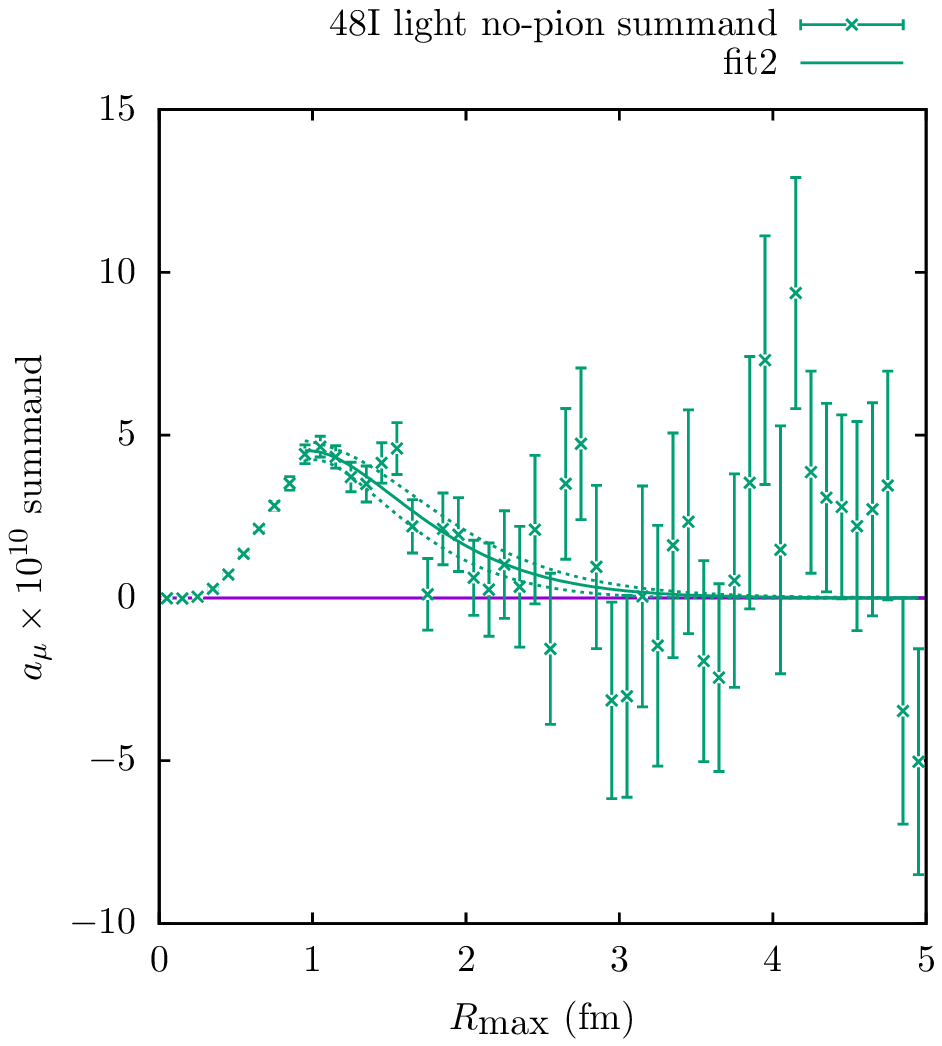}
\includegraphics[width=0.35\textwidth]{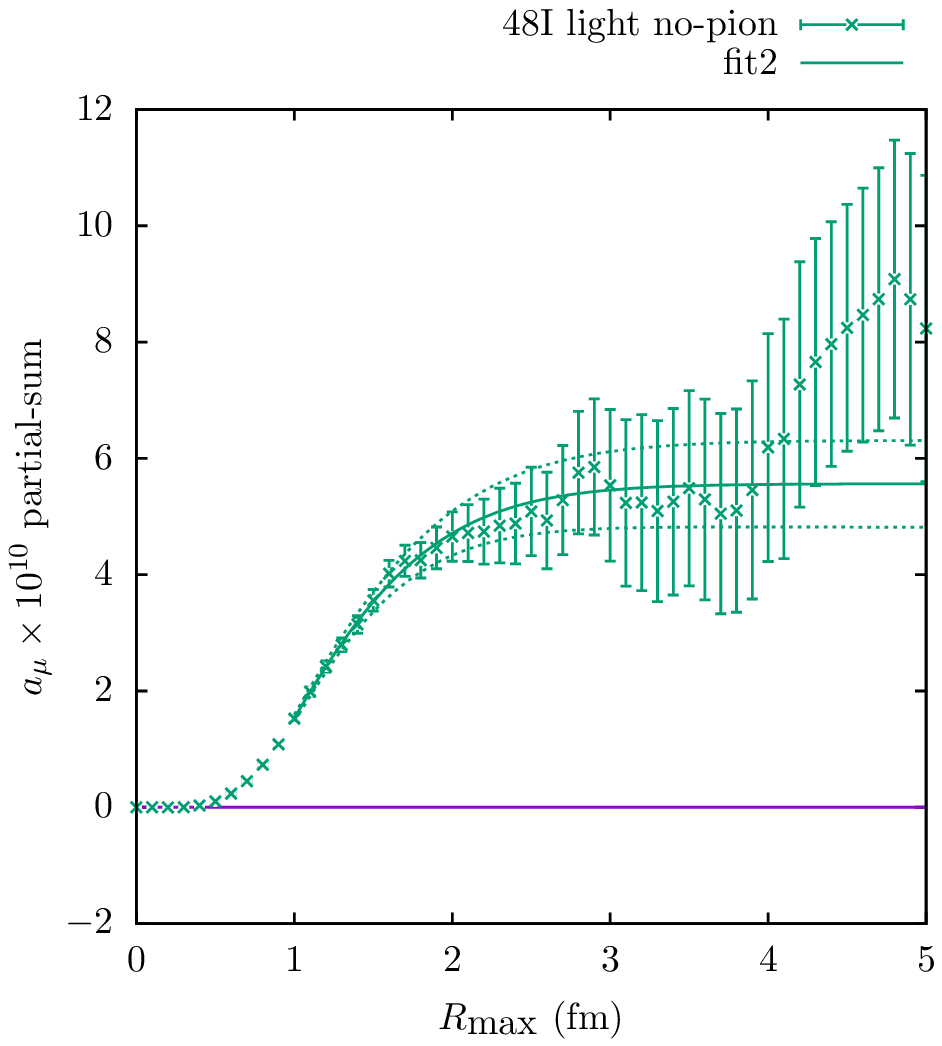}
\caption{\label{fig:48I-light-no-pion-2}Similar to Fig.~\ref{fig:48I-light-no-pion} but with the fit function in Eq.~(\ref{eq:fit-form-2}). The fit starts at $1.0~\mathrm{fm}$.}
\end{figure*}

\begin{figure*}[t]
\centering
\includegraphics[width=0.35\textwidth]{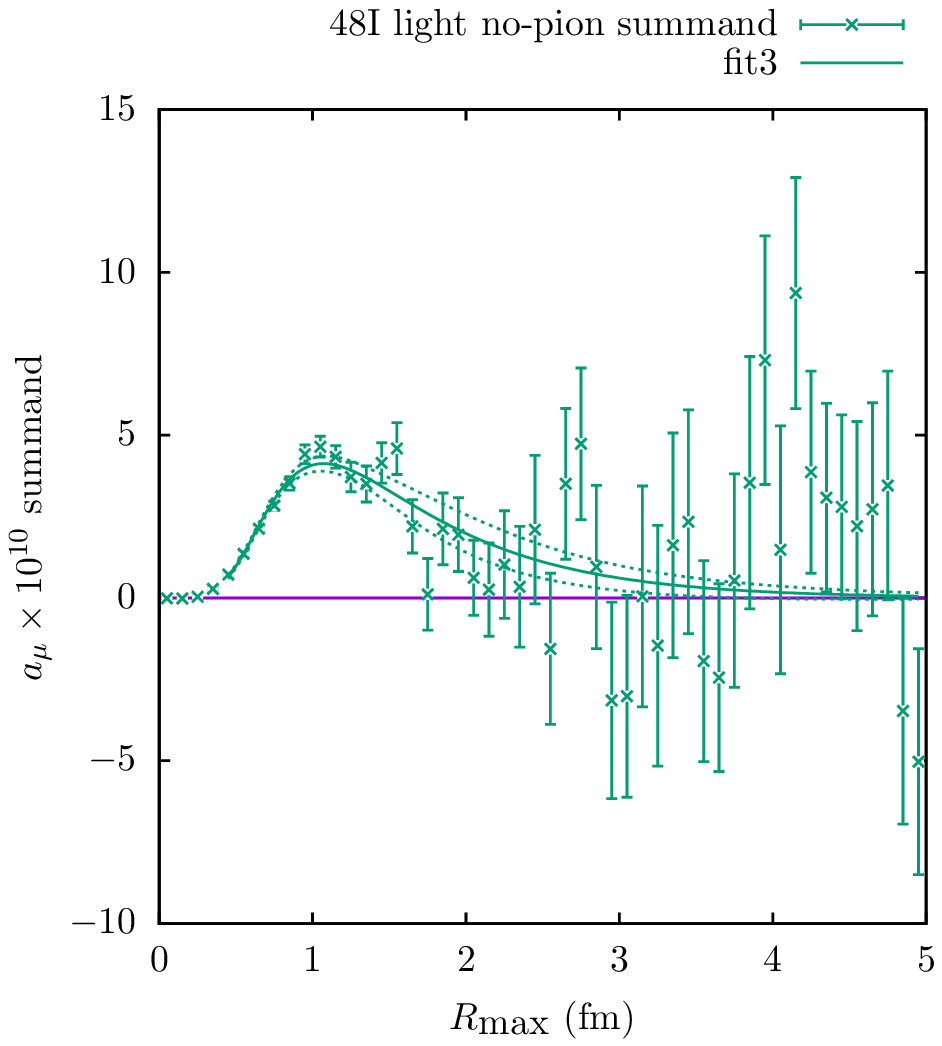}
\includegraphics[width=0.35\textwidth]{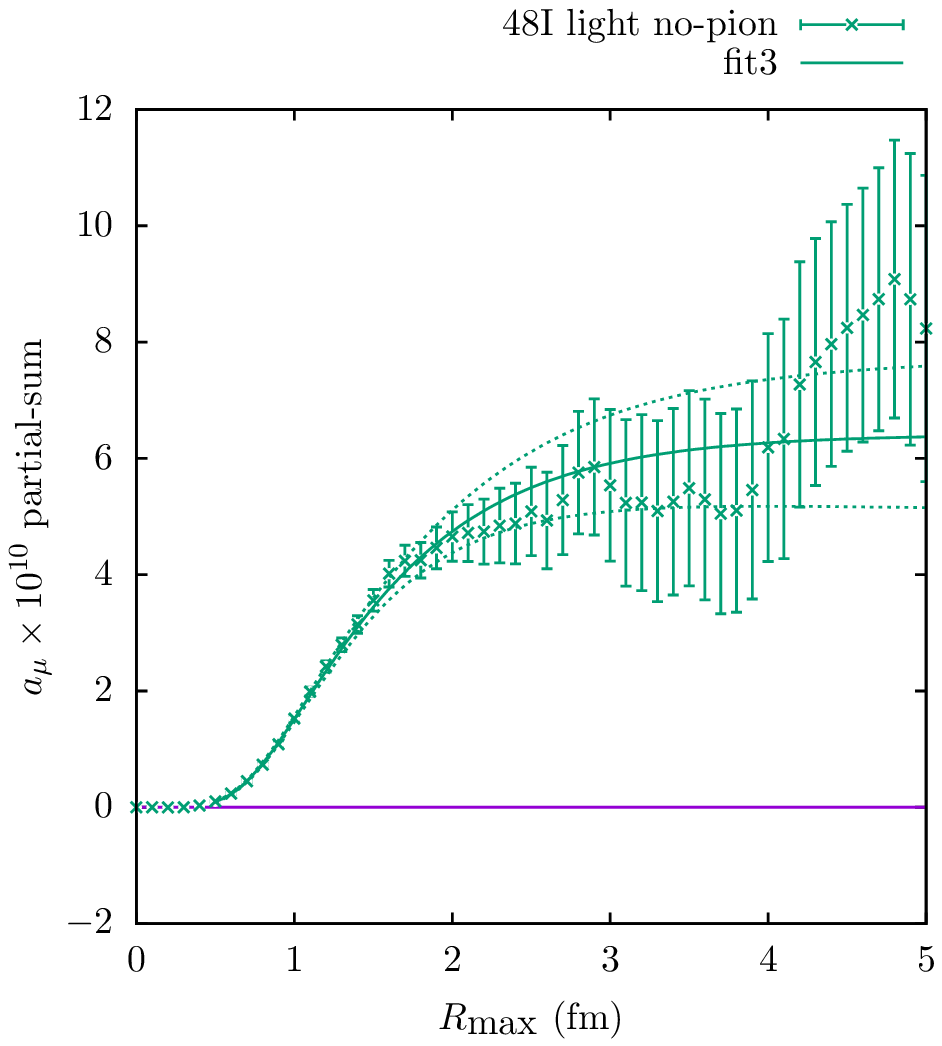}
\caption{\label{fig:48I-light-no-pion-3}Similar to Fig.~\ref{fig:48I-light-no-pion} but with the fit function in Eq.~(\ref{eq:fit-form-3}). The fit starts at $0.5~\mathrm{fm}$.}
\end{figure*}

\begin{figure*}[t]
\centering
\includegraphics[width=0.35\textwidth]{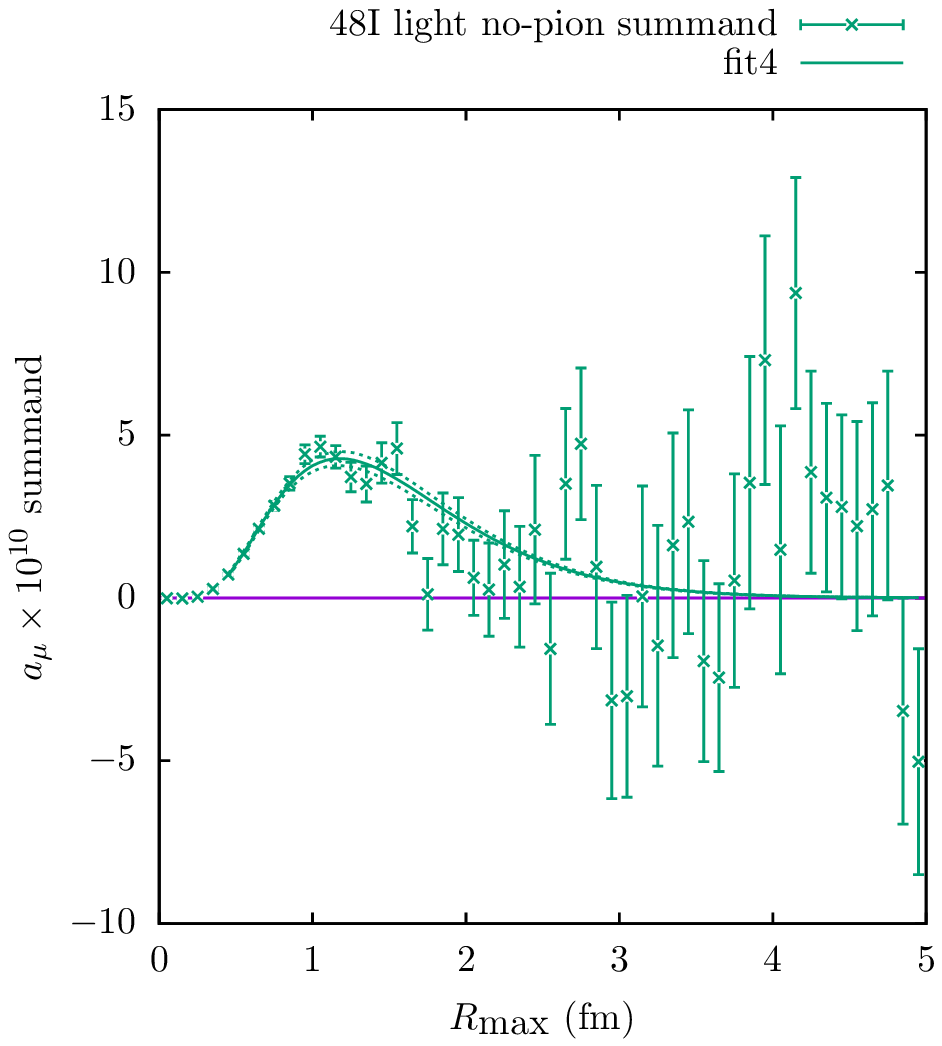}
\includegraphics[width=0.35\textwidth]{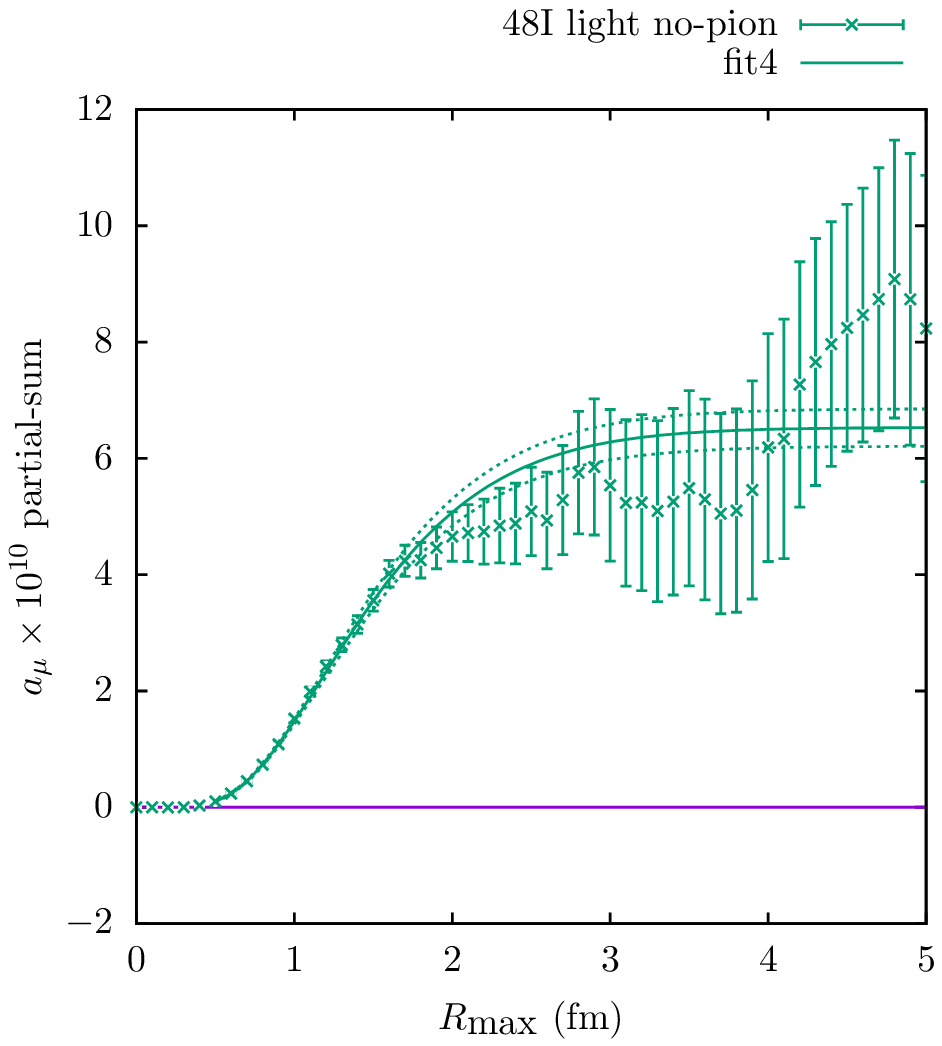}
\caption{\label{fig:48I-light-no-pion-4}Similar to Fig.~\ref{fig:48I-light-no-pion} but with the fit function in Eq.~(\ref{eq:fit-form-4}). The fit starts at $0.5~\mathrm{fm}$.}
\end{figure*}

\begin{figure*}[t]
\centering
\includegraphics[width=0.35\textwidth]{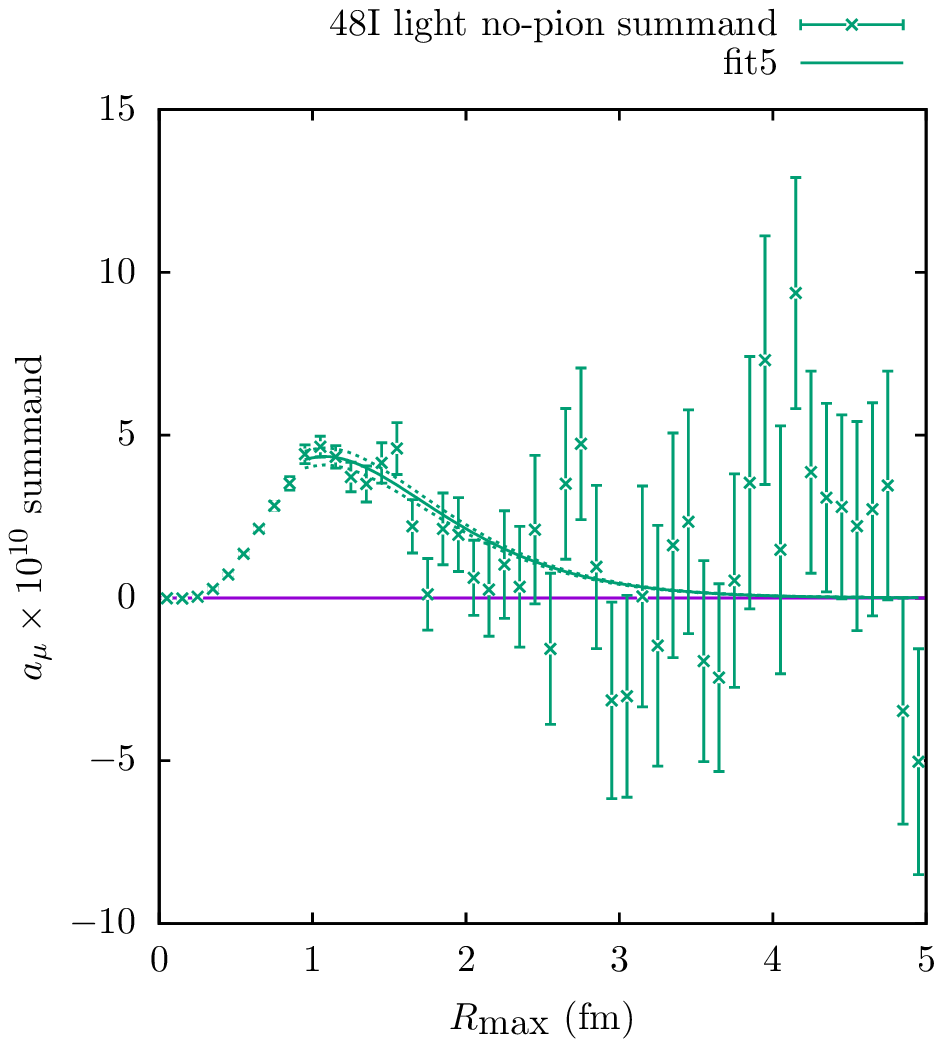}
\includegraphics[width=0.35\textwidth]{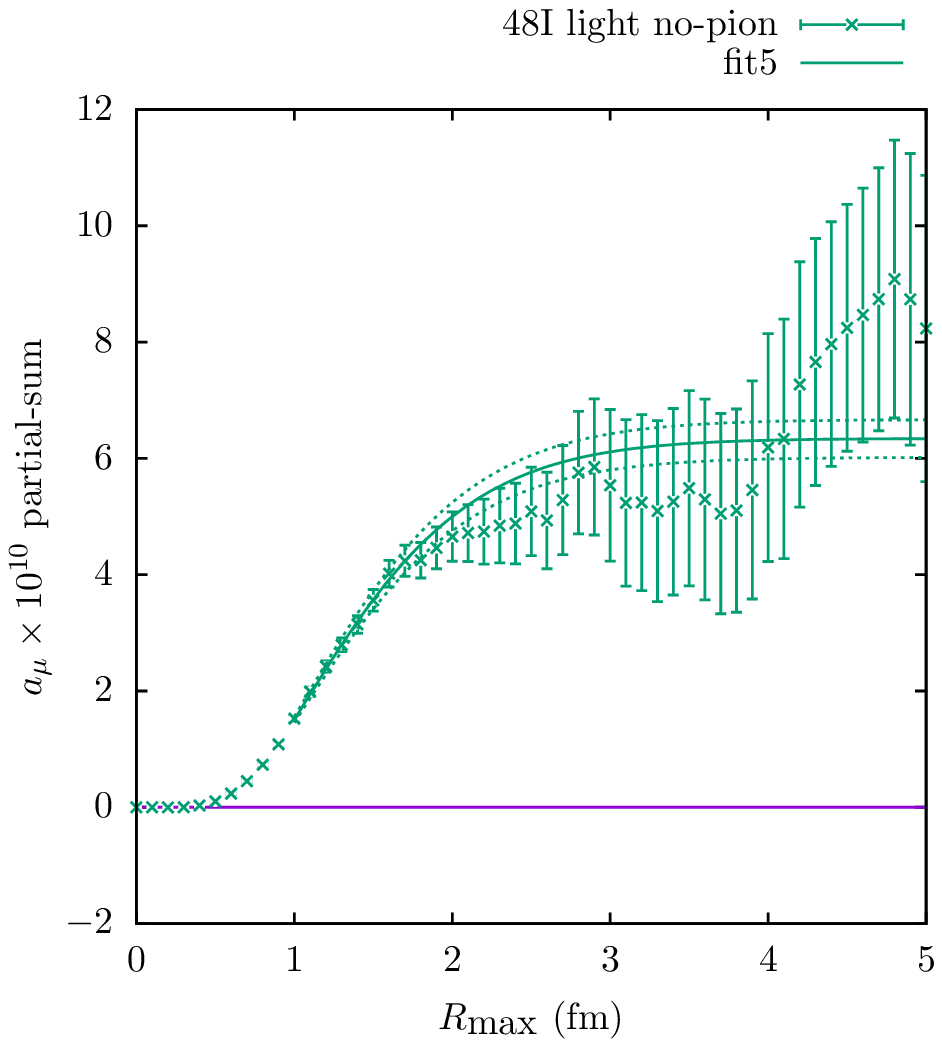}
\caption{\label{fig:48I-light-no-pion-5}Similar to Fig.~\ref{fig:48I-light-no-pion} but with the fit function in Eq.~(\ref{eq:fit-form-5}). The fit starts at $1.0~\mathrm{fm}$.}
\end{figure*}

\begin{figure*}[t]
\centering
\includegraphics[width=0.35\textwidth]{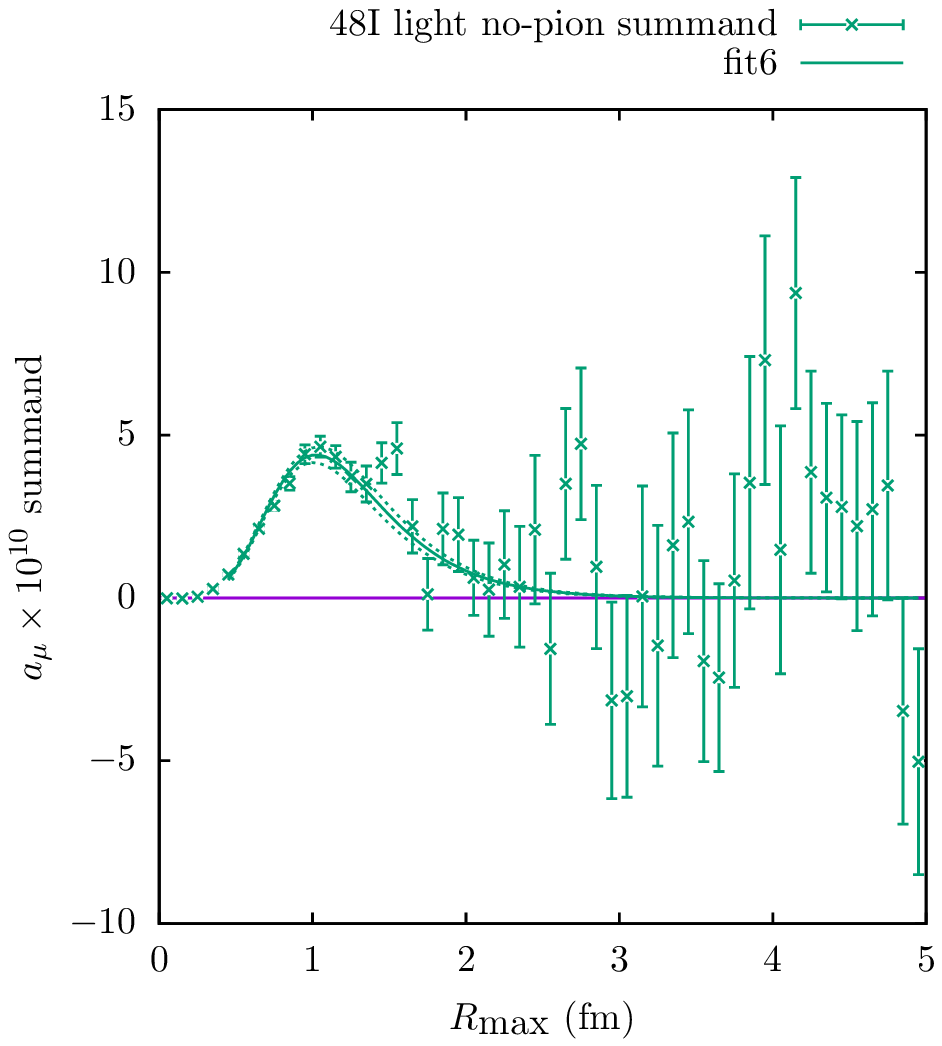}
\includegraphics[width=0.35\textwidth]{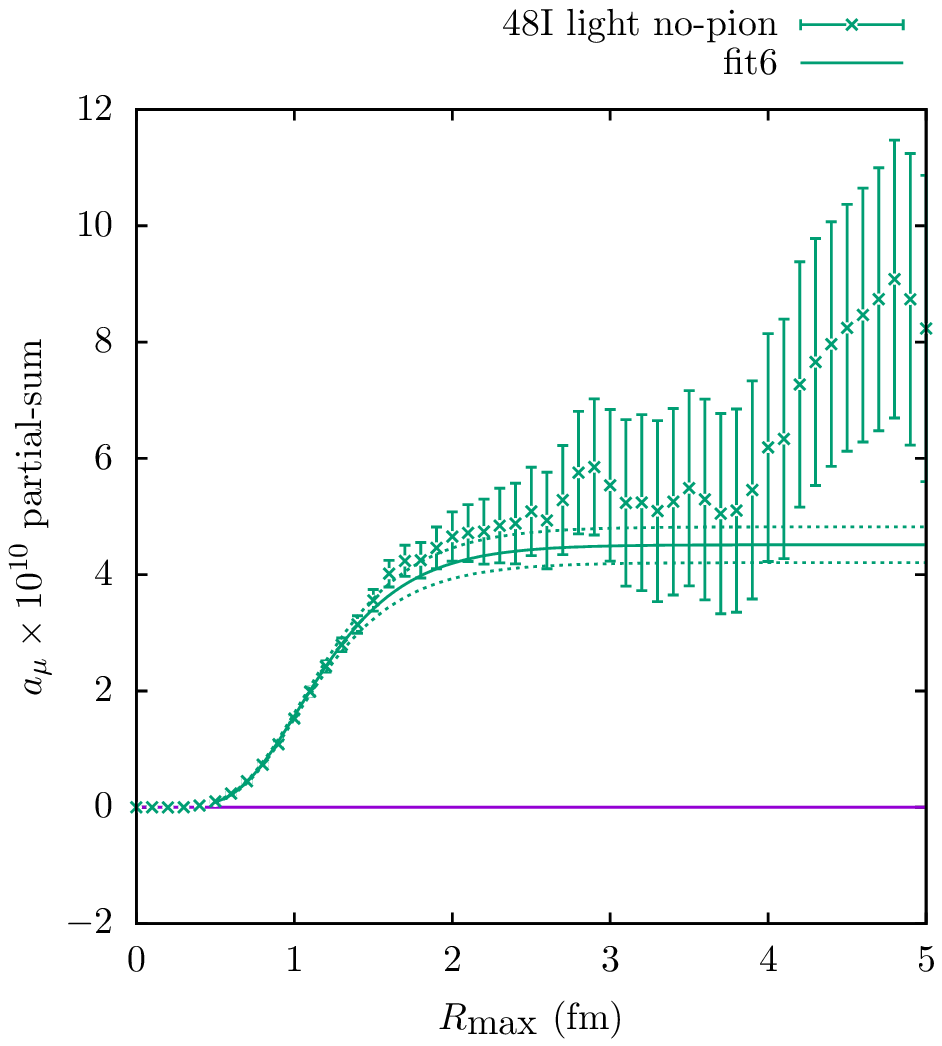}
\caption{\label{fig:48I-light-no-pion-6}Similar to Fig.~\ref{fig:48I-light-no-pion} but with the fit function in Eq.~(\ref{eq:fit-form-6}). The fit starts at $0.5~\mathrm{fm}$.}
\end{figure*}

\begin{figure*}[t]
\centering
\includegraphics[width=0.35\textwidth]{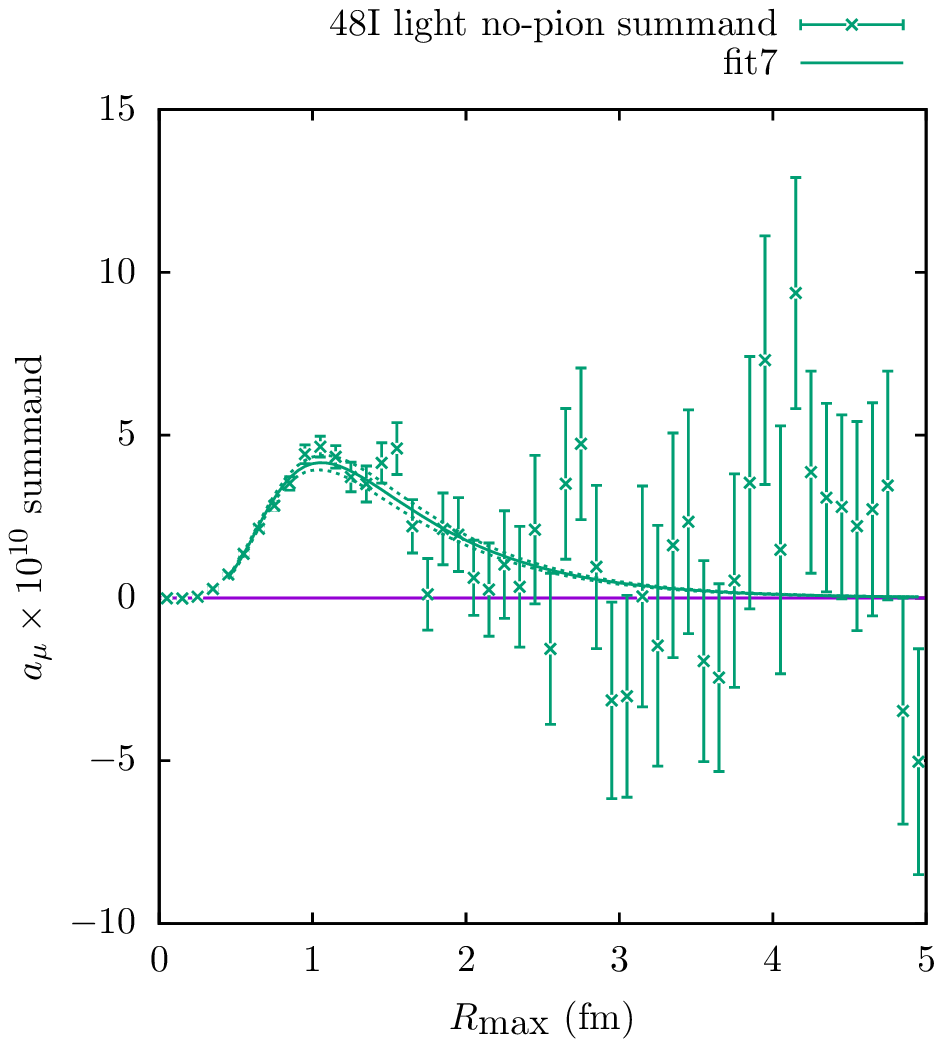}
\includegraphics[width=0.35\textwidth]{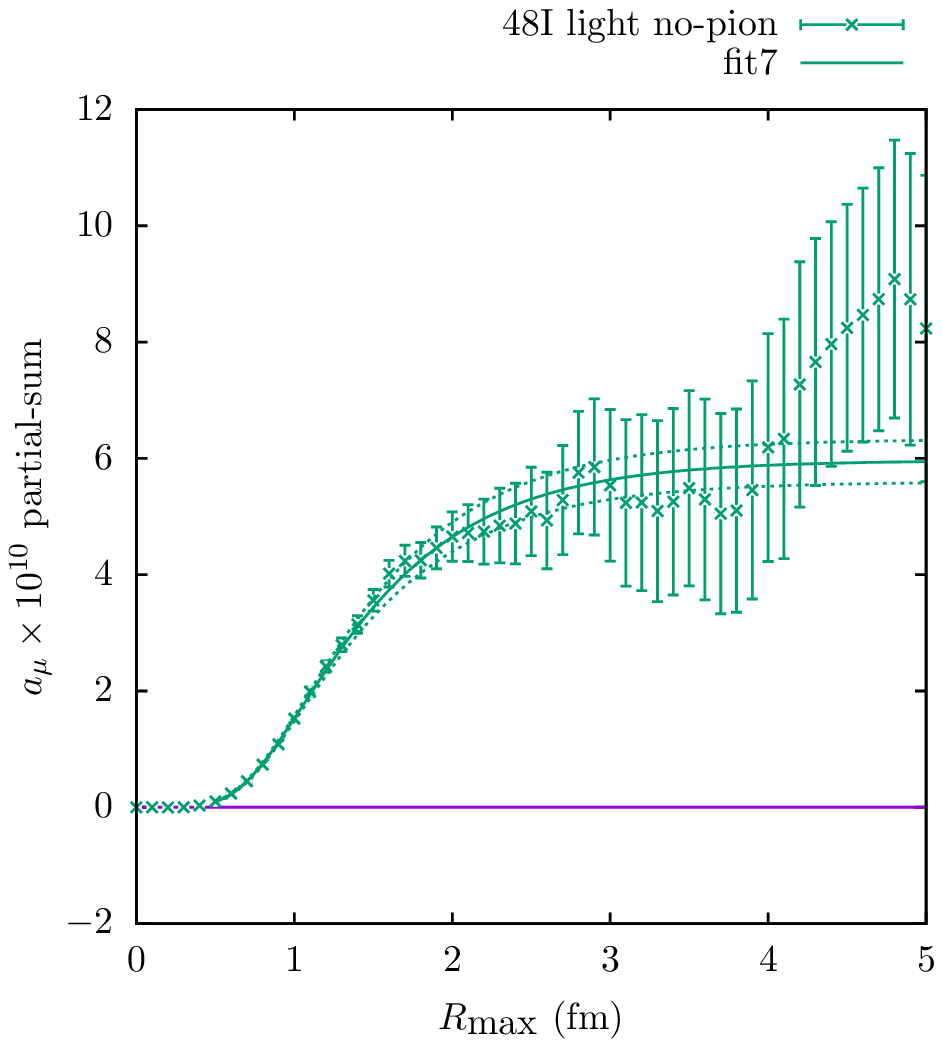}
\caption{\label{fig:48I-light-no-pion-7}Similar to Fig.~\ref{fig:48I-light-no-pion} but with the fit function in Eq.~(\ref{eq:fit-form-7}). The fit starts at $0.5~\mathrm{fm}$.}
\end{figure*}

In Figs.~\ref{fig:48I-light-no-pion-1},\ref{fig:48I-light-no-pion-2},\ref{fig:48I-light-no-pion-3},\ref{fig:48I-light-no-pion-4},\ref{fig:48I-light-no-pion-5},\ref{fig:48I-light-no-pion-6}, and \ref{fig:48I-light-no-pion-7}, we show the resulting fits along with the data.

\begin{figure*}[t]
\centering
\includegraphics[width=0.35\textwidth]{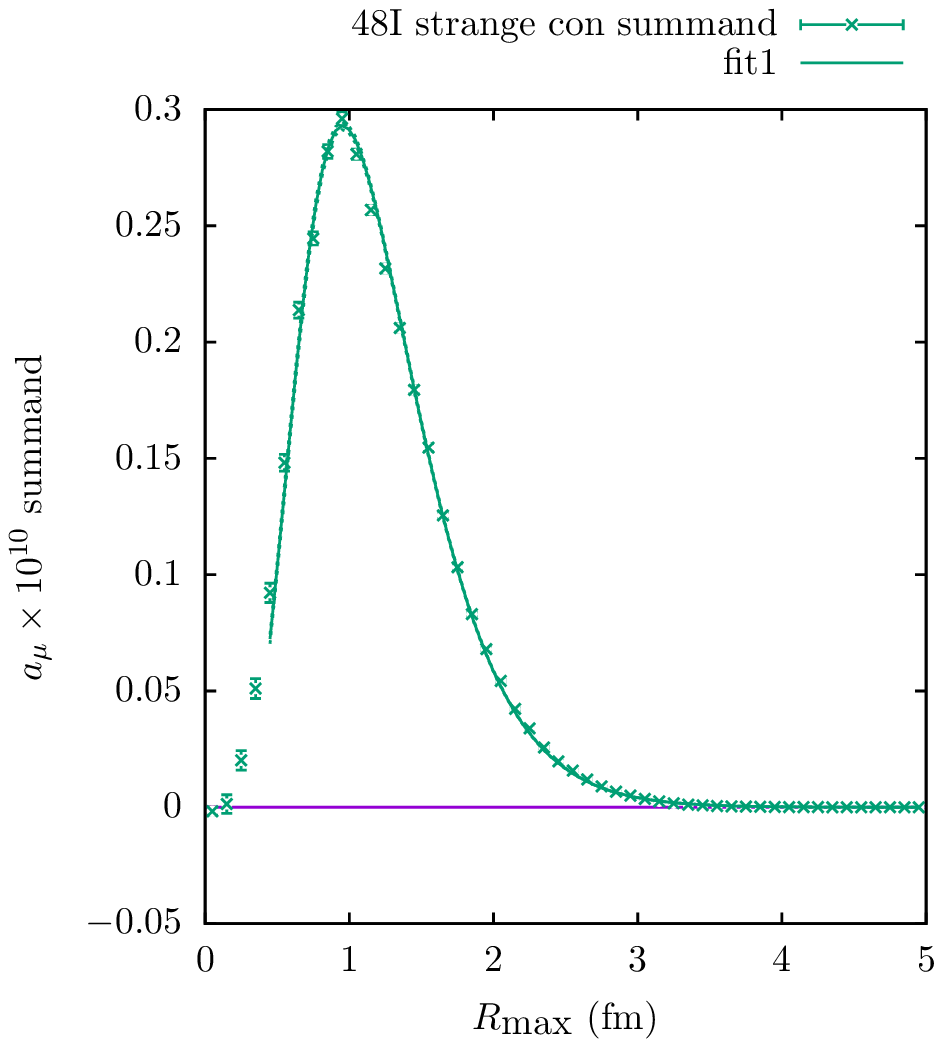}
\includegraphics[width=0.35\textwidth]{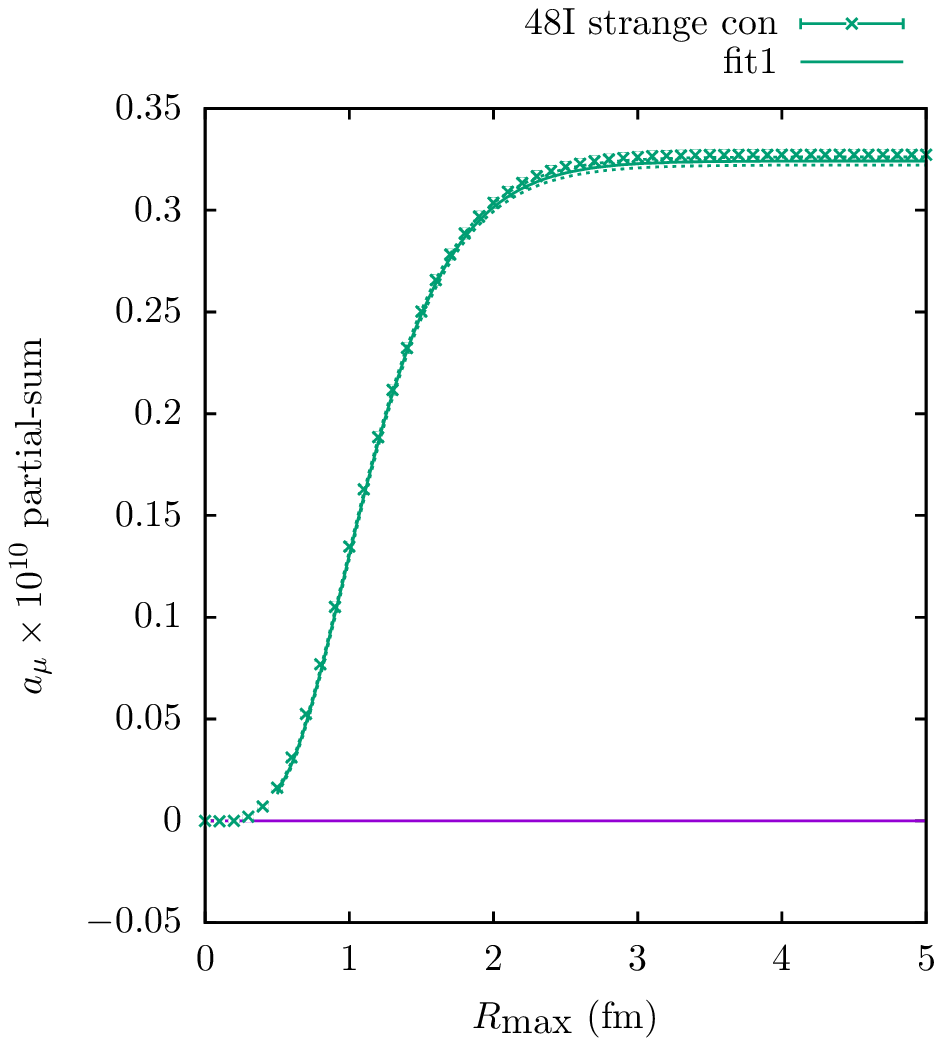}
\caption{\label{fig:48I-strange-con-1}Similar to Fig.~\ref{fig:48I-light-no-pion} but for the strange quark connected contriubtion with the fit function in Eq.~(\ref{eq:fit-form-1}). The fit starts at $0.5~\mathrm{fm}$.}
\end{figure*}

\begin{figure*}[t]
\centering
\includegraphics[width=0.35\textwidth]{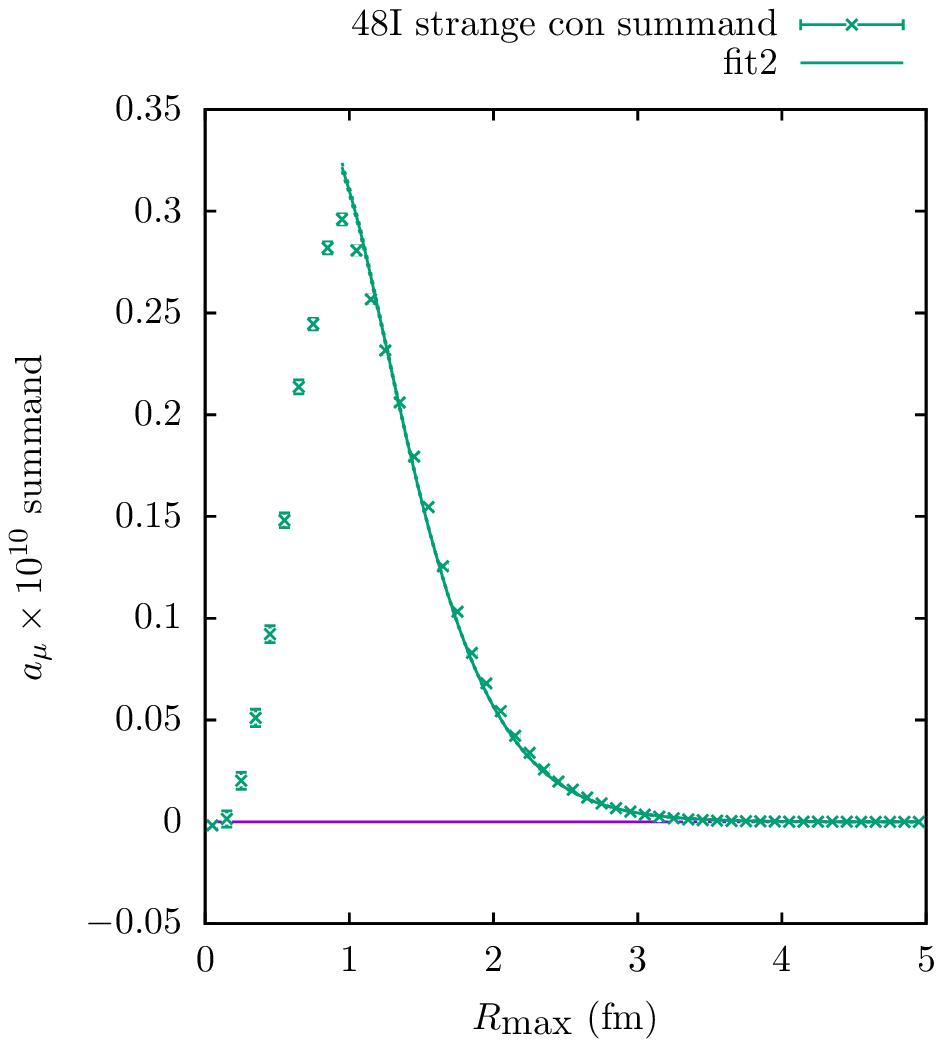}
\includegraphics[width=0.35\textwidth]{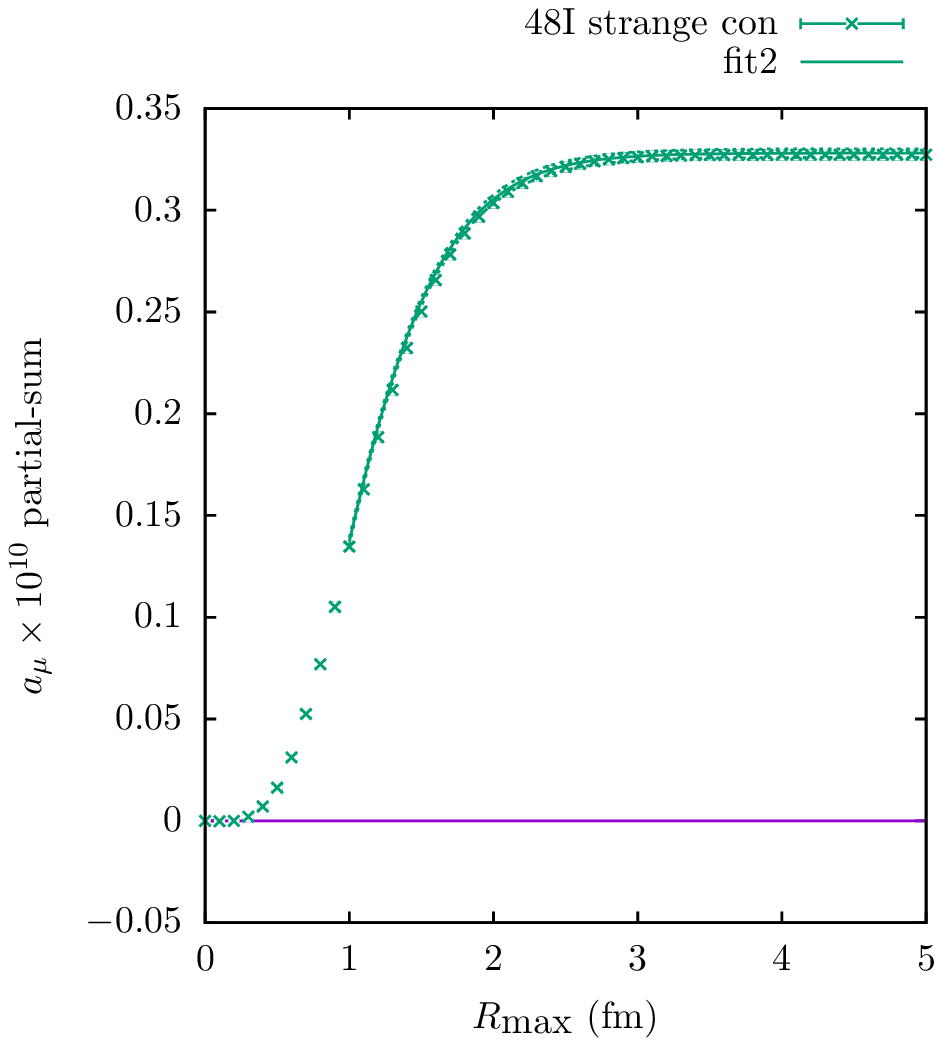}
\caption{\label{fig:48I-strange-con-2}Similar to Fig.~\ref{fig:48I-light-no-pion} but for the strange quark connected contribution  with the fit function in Eq.~(\ref{eq:fit-form-2}). The fit starts at $1.0~\mathrm{fm}$.}
\end{figure*}

\begin{figure*}[t]
\centering
\includegraphics[width=0.35\textwidth]{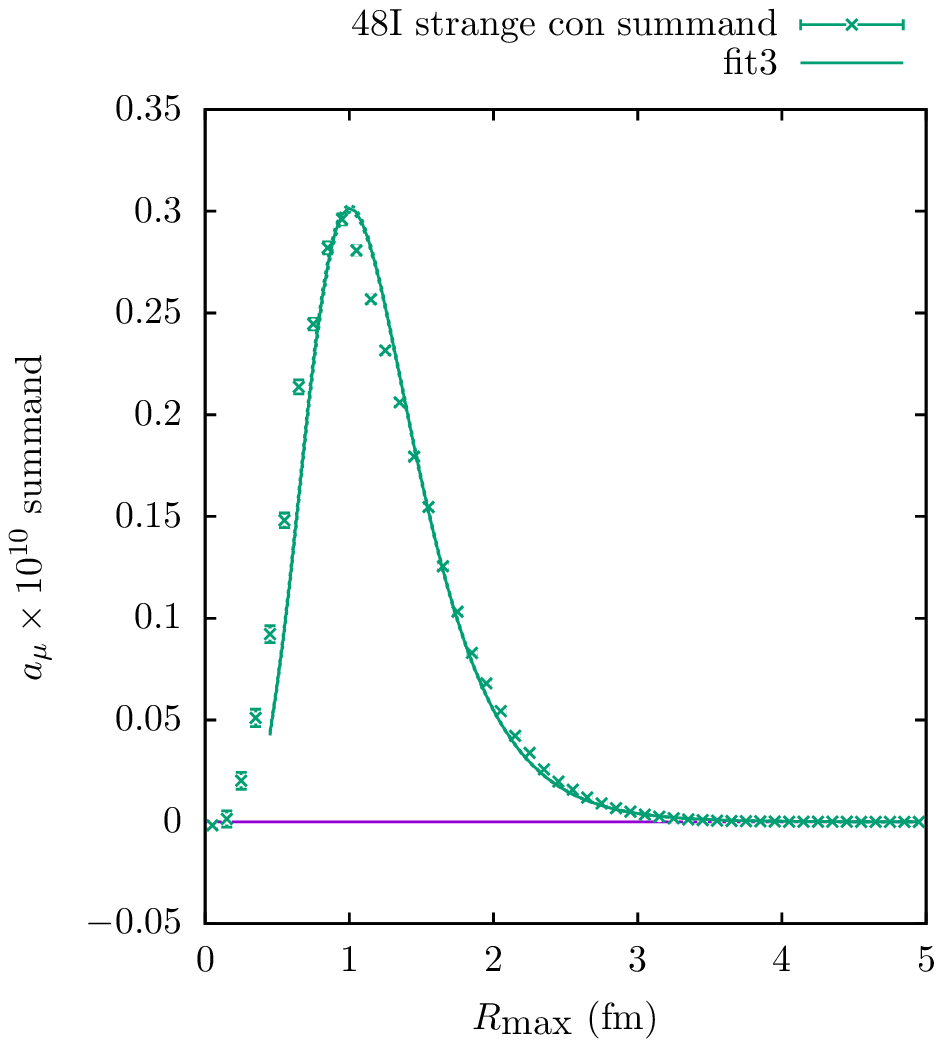}
\includegraphics[width=0.35\textwidth]{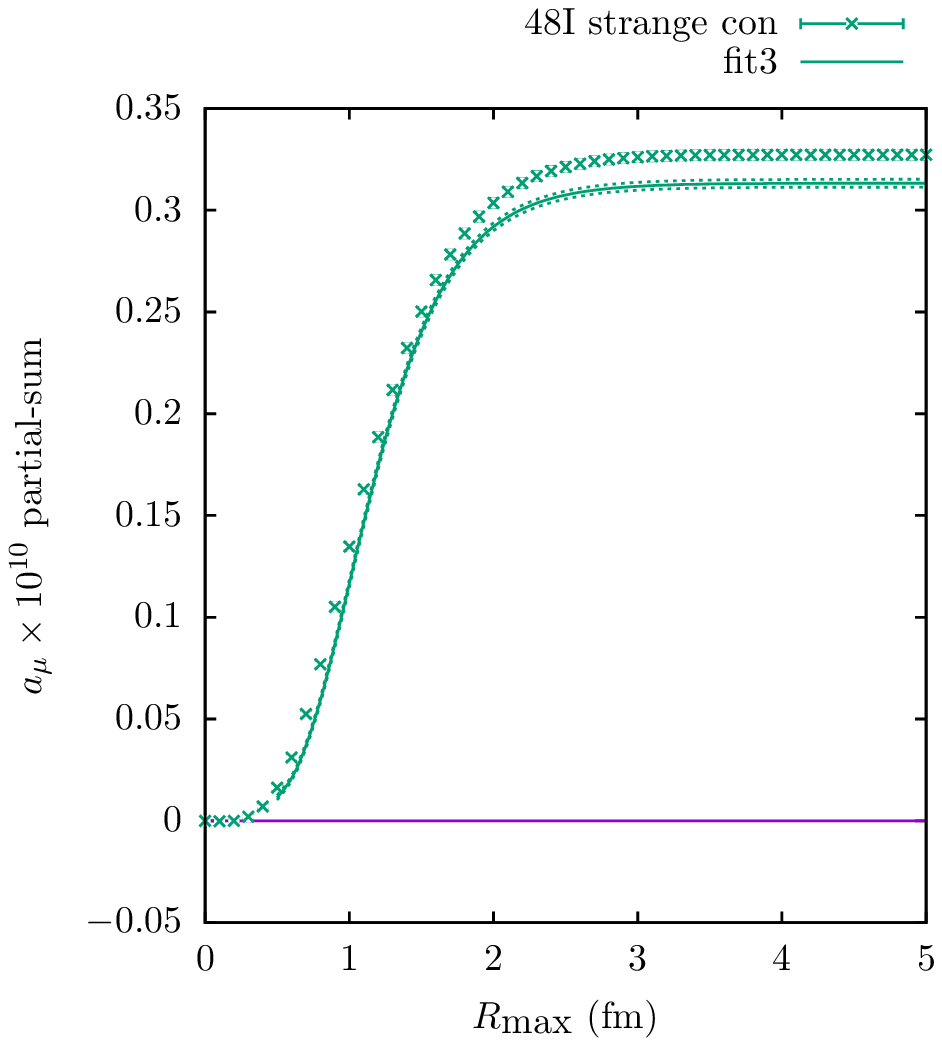}
\caption{\label{fig:48I-strange-con-3}Similar to Fig.~\ref{fig:48I-light-no-pion} but for the strange quark connected contribution with the fit function in Eq.~(\ref{eq:fit-form-3}). The fit starts at $0.5~\mathrm{fm}$.}
\end{figure*}

\begin{figure*}[t]
\centering
\includegraphics[width=0.35\textwidth]{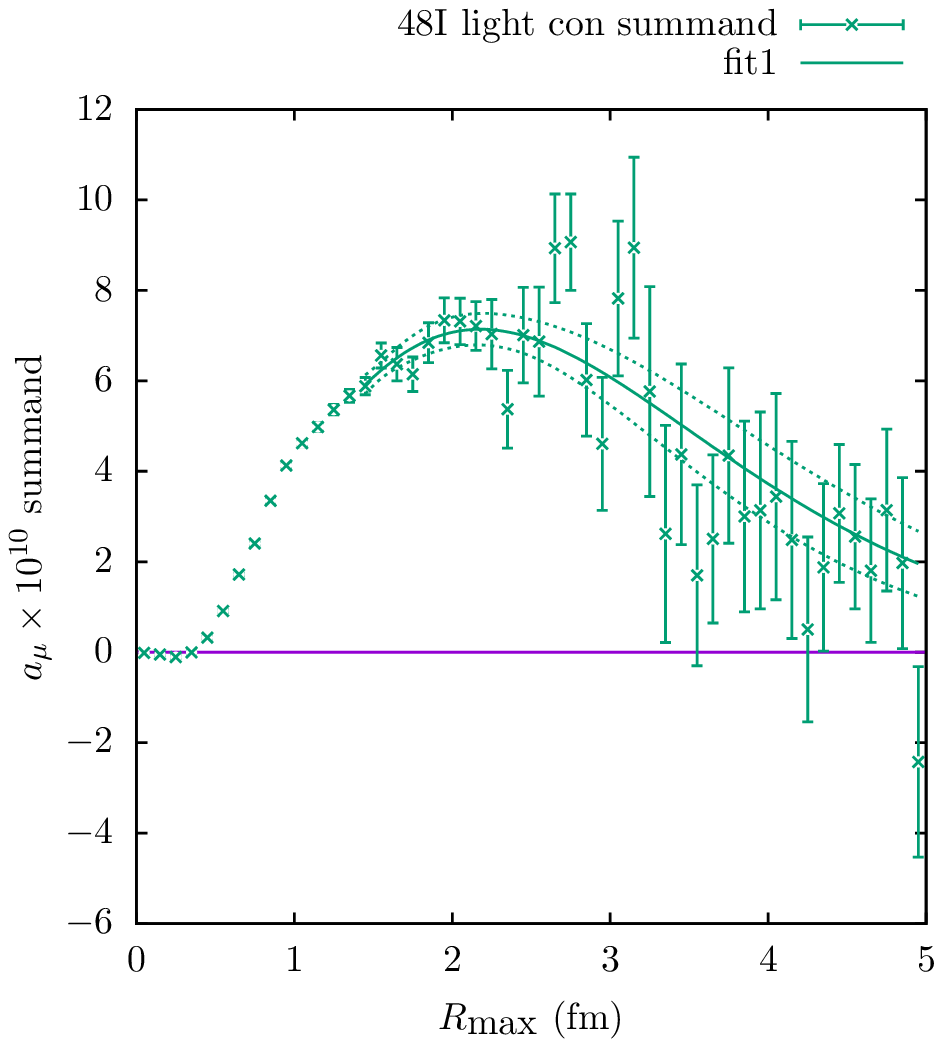}
\includegraphics[width=0.35\textwidth]{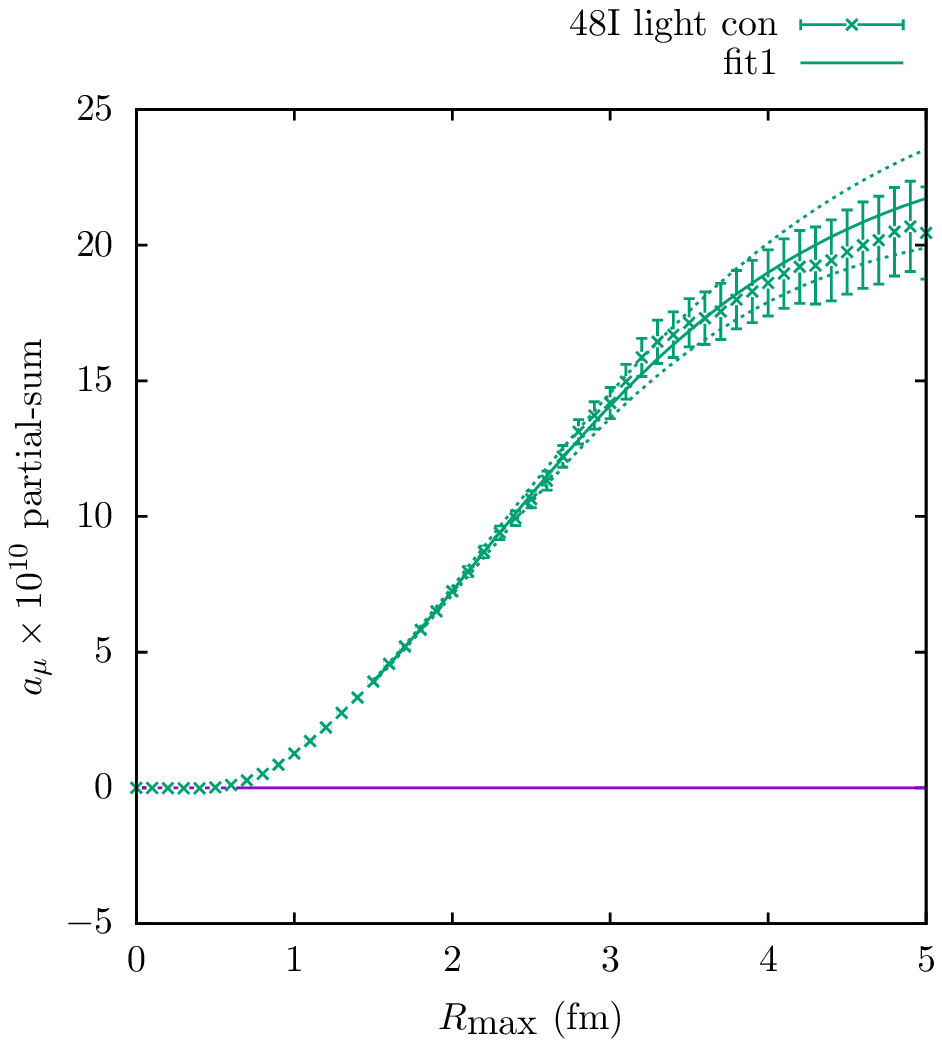}
\caption{\label{fig:48I-light-con-1}Similar to Fig.~\ref{fig:48I-light-no-pion} but for light quark connected contribution with the fit function in Eq.~(\ref{eq:fit-form-1}). The fit starts at $1.5~\mathrm{fm}$.}
\end{figure*}

\begin{figure*}[t]
\centering
\includegraphics[width=0.35\textwidth]{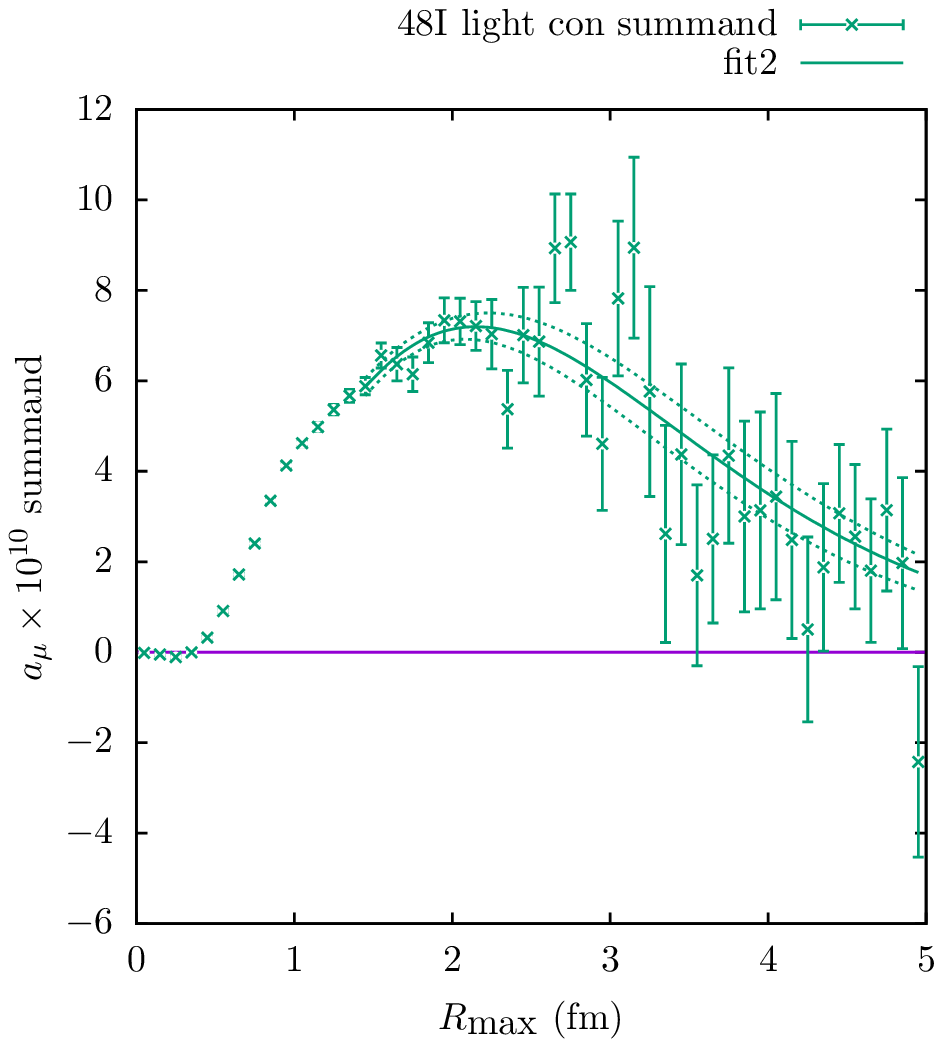}
\includegraphics[width=0.35\textwidth]{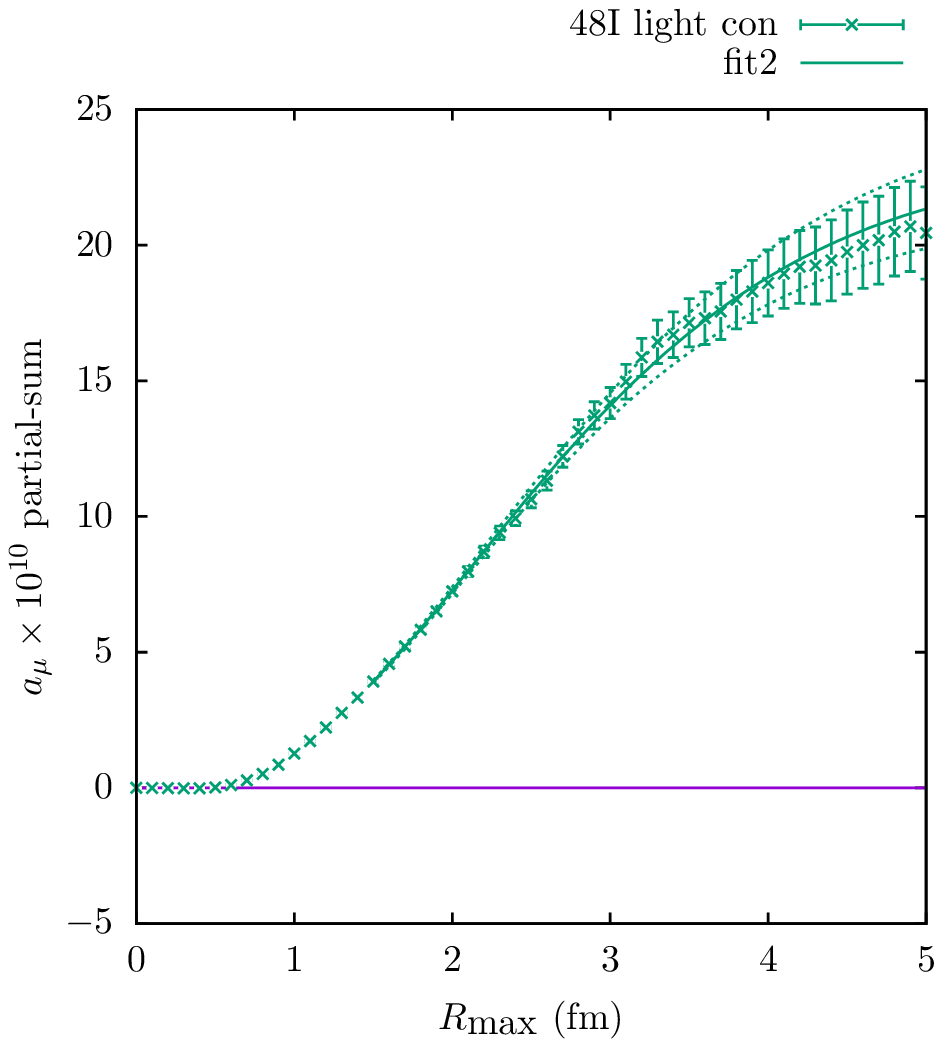}
\caption{\label{fig:48I-light-con-2}Similar to Fig.~\ref{fig:48I-light-no-pion} but for light quark connected contribution with a the fit function in Eq.~(\ref{eq:fit-form-2}). The fit starts at $1.5~\mathrm{fm}$.}
\end{figure*}

\begin{figure*}[t]
\centering
\includegraphics[width=0.35\textwidth]{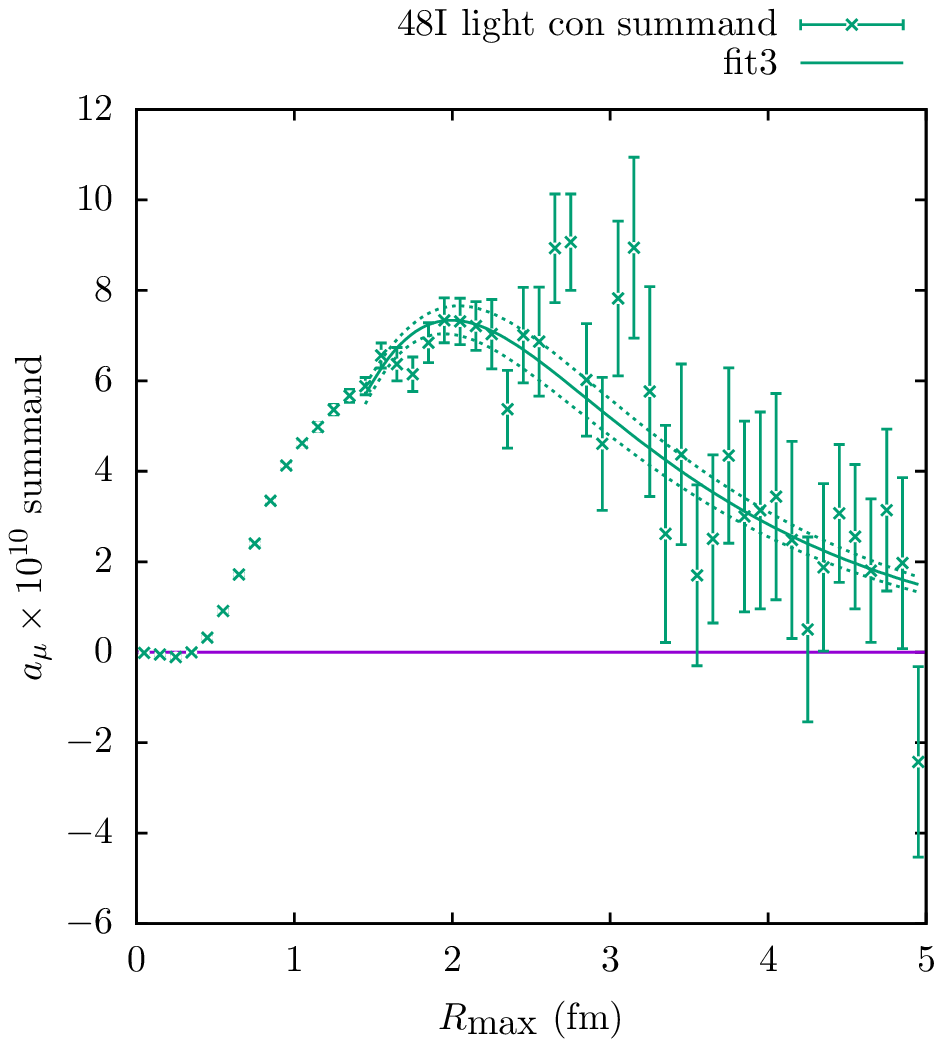}
\includegraphics[width=0.35\textwidth]{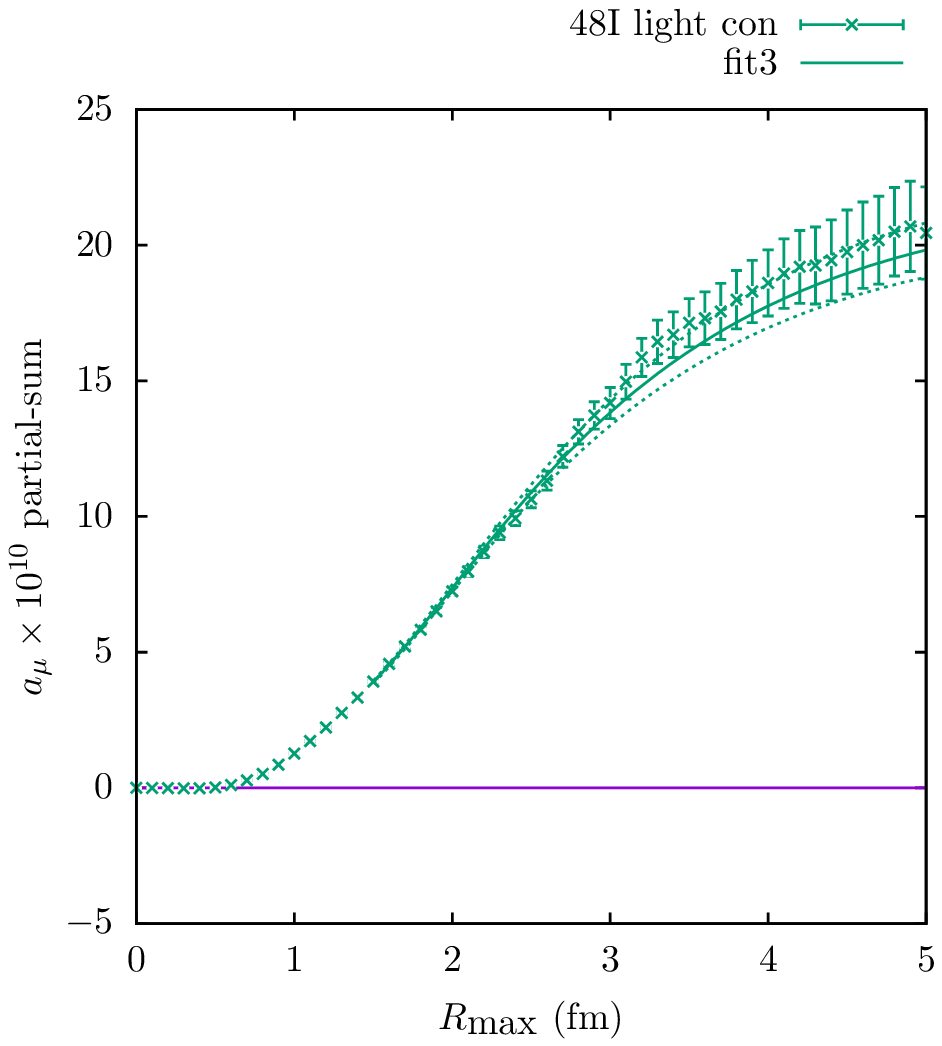}
\caption{\label{fig:48I-light-con-3}Similar to Fig.~\ref{fig:48I-light-no-pion} but for the light quark connected contribution with the fit function in Eq.~(\ref{eq:fit-form-3}). The fit starts at $1.5~\mathrm{fm}$.}
\end{figure*}

\begin{figure*}[t]
\centering
\includegraphics[width=0.35\textwidth]{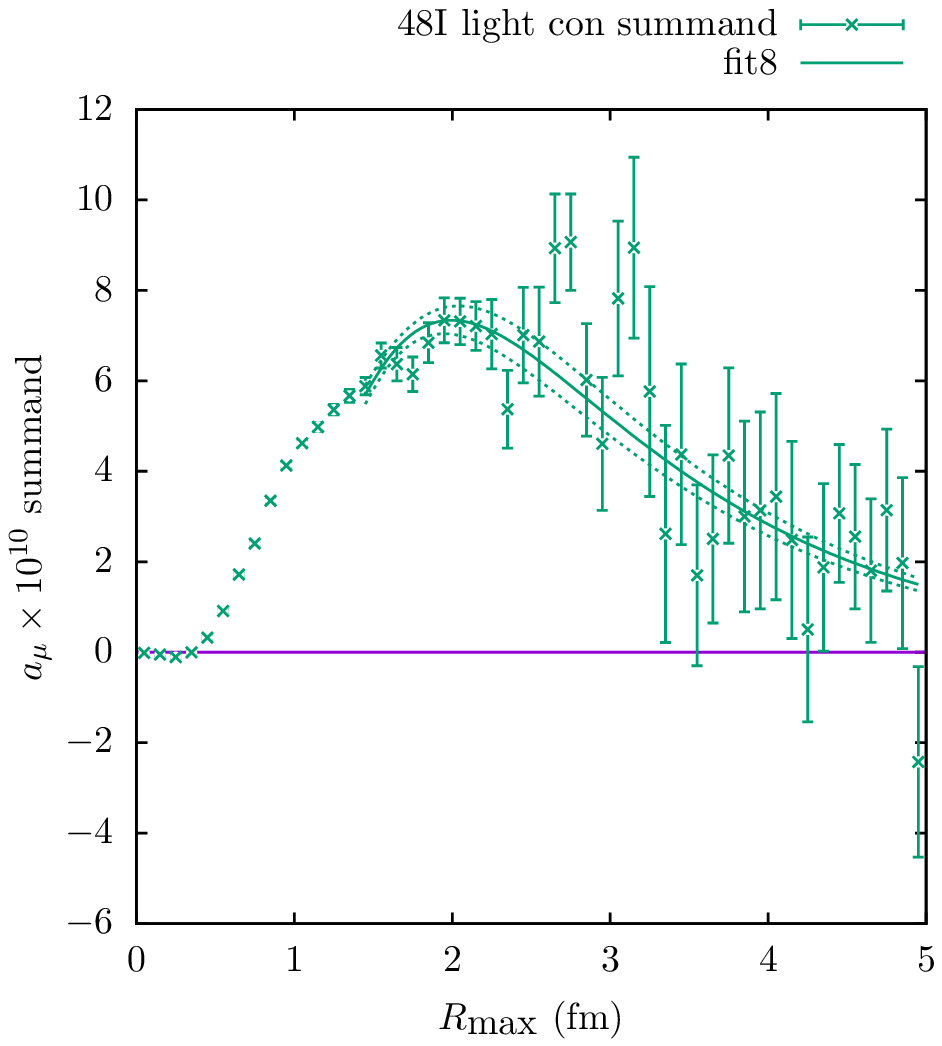}
\includegraphics[width=0.35\textwidth]{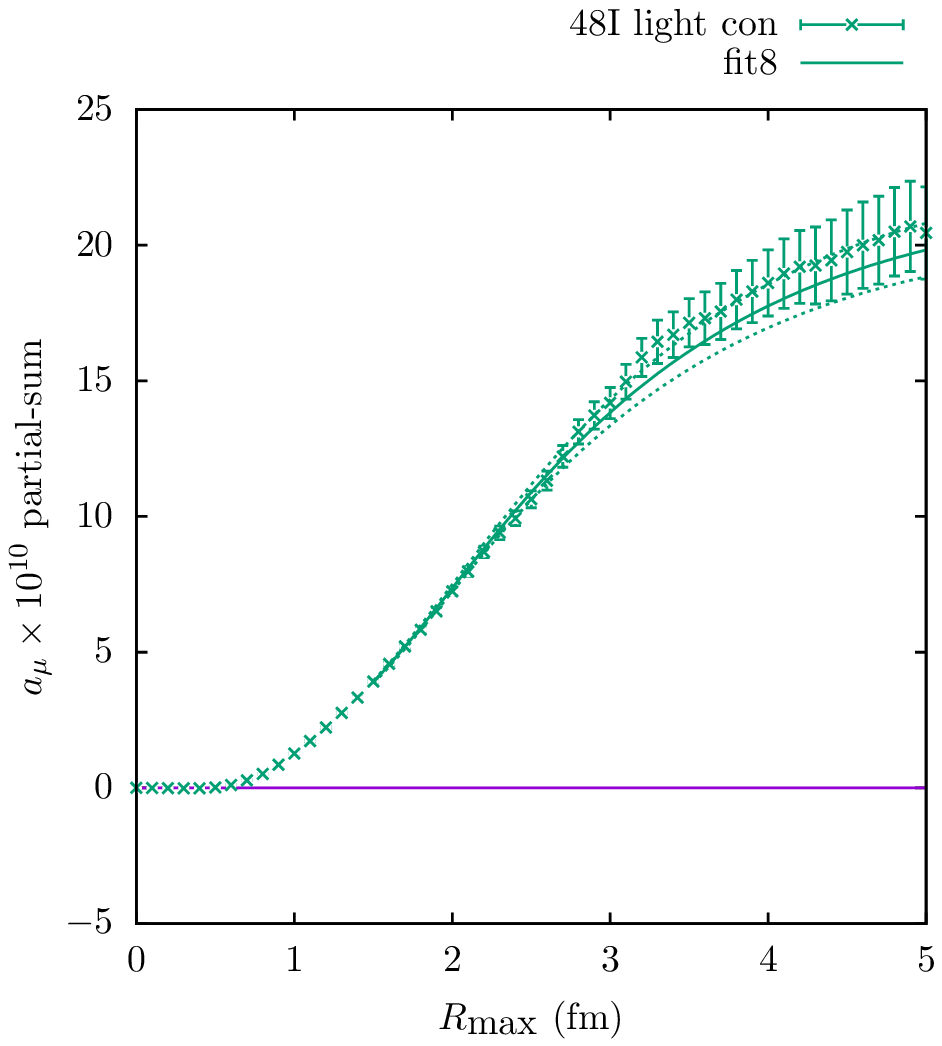}
\caption{\label{fig:48I-light-con-8}Similar to Fig.~\ref{fig:48I-light-no-pion} but for the light quark connected contribution with the fit function in Eq.~(\ref{eq:fit-form-8}). The fit starts at $1.5~\mathrm{fm}$.}
\end{figure*}

\begin{figure*}[t]
\centering
\includegraphics[width=0.35\textwidth]{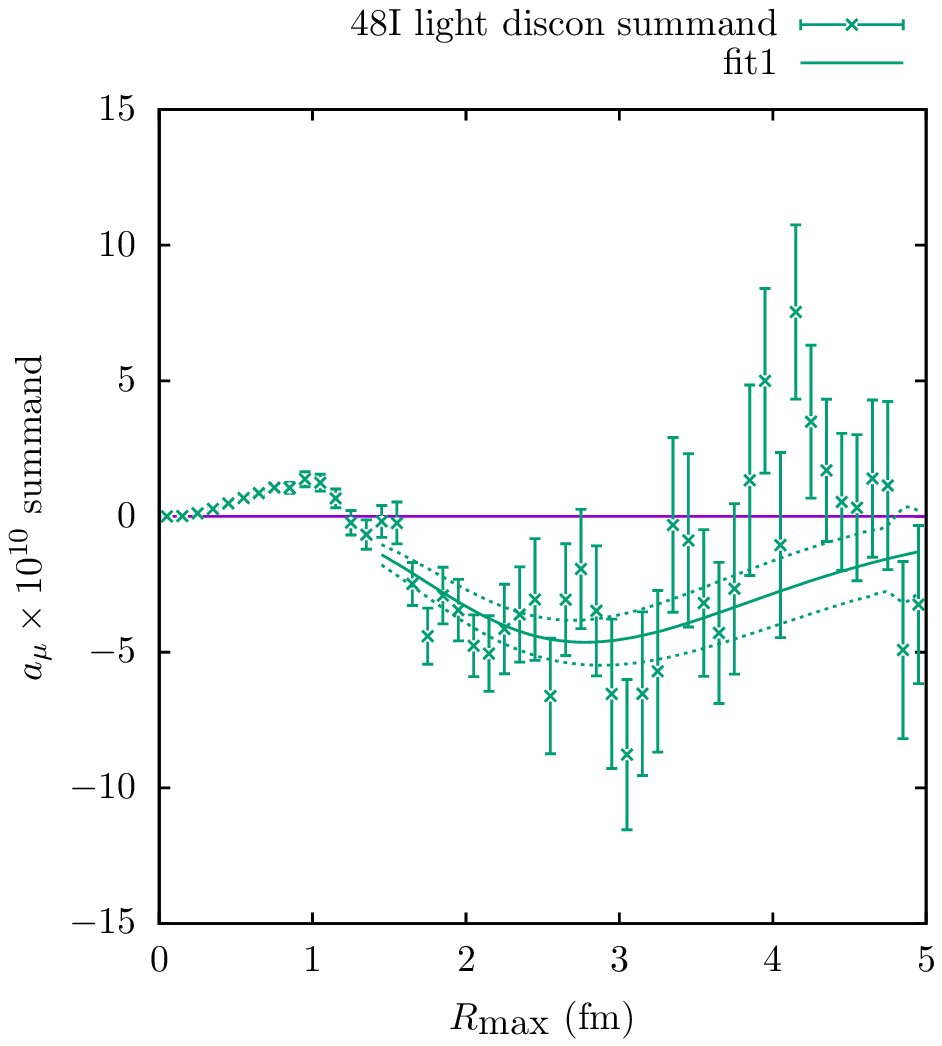}
\includegraphics[width=0.35\textwidth]{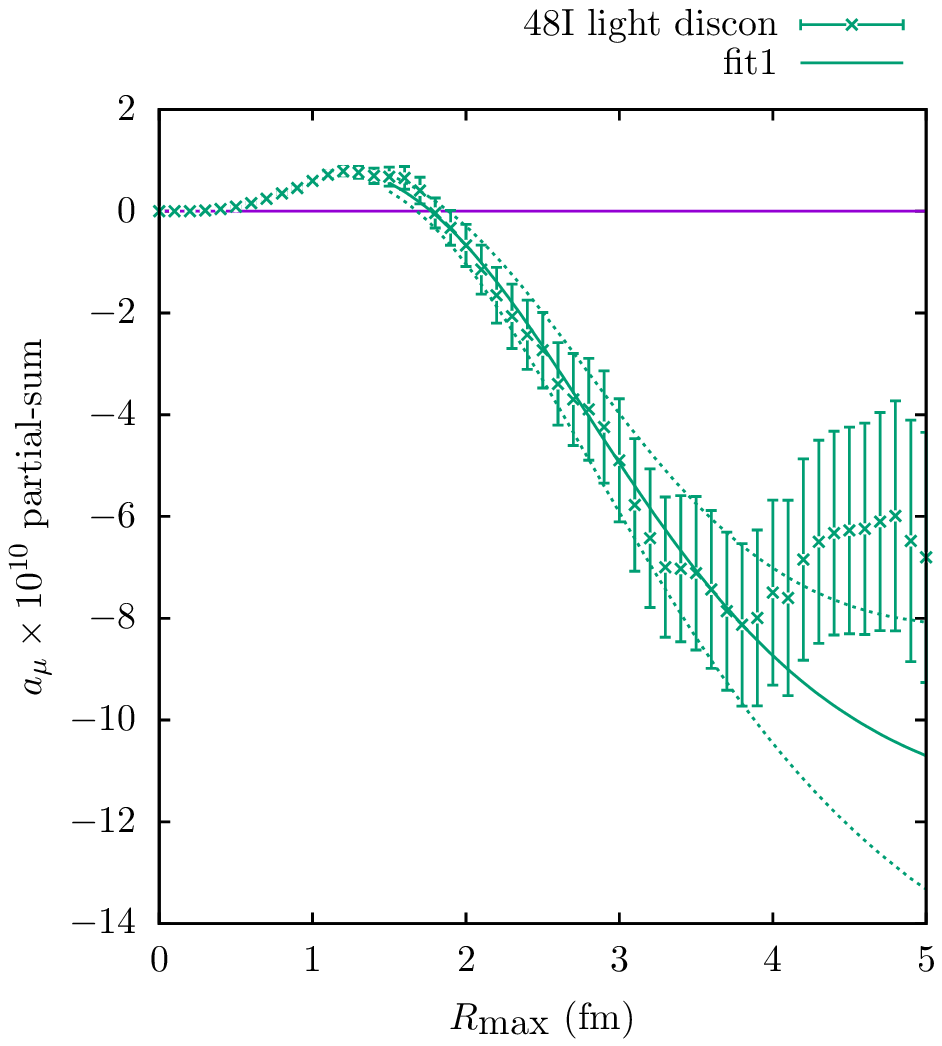}
\caption{\label{fig:48I-light-discon-1}Similar to Fig.~\ref{fig:48I-light-no-pion} but for the light quark disconnected contribution with the fit function in Eq.~(\ref{eq:fit-form-1}). The fit starts at $1.5~\mathrm{fm}$.}
\end{figure*}

\begin{figure*}[t]
\centering
\includegraphics[width=0.35\textwidth]{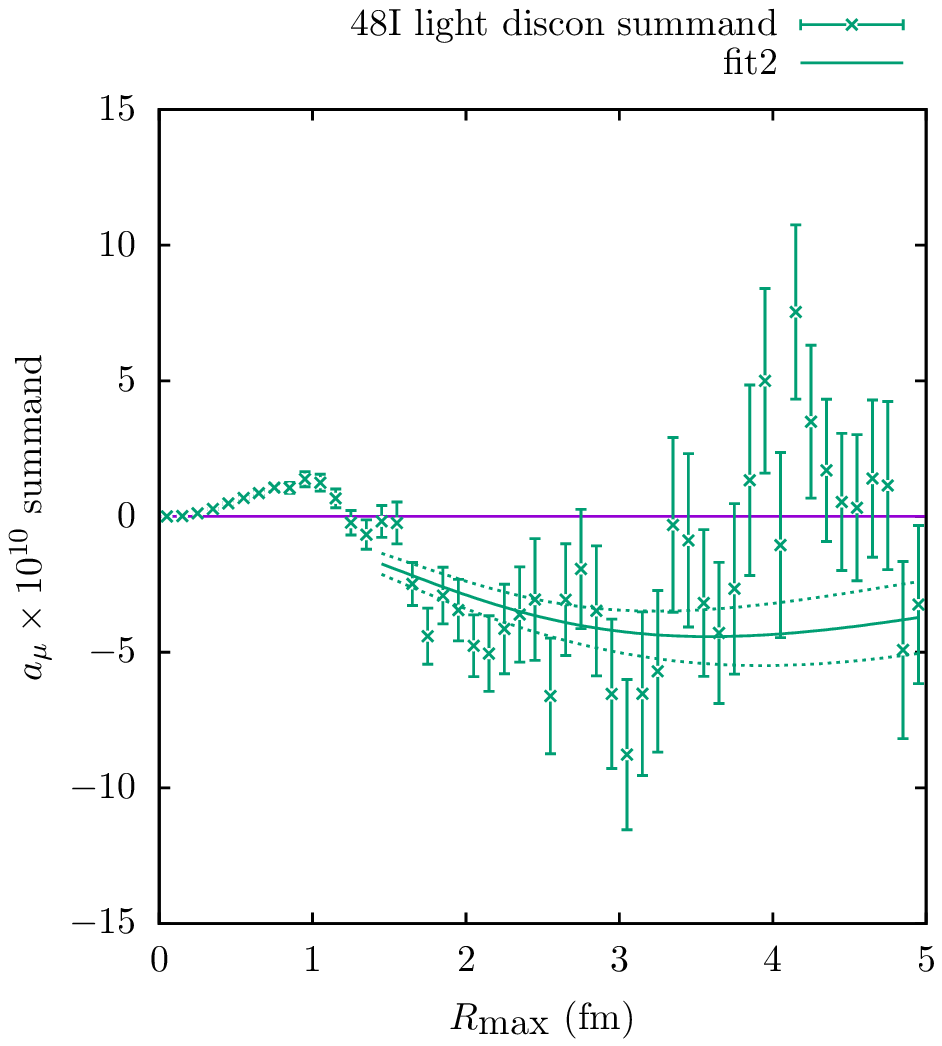}
\includegraphics[width=0.35\textwidth]{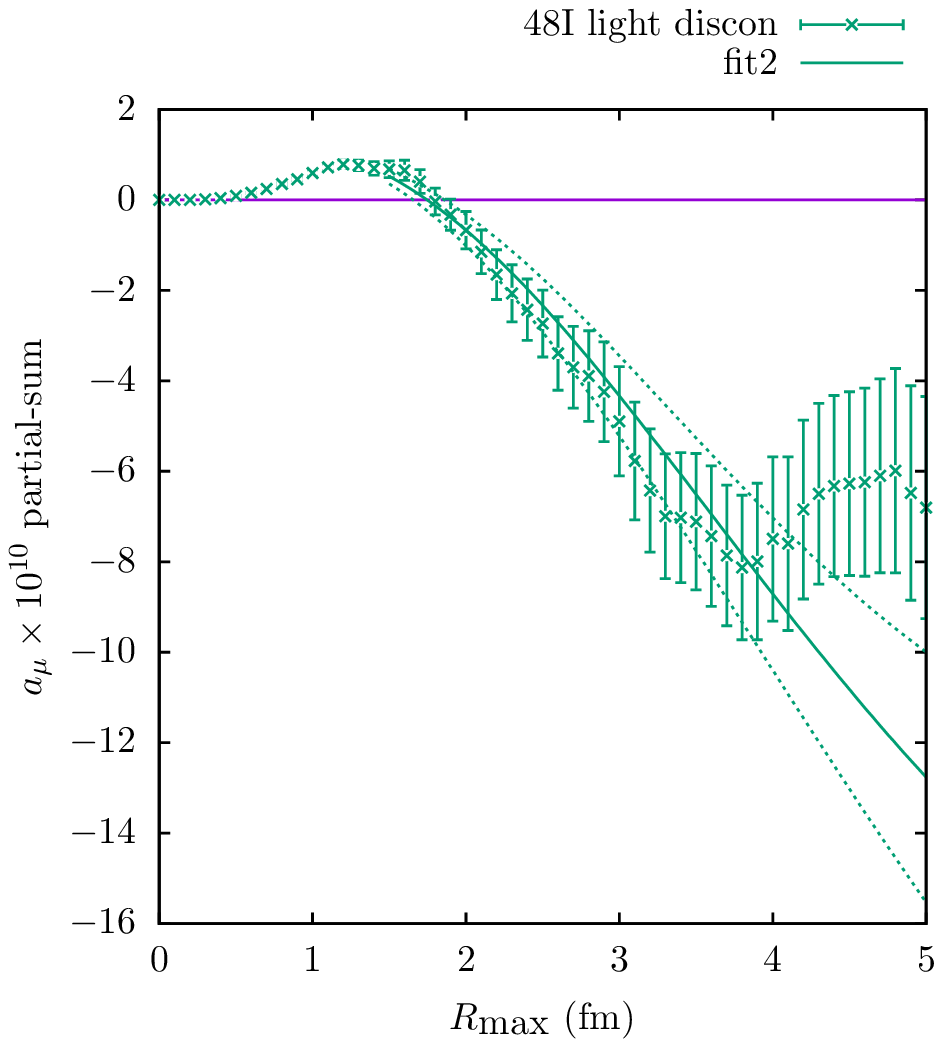}
\caption{\label{fig:48I-light-discon-2}Similar to Fig.~\ref{fig:48I-light-no-pion} but for the light quark disconnected conrtibution with the fit function in Eq.~(\ref{eq:fit-form-2}). The fit starts at $1.5~\mathrm{fm}$.}
\end{figure*}

\begin{figure*}[t]
\centering
\includegraphics[width=0.35\textwidth]{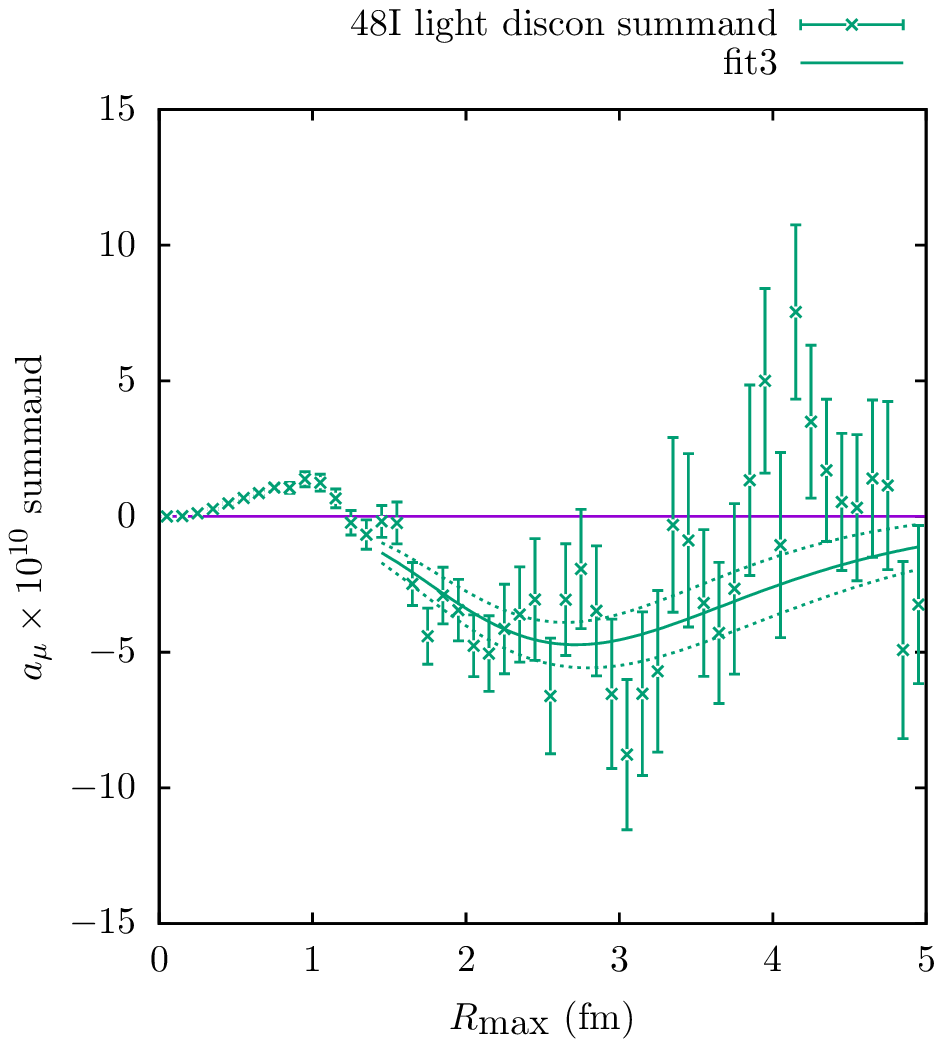}
\includegraphics[width=0.35\textwidth]{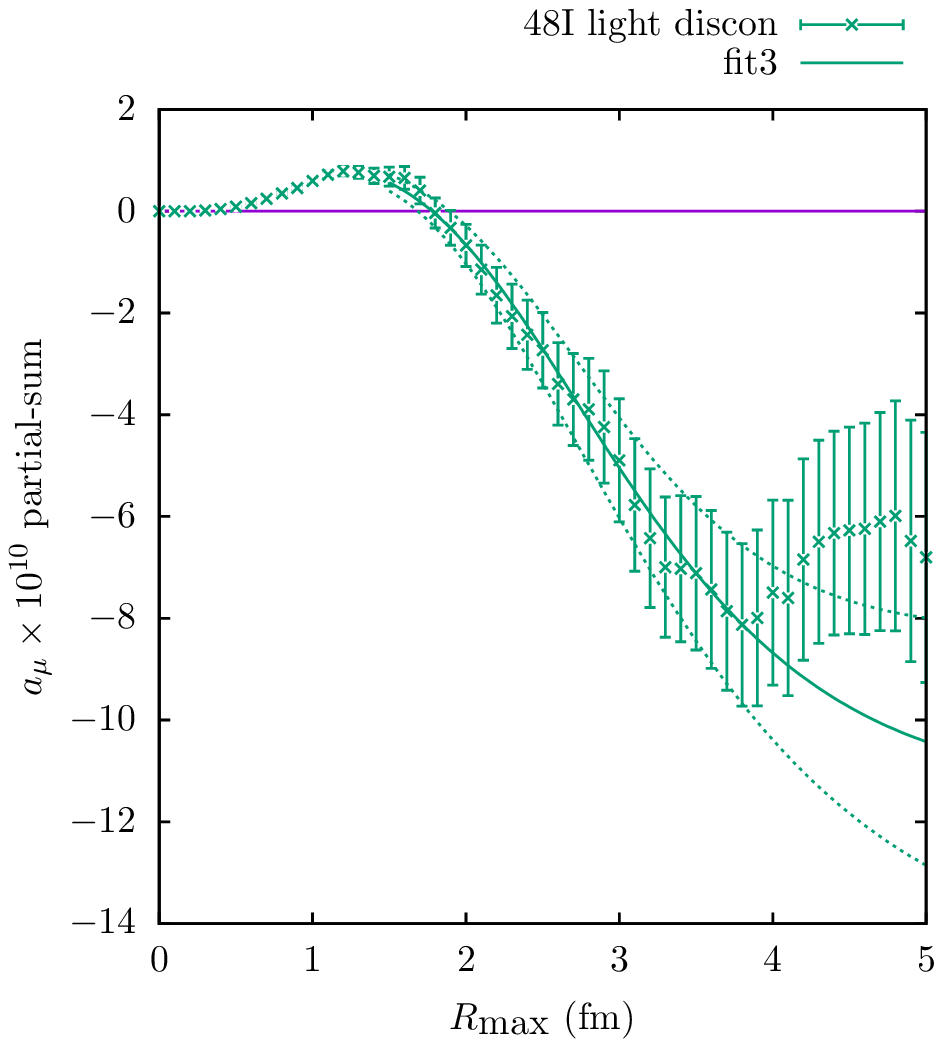}
\caption{\label{fig:48I-light-discon-3}Similar to Fig.~\ref{fig:48I-light-no-pion} but for the light quark disconnected contribution with the fit function in Eq.~(\ref{eq:fit-form-3}). The fit starts at $1.5~\mathrm{fm}$.}
\end{figure*}

\begin{figure*}[t]
\centering
\includegraphics[width=0.35\textwidth]{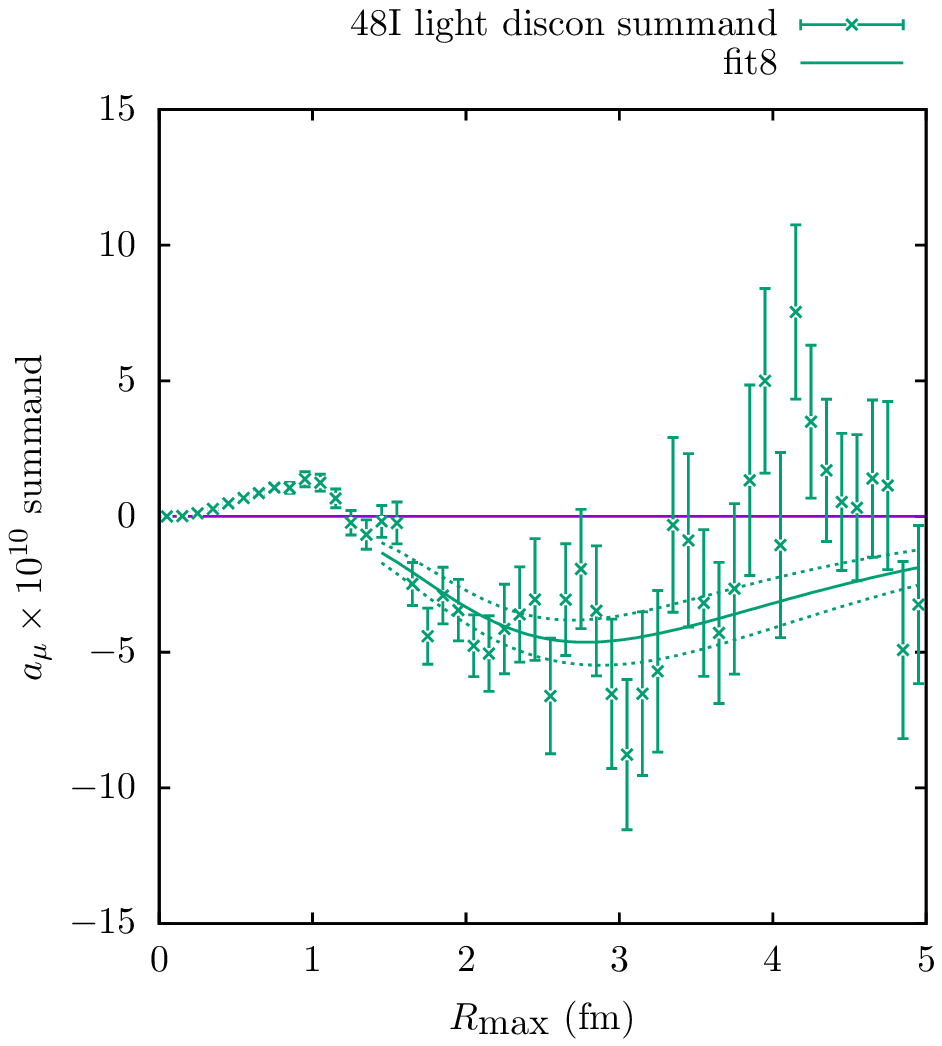}
\includegraphics[width=0.35\textwidth]{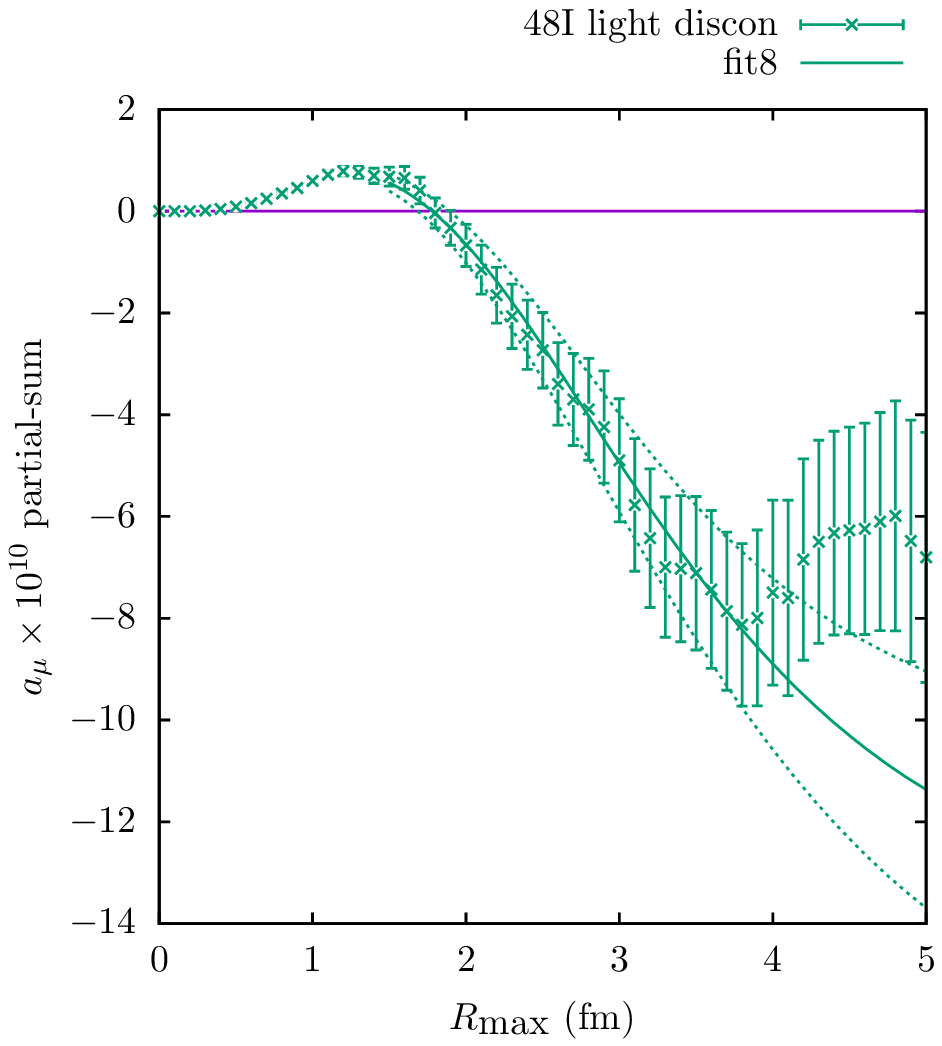}
\caption{\label{fig:48I-light-discon-8}Similar to Fig.~\ref{fig:48I-light-no-pion} but for the light quark disconnected contribution with the fit function in Eq.~(\ref{eq:fit-form-8}). The fit starts at $1.5~\mathrm{fm}$.}
\end{figure*}

We also apply these fit functions to different contributions,
including the strange quark connected diagram, the light quark connected diagrams only, and the light quark disconnected diagrams only.
Table~\ref{tab:48I-tail-fits} summarizes the fits.
The strange quark connected fits are displayed in Figs.~\ref{fig:48I-strange-con-1},\ref{fig:48I-strange-con-2},\ref{fig:48I-strange-con-3} and, for the light quark connected diagrams, in Figs.~\ref{fig:48I-light-con-1},\ref{fig:48I-light-con-2},\ref{fig:48I-light-con-3},\ref{fig:48I-light-con-8}. And finally,
the light quark disconnected results are shown in Figs.~\ref{fig:48I-light-discon-1},\ref{fig:48I-light-discon-2},\ref{fig:48I-light-discon-3},\ref{fig:48I-light-discon-8}.
The first fit function (fit1) defined in Eq.~(\ref{eq:fit-form-1}) fits the contribution from the strange quark connected diagrams,
which are statistically very precise, very well.
While the $\chi^2_\text{d.o.f.}$ for this fit is quite large, this is mostly due to the small
statistical error for the strange quark connected diagrams results.
The difference between the fit and data for these strange quark connected diagrams results
is much smaller than the needed precision for the fit for the light no-pion combinations.
Based on the comparison of qualities of all the fits performed,
we chose fit1 defined in Eq.~(\ref{eq:fit-form-1}) as the central value for the tail part of the light quark no-pion contribution
in our main calculation described in section~\ref{sec:light-contrib}.
For the light quark no-pion long-distance contributions, we observe roughly 100\%
variation about the central value (``fit1''), which is the reason
for the 100\% systematic uncertainty assigned to the contributions
``48I light no-pion $R_\text{max}>2.5\mathrm{fm}$''
and
``48I light no-pion $R_\text{max}>2\mathrm{fm}$''
in Table~\ref{tab:48I-light}.
Note that the ``fit3'' result for ``48I light no-pion'' has large statistical
uncertainty.
This is due to the fact that the data does not constrain the exponent $B$ very much,
in contrast to ``fit6'' and ``fit7'' which describe the data well.
Since the intermediate states should have at least energy $2 m_\pi$,
we view the results from ``fit7'' to be the upper bound for the fitting form ``fit3'',
which falls into the range of the results obtained with ``fit1'' and the 100\% systematic uncertainty based on the other fits.

From Tab.~\ref{tab:48I-tail-fits}, we observe that the fits to
the long distance parts of
``48I light con'' and ``48I light discon''
are very unreliable.
Note that the statistical error for the connected diagrams and disconnected diagrams
are almost independent.
If we fit the contributions from the connected and disconnected diagrams
individually and add the results,
the statistical error will approximately add up in quadrature.
The difficulty of fitting the contributions from the individual connected and disconnected diagrams is likely due to the long tail from $\pi^0$ exchange.
The large statistical and finite volume error in the long distance region make it hard to quantify the size of this long tail.
However, combining them to form the ``light no-pion'' contribution with Eq.~(\ref{eq:a-mu-no-pion})
cancels the long distance $\pi^0$ exchange contribution entirely, making the fit much more reliable.
Still, we found some fitting form dependence even after the cancellation of the $\pi^0$ exchange contribution,
which we accommodate with the 100\% systematic uncertainty estimated previously.

\begin{table*}[t]
\centering
\begin{tabular}{lc|c|c|c|ccc}
\hline
Contribution name & Form & fit range$/\mathrm{fm}$ & $\chi^2_\text{d.o.f.}$ & $p$-value & $A$ & $B$ & $C$ \\
\hline
48I light no-pion & fit1
            & 0.5 - 4.0 & 0.94 & 0.305
            & 130.57748 & 0.62982 & 0.66432 \\
48I light no-pion & fit2
            & 1.0 - 4.0 & 0.85 & 0.481
            & 102.07658 & 0.61515 & - \\
48I light no-pion & fit3
            & 0.5 - 4.0 & 1.20 & 0.110
            & 25.57747 & 0.24408 & 0.70365 \\
48I light no-pion & fit4
            & 0.5 - 4.0 & 1.04 & 0.274
            & 76.96882 & - & 0.55112 \\
48I light no-pion & fit5
            & 1.0 - 4.0 & 0.95 & 0.433
            & 69.98107 & - & - \\
48I light no-pion & fit6
            & 0.5 - 4.0 & 1.56 & 0.031
            & 275.93065 & - & 0.98944 \\
48I light no-pion & fit7
            & 0.5 - 4.0 & 1.18 & 0.160
            & 32.61837 & - & 0.73081 \\
\hline
48I strange con & fit1
            & 0.5 - 4.0 & 17.72 & 0.000
            & 18.07733 & 0.76754 & 0.64582 \\
48I strange con & fit2
            & 1.0 - 4.0 & 39.77 & 0.000
            & 13.62403 & 0.74617 & - \\
48I strange con & fit3
            & 0.5 - 4.0 & 79.03 & 0.000
            & 15.87426 & 0.53922 & 0.94749 \\
\hline
48I light con & fit1
            & 1.5 - 4.0 & 1.25 & 0.088
            & 12.77979 & 0.26616 & -0.59554 \\
48I light con & fit2
            & 1.5 - 4.0 & 1.20 & 0.168
            & 14.56237 & 0.27545 & - \\
48I light con & fit3
            & 1.5 - 4.0 & 1.67 & 0.024
            & 50.98630 & 0.13900 & 1.34340 \\
48I light con & fit8
            & 1.5 - 4.0 & 1.60 & 0.042
            & 50.98629 & - & 1.34340 \\
\hline
48I light discon & fit1
            & 1.5 - 4.0 & 1.17 & 0.162
            & -43.31953 & 0.32417 & 2.85847 \\
48I light discon & fit2
            & 1.5 - 4.0 & 1.48 & 0.060
            & -1.92375 & 0.16497 & - \\
48I light discon & fit3
            & 1.5 - 4.0 & 1.12 & 0.212
            & -495.03384 & 0.22956 & 2.81122 \\
48I light discon & fit8
            & 1.5 - 4.0 & 1.10 & 0.210
            & -72.47126 & - & 2.18553 \\
\hline
\end{tabular}
\vspace{0.3cm}
\\
\begin{tabular}{lc|rrr}
\hline
Contribution name & Form & $R_\text{max}>4\mathrm{fm}$ & $R_\text{max}>2.5\mathrm{fm}$ & $R_\text{max}>2.0\mathrm{fm}$ \\
\hline
48I light no-pion & fit1
            & $0.01(0.02)$ & $0.32(0.24)$ & $0.89(0.48)$ \\
48I light no-pion & fit2
            & $0.01(0.02)$ & $0.32(0.21)$ & $0.85(0.40)$ \\
48I light no-pion & fit3
            & $0.15(0.27)$ & $0.92(0.76)$ & $1.67(1.02)$ \\
48I light no-pion & fit4
            & $0.03(0.00)$ & $0.63(0.04)$ & $1.46(0.09)$ \\
48I light no-pion & fit5
            & $0.03(0.00)$ & $0.58(0.04)$ & $1.34(0.08)$ \\
48I light no-pion & fit6
            & $0.00(0.00)$ & $0.09(0.01)$ & $0.32(0.05)$ \\
48I light no-pion & fit7
            & $0.08(0.01)$ & $0.66(0.06)$ & $1.31(0.12)$ \\
\hline
48I strange con $\times 1000$ & fit1
            & $0.06(0.00)$ & $5.90(0.06)$ & $22.72(0.15)$ \\
48I strange con $\times 1000$ & fit2
            & $0.08(0.00)$ & $6.13(0.05)$ & $22.64(0.14)$ \\
48I strange con $\times 1000$ & fit3
            & $0.10(0.00)$ & $5.79(0.05)$ & $21.40(0.14)$ \\
\hline
48I light con & fit1
            & $4.96(2.11)$ & $13.14(3.04)$ & $16.69(3.08)$ \\
48I light con & fit2
            & $4.43(1.06)$ & $12.40(1.86)$ & $15.96(2.00)$ \\
48I light con & fit3
            & $4.14(0.48)$ & $11.01(1.00)$ & $14.54(1.17)$ \\
48I light con & fit8
            & $4.14(0.39)$ & $11.01(0.93)$ & $14.54(1.12)$ \\
\hline
48I light discon & fit1
            & $-3.13(4.06)$ & $-9.21(5.04)$ & $-11.20(5.07)$ \\
48I light discon & fit2
            & $-13.23(6.56)$ & $-19.61(7.61)$ & $-21.27(7.70)$ \\
48I light discon & fit3
            & $-2.76(2.33)$ & $-8.73(3.50)$ & $-10.78(3.57)$ \\
48I light discon & fit8
            & $-5.22(1.84)$ & $-11.46(3.08)$ & $-13.46(3.24)$ \\
\hline
\end{tabular}
\caption{\label{tab:48I-tail-fits}The long distance part of the
$a_\mu^\text{no-pion}\times 10^{10}$
as defined in Eq.~\ref{eq:a-mu-no-pion} from different fits to the 48I ensemble data.
The numbers in parentheses are statistical uncertainties.
We perform uncorrelated fit for the summand.
The minimized $\chi^2_\text{d.o.f.}$ is shown in the table.
We also calculated the $p$-value using the method invented in Ref.~\cite{Christ:2024nxz},
which allows a correct $p$-value determined for uncorrelated fits.
In addition to the light no-pion contribution, we also perform the relevant fit to
the data from
stanage quark connected diagrams,
light quark connected diagrams,
and light quark disconnected diagrams.
The contributions from the 48I strange quark connected diagrams are
much smaller compared with other quantities.
We multiply them by a factor of $1000$ to fit them in this table.
}
\end{table*}

\clearpage

\bibliography{ref}
    
\end{document}